# Electrical control of magnetism by electric field and current-induced torques

Albert Fert[1,2,3], Ramamoorthy Ramesh[4,5,6], Vincent Garcia[1], Fèlix Casanova[7,8] & Manuel Bibes[1]*

[1] Unité Mixte de Physique, CNRS, Thales, Université Paris-Saclay, 91767 Palaiseau, France.

[2] Dept. Advanced Polymers and Materials: Physics, Chemistry and Technology, Faculty of Chemistry, University of Basque Country (UPV/EHU), 20018 Donostia-San Sebastian, Basque Country, Spain

[3] Donostia International Physics Center (DIPC), 20018 Donostia-San Sebastian, Basque Country, Spain

[4] Department of Materials Science and Engineering, University of California, Berkeley, CA 94720, USA.

[5] Department of Physics, University of California, Berkeley, CA 94720, USA.

[6] Materials Sciences Division, Lawrence Berkeley National Laboratory, Berkeley, CA 94720, USA.

[7] CIC nanoGUNE BRTA, 20018 Donostia-San Sebastian, Basque Country, Spain.

[8] IKERBASQUE, Basque Foundation for Science, 48013 Bilbao, Basque Country, Spain.

* manuel.bibes@cnrs-thales.fr

# Abstract

The remanent magnetization of ferromagnets has long been studied and used to store binary information. While early magnetic memory designs relied on magnetization switching by locally generated magnetic fields, key insights in condensed matter physics later suggested the possibility to do it by electrical means instead. In the 1990s, Slonczewzki and Berger formulated the concept of current-induced spin torques in magnetic multilayers through which a spin-polarized current generated by a first ferromagnet may be used to switch the magnetization of a second one. This discovery drove the development of spin-transfer-torque magnetic random-access memories (STT-MRAMs). More recent fundamental research revealed other types of current-induced torques named spin-orbit-torques (SOTs) and will lead to a new generation of devices including SOT-MRAMs and skyrmion-based devices. Parallel to these advances, multiferroics and their magnetoelectric coupling, first investigated experimentally in the 1960s, experienced a *renaissance*. Dozens of multiferroic compounds with new magnetoelectric coupling mechanisms were discovered and high-quality multiferroic films were synthesized (notably of $BiFeO_3$), also leading to novel device concepts for information and communication technology such as the MESO transistor (MESO stands for magneto-electric spin-orbit). The story of the electrical switching of magnetization, which we review in this article, is that of a dance between fundamental research (in spintronics, condensed matter physics, and materials science) and technology (MRAMs, MESO, microwave emitters, spin-diodes, skyrmion-based devices, components for neuromorphics, etc). This *pas de deux* led to major scientific and technological breakthroughs over the last decades (such as the conceptualization of pure spin currents, the observation of magnetic skyrmions, or the discovery of spin-charge interconversion effects). As a result, this field has not only propelled MRAMs into consumer electronics products but also fueled discoveries in adjacent research areas such as ferroelectrics or magnonics. In this review, we cover recent advances in the control of magnetism by electric fields and by current-induced torques. We first review fundamental concepts in these two directions, then discuss their combination, and finally present various families of devices harnessing the electrical control of magnetic properties for various application fields. We conclude by giving perspectives in terms of both emerging fundamental physics concepts and new directions in materials science.

# Table of contents

# 1. Introduction

## 1.1. Macroscale perspective

The macro-systems perspective for this article is based on the field of information technologies. Microelectronics components and systems form an ever-increasing backbone of our society, pervading many parts of our daily life, for example through a host of consumer electronics systems, providing sensing, actuation, communication, and processing and storage of information. All of these are built upon an approximately $470B/year global market that is exponentially growing at a pace of 10-15% annually [1,2]. Many of such components likely started as materials physics research ideas, often first discussed within the confines of physics and materials conferences worldwide. A few emerging global phenomena will likely completely change this microelectronics landscape. The first among them is the "Internet of Things" (IoT), which is the network of physical devices, transportation systems, appliances, and other items embedded with electronics for sensing/actuating, computing, storage and communications functions, illustrated in **Figure 1.** As an example, a modern automobile has a large number of sensing, communicating and computing components embedded and this is only going to increase; for example, the emergence of autonomous vehicles will require orders of magnitude higher levels of computing, with sustainable power consumption.

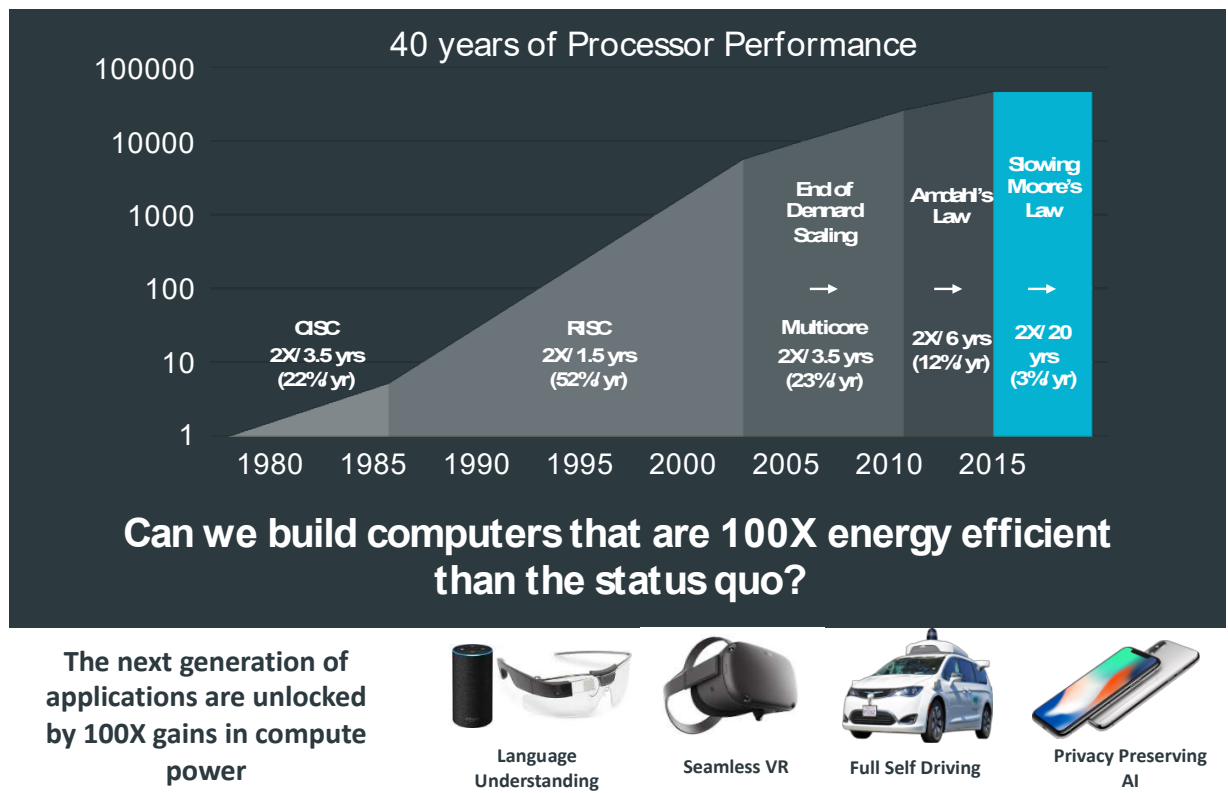

*Figure 1. A schematic illustrating the emergence of the "Internet of Things" and Machine Learning/ Artificial Intelligence as macroscale drivers for the Beyond Moore's Law R&D. Describes the leveling off the various scaling laws (Dennard's Law states that as the dimensions of a device go down, so does power consumption; Amdahl's law is a principle that states that the maximum potential improvement to the performance of a system is limited by the portion of the system that cannot be improved) as a function of time, leading to the end of Moore's Law.*

The second major phenomenon is the emergence of machine learning (ML) / artificial intelligence (AI), that is taking the technology world by storm. It uses a large amount of computing and data analytics which, in turn, provides the system the ability to "learn" and do things better without human intervention. Of relevance to us is the fact that microelectronic components are critical underpinnings for this field.

We can now ask the question: how do these macroscale phenomena relate to microelectronics and, more importantly, to new materials and physics underpinning them? Stated differently, what can *materials physics* do to enable this coming paradigm shift? To put this into perspective, we now need to look at the fundamental techno-economic framework that has been driving the microelectronic field for more than five decades. The well-known "Moore's Law" [3], the techno-economic principle that has so far underpinned the field of microelectronics through the scaling of CMOS-based transistors is displayed in **Figure 2** (CMOS stands for complementary metal oxide semiconductor). Broadly, it states that the critical dimensions of the CMOS transistor shrink by 50% every 18-24 months. At their inception, CMOS transistors were "macroscopic" with the critical dimension well over 1 µm and Dennard scaling provided a path to shrinking such transistors, while keeping the power density constant [4]. Today, this power scaling is no longer possible while the critical dimensions of modern transistors have entered sub-10 nm scales, the point at which both the fundamental science (*i.e.*, classical electron dynamics) is no longer sufficient to adequately describe the physics of the transistor and ever more complex manufacturing issues must be addressed. Therefore, in the past decade or so, there has been an ever-increasing sense that something has to be done about this issue [5–9].

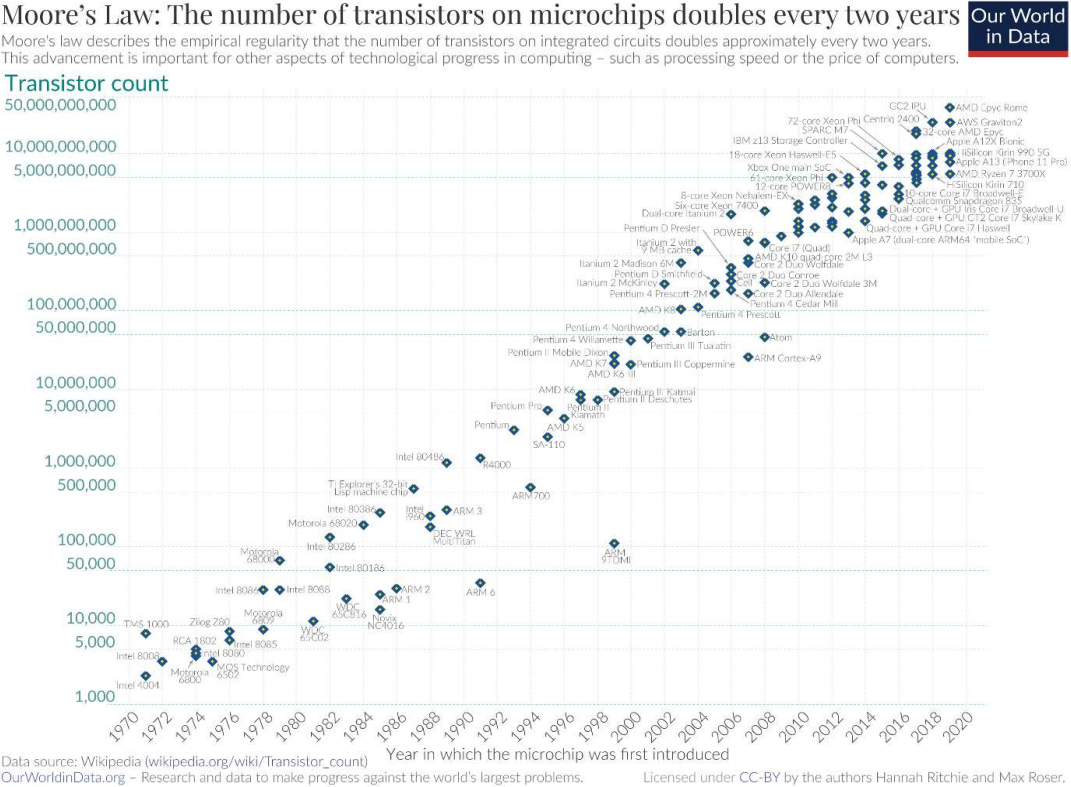

*Figure 2. Moore's law: evolution of the number of transistors per chip over time* [10].

What is needed to mitigate this major issue now is a paradigm shift similar to the introduction of CMOS technology to replace bipolar transistors in the 1990s [11–13], cf **Figure 3**. One can explore many pathways to address this impending crisis. In some sense, this is a matter of perspective: circuit design engineers may prefer to go to specialized architectures [14] or pivot from the conventional boolean or von Neumann architecture into a neuromorphic architecture [15]. Another pathway could be to go away from highly deterministic computing (which tolerates errors at the scale of $10^{-10}$ to $10^{-12}$) to more of a stochastic computing. The third way overtly involves "Quantum Materials", materials in which quantum mechanical effects such as exchange interaction or spin-orbit coupling directly lead to exotic physical phenomena (to start with magnetism, ferroelectricity, multiferroic behavior, and more recently topological behavior arising from band topology). We get to this after a short description of another looming challenge, namely energy or more specifically, energy efficiency in computing and how it impacts the global energy consumption in microelectronic systems.

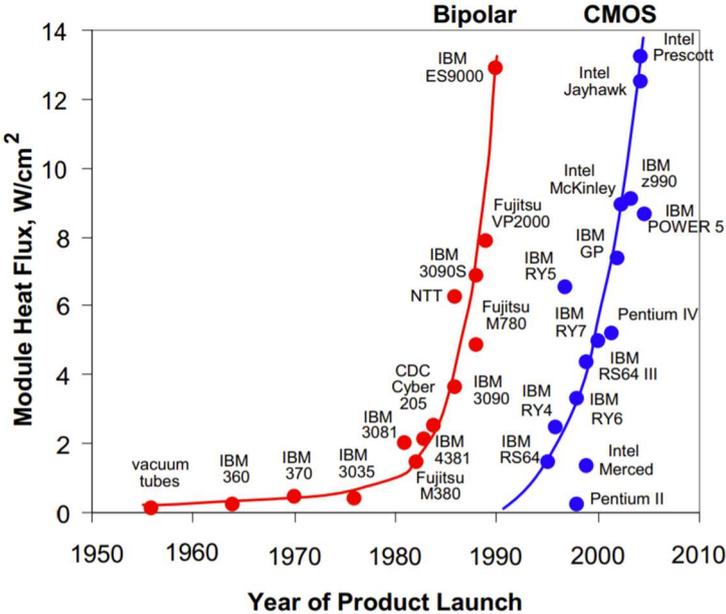

*Figure 3. Heat output over time for bipolar and CMOS transistor chips [11].*

In today's CMOS transistor, the energy consumed per logic operation is of the order of 10-100 pJ for a typical 32-transistor logic circuit. It is noteworthy that at the single transistor level, the energy consumption in state of the art transistors is ~50 aJ; however, the design of logic circuits involving a large number of such transistors leads to the eventual energy/logic operation. In this sense, a reduction in the number of transistors required to perform logic operations and/or moving to capactive elements (as in magneto-electric spin orbit, MESO, devices [8], discussed Section 5.1.4) could also reduce the number of building blocks required to perform the logic operations. If we assume that there is no change to this number in the near future, and at the same time the demand for and consumption of microelectronic components in the Internet of Things, Artificial Intelligence (AI) and machine learning is predicted to increase, the total energy consumption in all of microelectronics could grow to ~20% of primary energy by 2030, cf **Figure 4**. At this scale, microelectronics would become a significant part of the worldwide energy consumption and thus deserves to be addressed from the energy efficiency perspective as well.

The end of the conventional Si-CMOS based Moore's law thus emerges as a fantastic opportunity to explore pathways for Beyond Moore's Law architectures. Indeed, the past decade has witnessed innovations at multiple levels. Particularly, there have been a large number of fundamental physics-based innovations in spintronics and spin-based devices. Thus, if pathways are found to reduce their energy consumption, notably to control magnetization, then this presents an exciting opportunity to create the next generation of computing paradigms. This includes logic-in-memory architectures departing from Von Neuman's archictures by embedding memory and logic, thereby removing the energy-costly transfer of data beween separated memory and computing units.

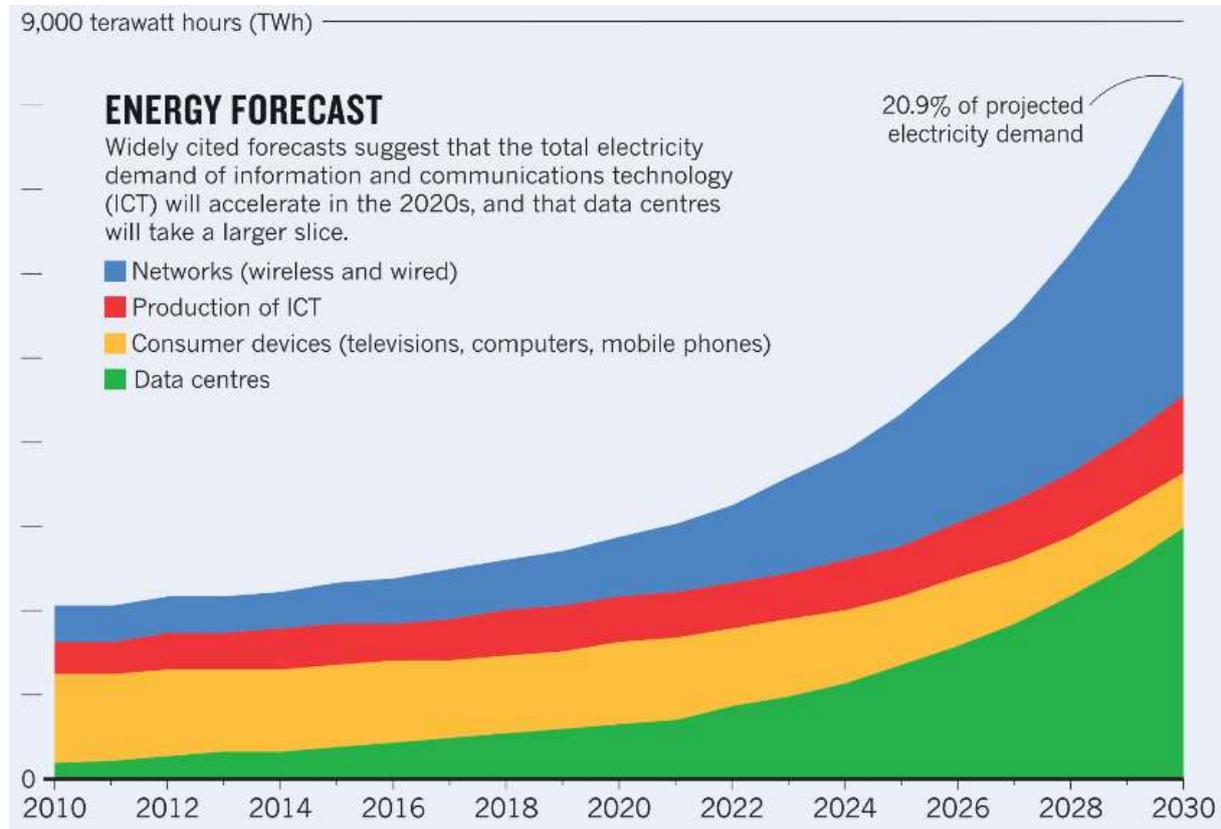

*Figure 4. Energy consumption of information and communication technology systems over time* [16]

## 1.2. The need for a paradigm shift and for new materials

We begin the exploration of new materials physics by going back to the fundamentals of CMOS devices. CMOS transistors utilize a gate voltage to control the flow of current between the source and the drain. By adjusting the energy bands in the semiconducting channel, the gate voltage either permits the movement of electrons (the 'on' state) or obstructs it (the 'off' state). However, the electron energies from the source are spread out at finite temperatures. Consequently, there exists a finite density of electrons with sufficiently high energy to surpass the barrier that would otherwise impede their journey between the source and drain in the 'off' state. This leakage current leads to energy wastage. According to fundamental thermodynamic principles, reducing this current by a factor of 10 necessitates increasing the barrier by approximately 60 meV at room temperature [9]. However, to prevent energy wastage caused by leakage current, the current must be reduced by a factor of at least 100 000, requiring a minimum barrier of 300 meV. Consequently, a minimum gate voltage of at least 300 mV becomes necessary. This minimum gate voltage establishes a lower limit on switching energy. This limitation is referred to as 'Boltzmann's tyranny,' named after Ludwig Boltzmann, who elucidated the spreading of particle energies due to temperature. Boltzmann's tyranny is believed to restrict the extent to which the operating gate voltage can be reduced for a transistor, irrespective of the material used.

In recent years, the community realized that this Boltzmann's tyranny needs to be addressed – setting the stage for new materials and new phenomena, with a view towards designing entirely new computing building blocks to replace CMOS transistors operating at low voltage and dissipating much less power. One proposed pathway identifies the broad class of quantum materials, for instance materials exhibiting a metal-to-insulator transition [17] or those possessing a ferroic order such as ferromagnets or ferroelectrics. In these compounds the exchange energy (in ferromagnets) or the dipolar energy (in ferroelectrics) makes the spins or the dipoles align collectively *without the need* for an external source of energy (such as an applied field). Thus, if one could use a spontaneous magnetic/dipole moment as the primary order parameter rather than electronic charge in a CMOS device, one could take advantage of such internal collective order to reduce the energy consumption. Indeed, this is the premise behind two recent research articles [2,8], where the rudiments of a possible magneto-electric spin orbit (MESO) coupled memory-logic device are discussed. As we will see in this review article, harnessing the electric-field control of magnetism offers promising opportunities to realize ultralow power, beyond-CMOS computing devices.

### 1.3. Magnetism and spintronics

While magnetic phenomena have been known since ancient times, spintronics is a relatively new field of electronics that not only acts on the charge of electrons, but also their spin. The field of spintronics was initially sparked by the discovery of the giant magnetoresistance (GMR) in magnetic multilayers in 1988 [18,19], which introduced new concepts for utilizing spin-polarized currents and demonstrated potential applications for spin-based technology. In the early days of spintronics, spin-polarized currents were generated by utilizing the influence of the orientation of spin on the transport properties of electrons in ferromagnetic conductors. This influence, which was first suggested by Mott [20], had been experimentally demonstrated and theoretically described a decade before the GMR discovery [21–23]. This method of generating spin-polarized currents was used in "classical spintronics" during the first decade after the GMR discovery. Major advancements during this time included the discovery of the tunneling magnetoresistance and spin transfer torque. Additionally, important concepts such as spin accumulation and pure spin current (a current of spin without a current of charge) were introduced. In more recent times, it has become possible to produce spin-polarized currents and pure spin currents without using magnetic materials by utilizing spin-orbit interactions in non-magnetic materials, which is known as spin-orbitronics. Today, spintronics is expanding in various directions, with promising new areas of research including spintronics with topological systems, such as the interface states of topological insulators, and spintronics with magnetic skyrmions.

The idea that magnetism could be used to store digital information dates back to the 1950s and the development of soft-core ferrite-based memories [24]. In these destructive read-out devices, magnetic tori made of ferrites were organized into an array and magnetized in one or the other direction by the magnetic field produced by currents running in two perpendicular electrical wires passing through each torus. This technology remained the dominant random-access computer memory until the introduction of semiconductor memory in the late 1960s which allowed for both an increase in density and a decrease in cost. Magnetic disk technology appeared in the 1960s as well and led to the development of hard-disk drives and floppy disks. The write process involved passing a current into an electromagnetic write-head, generating a local magnetic field. Initially, the read-out process was based on magnetic induction but, in 1990, IBM introduced read heads relying on anisotropic magnetoresistance (AMR), pioneering a new method to sense magnetization (*M*) through its influence on electrical transport. The discovery of GMR in 1988 [18,19] prompted the development of GMR-based read heads that replaced AMR-based ones in 1997, marking the beginning of spintronic-based technologies. However, magnetic information writing continued to rely on the generation of local magnetic field by electrical current. The Oersted field produced by current running through perpendicular current lines – as in soft-core memories – was also the method used to write information in the first prototype of magnetic random-access memories (MRAMs) that was announced in 1995 [25] and released in 2006. In today's generation of the STT-MRAMs, on the market since 2019, writing has become purely electrical thanks to the use of the Spin Transfer Torque (STT) mechanism for the conversion of spin-polarized current into torques acting on the magnetization. Several companies have announced that replacing the Flash memories by STT-MRAMs in computers or phones reduces the energy consumption and increase the speed by large factors. Commercial products are already on the market. The next generation will be the SOT-RAMs which exploit pure spin current induced by Spin-Orbit (SO) in heavy or topological materials and the resulting Spin-Orbit Torques (SOT), see 5.1.1.

With this as the technological background, in this article we review the efforts in the endeavor focusing on controlling magnetism not by magnetic field but by electrical means, namely voltage and electric current. Research in this field has been fueled by advances in condensed matter physics and materials science, along directions that remained parallel for several decades. As we will see, the research on multiferroics and magnetoelectrics started in the 1960s but remained rather confidential for 40 years, while spintronics brilliantly entered the stage with the discovery of GMR in 1988. Both fields developed nearly without interacting until the 2000s and the rediscovery of multiferroic materials and magnetoelectric coupling. Magnetoelectric coupling precisely aims to achieve an electrical control of magnetization, mostly using multiferroics, and its revival prompted the development on voltage-controlled magnetic anisotropy in classical spintronic devices such as magnetic tunnel junctions (MTJs), not involving magnetoelectric or multiferroic materials *per se*.

### 1.4. Magnetoelectric coupling and multiferroics

Multiferroics exhibit more than one primary ferroic ordering (*i.e.*, ferromagnetism, ferroelectricity, ferroelasticity, or ferrotoroidicity) in the same phase [26], cf.**Figure 5**. This terminology is usually extended to include other types of order such as antiferromagnetism as well as composites of individual ferroics, and is most often used to refer specifically to magnetoelectric [27] materials combining ferroelectric and magnetic behavior in a single phase. The co-existence of ferroic orders can lead to coupling between them, so that one ferroic property can be manipulated with the conjugate field of the other [28]. A good example of a multiferroic is the case of ferromagnetic shape memory alloys (FSMA), which exhibit ferromagnetism along with a spontaneous strain [29]. In contrast, the coexistence of spin and charge orders (particularly ferromagnetism and ferroelectricity) is challenging, since ferroelectricity requires an insulator while typical ferromagnets require electronic exchange interactions [30]. Many insulating magnets are either antiferromagnets or ferrimagnets (driven by super-exchange interactions); ferrimagnets are antiferromagnets with uncompensated magnetic sublattices and thus possess a finite magnetization. Therefore, progress in multiferroic research requires (i) understanding the electronic structure at the most fundamental level, (ii) new material chemistries to implement them, (iii) the development of new tools to compute and characterize the novel properties associated with the coupled behaviors and (iv) new approaches to synthesize such materials with atomic-scale precision. When this is successful, it presents possible routes to entirely new device architectures [31–33], as exemplified by Intel's MESO device [8]. The field of multiferroics is now vast and we would direct the reader to other recent reviews with different emphases [34–42] to complement what we present in this article.

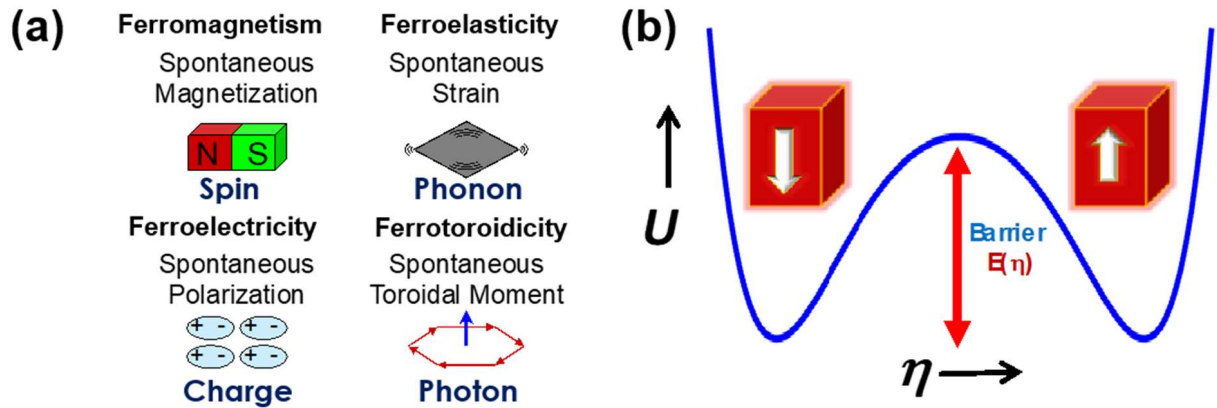

*Figure 5. Fundamental taxonomy of solid-state order parameters. (a) Emergence of ferromagnetism due to spontaneous time reversal symmetry breaking; ferroelectricity due to spontaneous spatial inversion symmetry breaking; ferroelasticity which is characterized by a spontaneous strain and ferrotoroidicity which breaks both time and spatial inversion symmetry [43]. Coexistence of at least two order parameters defines multiferroics and coupling between them leads to magnetoelectricity, piezoelectricity and piezomagnetism. (b) Scheme of a classical double-well energy U landscape that characterizes the emergence of the order parameters (here $\eta$) described in (a); switching between equivalent states requires overcoming an energy barrier $E(\eta)$, often described as the Landau barrier.*

There are now many established routes to circumvent the "contra-indication" between ferroelectricity (associated with ionic species with empty *d*-orbitals) and magnetism (associated with partially filled *d* orbitals) [30]. Although there are several known multiferroics, there is still a dearth of technologically viable multiferroics, *i.e.*, those that can be manipulated at room temperature and exhibit strong coupling between spin and charge degrees of freedom. Thus, there should be no doubt that a more diverse palette of new materials with robust room-temperature coupling of magnetism and ferroelectricity is still urgently needed and indeed should be the focus of interdisciplinary research. **Table 1** summarizes five main physical principles that have led to the discovery of multiferroics. Of these, the two most studied are multiferroics in which the polar order comes from one of the crystal sites and the magnetic order is built into the other chemical site, as exemplified by $BiFeO_3$. The second type, which has received considerable interest from the physics community, is based on a polar order emerging as a consequence of a magnetic transition, as for manganites [44]. An emerging third pathway is via the power of heteroepitaxy and superlattice design [45]. In this regard, although there were numerous attempts in the past to synthesize complex crystal symmetries to induce multiferroic behavior, this has not been extensively revisited in recent years. There appears to be a significant opportunity to "design" multiferroic behavior by selecting magnetic materials with low symmetry and then induce inversion symmetry breaking through heterophase epitaxy. We will use these as examples to explore both the fundamental materials physics of coupling as well as the potential for future applications (see Section 5).

| Pathway | Fundamental Mechanism | Example Systems | Type of Magnetic Order |
|---|---|---|---|
| A-site driven | Stereochemical activity of lone pairs on A-site leads to ferroelectricity; magnetism from B-site | $BiFeO_3$<br>$BiMnO_3$ | Antiferromagnet<br>Ferromagnet |
| Geometrically Driven | Long range dipole-dipole interactions and oxygen rotations breaks inversion symmetry | $YMnO_3$<br>$BaNiF_4$<br>$LuFeO_3$ | Antiferromagnet<br>Antiferromagnet<br>Antiferromagnet |
| Charge ordering | Non-centrosymmetric charge ordering leads to ferroelectricity in magnetic materials (e.g., Vervey transition) | $LuFe_2O_4$ | Ferrimagnet |
| Magnetic Ordering | Magnetic field driven Ferroelectricity induced by a lower symmetry ground state | $TbMnO_3$<br>$DyMnO_3$ | Antiferromagnet<br>Antiferromagnet |
| Atomically Designed Superlattices | Still under investigation; likely lattice mediated | $LuFeO_3 - LuFe_2O_4$ | |
| Vertical Epitaxial Nanocomposites | Coupling mediated by 3-D interfacial epitaxy, e.g., Spinel-Perovskite | $CoFe_2O_4$-$BiFeO_3$<br>$NiFe_2O_4$-$BiFeO_3$<br>$CoFe_2O_4$-$BaTiO_3$ | Ferrimagnet-Antiferromagnet |

*Table 1. This table summarizes the various identified mechanisms for creating multiferroics and magnetoelectrics. For generalities on oxides and their structural and electronic properties, we refer the readers to [46].*

## 1.5. Content of this article

We start this review by covering advances on the control of magnetism by electric field (Section 2) using magnetoelectric effects within multiferroics (Section 2.1), strain-driven magnetoelectric coupling in composites and multilayers (Section 2.2), and electric field-effect using dielectrics, ferroelectrics or ionic liquids (Section 2.3). Then, more recent progress the electric-field control of magnetism are dedicated to 2D magnets (Section 2.4), magnetic skyrmions (Section 2.5) and magnons (Section 2.6). The next section (Section 3) is devoted to the control of magnetism by current-induced torques. We start by recalling the definition and generation of spin currents (Section 3.1), then introduce spin-transfer torques (Section 3.2) and spin-orbit torques (Section 3.3) for magnetization switching. We also discussed specific systems and application of particular interest such as the current induced motion of domain walls (Section 3.4), skyrmions (Section 3.5) and the control of magnetism by current-induced torques in the recently discovered two-dimensional ferromagnets (Section 3.6). In Section 4 we cover the combined use of electric-field and current-induced torques. Finally, Section 5 reviews advances in devices harnessing the electrical control of magnetism including devices for logic and memory such as MRAMs and the MESO transistor (Section 5.1), spin-torque nano-oscillators and spin-diodes (Section 5.2) and devices based on domain walls and skyrmions (Section 5.3). We end by giving perspectives for this vast and vibrant field (Section 6).

## 2. Control of magnetism by electric field

### 2.1. Electric field control of magnetism in multiferroics

#### 2.1.1. Single phase multiferroics

##### 2.1.1.1. $BiFeO_3$

Of the known multiferroics, bismuth ferrite, $BiFeO_3$, remains arguably the most important, and certainly the most widely studied, with more than 6000 papers published over the last decade. The establishment of its large (90-100 µC/cm$^2$) ferroelectric polarization, combined with magnetic ordering well above room temperature [47] has spawned an intense research effort that continues to unveil fascinating new physics and potential new applications [48].

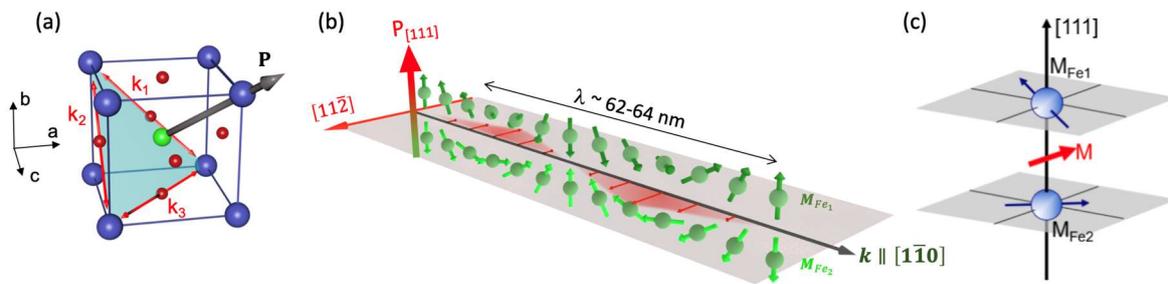

*Figure 6.* (a) Sketch of the $ABO_3$ perovskite unit cell of $BiFeO_3$. The Bi atoms are at the corners of the cell (A site), the Fe atom is at the center of the cell (B site) and the oxygen atoms form an octahedron around the Fe. The polarization points along <111>. The three corresponding propagation directions ($k_1$, $k_2$, $k_3$) are contained in the (111) plane. (b) Sketch of the spin cycloid, in which antiparallel spins are rotating in a plane defined by the polarization, P, and the propagation vector, k. A small canting, perpendicular to the cycloidal plane and varying in space, forms a coupled spin density wave (propagating in the grey plane). (c) Small canted moment resulting from the Dzyaloshinskii-Moriya (see section 2.5 for the definition) interaction [49].

$BiFeO_3$ formally belongs to the perovskite family of oxides, albeit rhombohedrally distorted from the cubic prototypical structure with *R3c* crystal symmetry in which the spontaneous polarization points along the eight equivalent <111> (**Figure 6**). While there was considerable debate in the early days regarding the magnitude of the spontaneous polarization [50] (due to the difficulty to make high-quality single crystals), it is now well established to be 90-100 µC/cm$^2$ both in films and single crystals [47,51] and confirmed theoretically [52,53]. In parallel with the scientific debate on the ferroelectric properties, there was an equal degree of debate as to the state of magnetism, particularly since it is complicated. Although the dominant super-exchange interaction stabilizes a G-type (ferromagnetic coupling in a {111} plane and antiferromagnetic coupling perpendicularly to this plane) antiferromagnetic structure [54], the magnetic structure is quite a bit more sophisticated. As a consequence of the antisymmetric magnetoelectric interaction [55], the spins are forced to rotate in an incommensurate spin cycloid (62-64 nm, **Figure 6**a in green), in a plane containing the polarization and the propagation vector (along the three high-symmetry <1-10> of the (111) plane) [56,57]. A second Dzyaloshinski-Moriya interaction, arising from the antiphase rotations of the oxygen octahedra along the <111> polarization direction (**Figure 6**c), favors an additional canting perpendicular to the cycloidal plane. This small canting is varying in space in the form of a spin-density wave (**Figure 6**b in red) locked to the spin cycloid, which gives rise to zero net magnetization [58].

In $BiFeO_3$ single crystals, this canted moment does not exhibit a macroscopically measurable magnetic moment until the spin cycloid is broken, e.g. through the application of a magnetic field of ~16-18 T [59]. While initially considered to not exist in thin films [49,60,61], there was experimental evidence over the last decade that the spin cycloid is preserved for moderate epitaxial strain in $BiFeO_3$ thin films using macroscopic averaging techniques [62,63] or scanning NV (NV stands for nitrogen-vacancy color center in diamond) magnetometry [64,65]. In addition, varying the epitaxial strain is a fantastic tool to control the antiferromagnetic textures in $BiFeO_3$ thin films from bulk-like to exotic cycloids, or pseudo-collinear G-type orderings [65,66]. On top of this, domain walls can play a key role in the emergence of a magnetic moment, which typically manifests in the form of a spin glass [67].

Understanding electric-field control of antiferromagnetism in $BiFeO_3$ thin films requires probing antiferromagnetism using X-rays, neutrons, second harmonic generation (SHG) or scanning NV magnetometry. Such studies of $BiFeO_3$ have shown that when the polarization state switches with the application of an electric field, there is a corresponding rotation of the magnetic order [57,64,68,69]. As illustrated in **Figure 7**a-b, this change can be spatially probed using a combination of piezoresponse force microscopy (PFM, to image the ferroelectric order) and X-ray magnetic linear dichroism (XMLD) photoemission electron microscopy (PEEM) (to image the antiferromagnetic order) (T. Zhao et al. 2006). SHG shows that in the canted antiferromagnetic state (large compressive strain), a single ferroelectric domain can either correspond to multiple submicron antiferromagnetic domains or to single domains, depending on the switching path (**Figure 7**c-d) [69]. Scanning NV magnetometry revealed that the electric field enables a deterministic control of antiferromagnetic domains in the cycloidal state (**Figure 7**e-h). It is interesting to note that there has been little detailed work on a full understanding of the dynamics of the manipulation of the antiferromagnetic state by an electric field – with most studies assuming the magnetic order merely follows that of the polar order, but not clarifying that pathway. This is an opportunity for future ultrafast dynamics research, since the antiferromagnetic resonance frequencies are in the several hundred GHz range and $BiFeO_3$ has electromagnons in the 600 GHz to 1 THz range [70–72], cf subsection 2.6.2. Given the current surge in interest in antiferromagnetic spintronics [73], such insulating multiferroics would also garner more interest especially through the use of nonlocal spin transport.

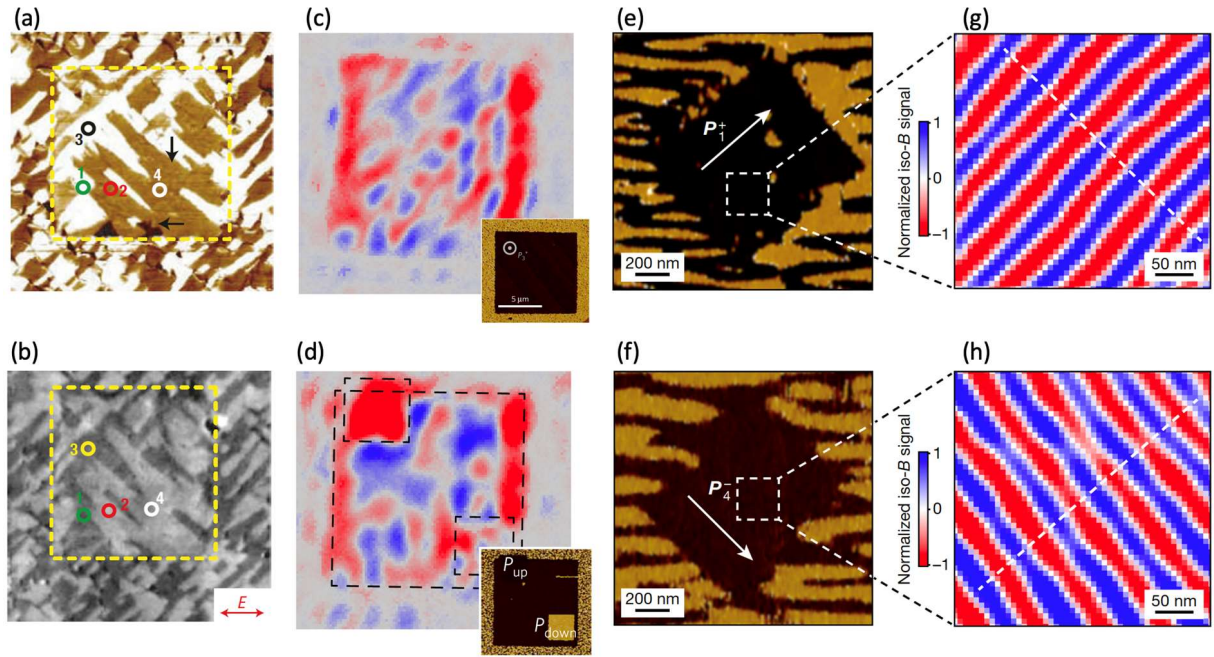

*Figure 7. Electric-field control of antiferromagnetism in BiFeO$_3$. (a) In-plane PFM and (b) XMLD-PEEM on a central area that has been electrically switched (T. Zhao et al. 2006). (c) Reconstructed antiferromagnetic configurations from SHG images in a single ferroelectric domain and (d) after switching in-plane (top left) and out-of-plane (bottom right) [69]. (e-f) In-plane PFM and (g-h) corresponding scanning NV magnetometry images of two different single ferroelectric domains defined by applying an electric field to the PFM tip [64].*

While first-principles density functional theory (DFT) calculations remain central for understanding and predicting the properties of multiferroics, second-principles calculations with embedded model Hamiltonians are proving increasingly valuable in the study of larger systems, for example heterostructures, domain walls and defects, as well as longer timescales in molecular dynamics. They have been applied to describe structural phase transitions of prototypical ferroelectrics [74,75] and recent extensions to include additional lattice degrees of freedom [76], as well as magnetic interactions [77], have extended their applicability to multiferroics. For example, an effective Hamiltonian consisting of a lattice part incorporating ferroelectric distortions, octahedral rotations and strain, a contribution from the interaction of the magnetic moments with each other, and coupling between the magnetic moments and the lattice, has been shown to accurately reproduce the crystal and magnetic structures of bulk BiFeO$_3$ [77]. On a larger length scale, a Landau-Ginzburg thermodynamic potential that includes both polar and antipolar distortions and their coupling to magnetism has been successful in reproducing the bulk behavior of BiFeO$_3$ and offers great promise for predicting properties in thin film heterostructures and nanostructures [78]. Multi-scale approaches that allow treatment of the electronic and lattice degrees of freedom on the same footing [79] could lead to vastly enhanced system size and accuracy when combined with improved tools for generating effective potentials using input from first principles (Wojdel et al. 2013). Modeling of the dynamics of ferroelectric switching [81] and its effect on magnetic order [82], both of which are on time- and length-scales that are far outside the ranges accessible using density functional methods, has now become feasible. Such models in combination with molecular dynamics start to allow calculation of dynamical magnetoelectric responses in the terahertz region [83], which is particularly timely as it coincides with advances in experimental methods for generating terahertz radiation [84]. Finally, the possibility of magnetoelectric multipole as an order parameter for phase transitions that break both space-inversion and time-reversal [85,86] seems intriguing, although not fully explored experimentally.

### 2.1.1.2. Manganites

Multiferroic perovskite manganites can be classified into three families: (i) $BiMnO_3$ and related phases, (ii) orthorhombic rare-earth manganites $RMnO_3$ and (iii) hexagonal manganites. Some materials from the second family can be metastable members of the third one and vice-versa.

$BiMnO_3$ (BMO) is a monoclinic perovskite first synthesized in Japan and the Soviet Union in the 1960s [87,88]. BMO was soon recognized as a ferromagnetic insulator with a $T_{CM}$ of about 105 K [87–89]. This ferromagnetic behavior was unexpected because the similar compound $LaMnO_3$ (the ionic radii of $Bi^{3+}$ and $La^{3+}$ ions are 1.24 and 1.22 Å, respectively) [90]) is an A-type (ferromagnetic coupling in a {001} plane and antiferromagnetic coupling perpendicularly to this plane) antiferromagnet [91]. In fact, while the Jahn-Teller effect lifts the degeneracy of the $e_g$ states in both compounds, the presence of stereochemically active $6s^2$ lone pairs on the Bi ions [92] triggers a peculiar three-dimensional orbital ordering of the Mn $d_{x2-z2}$ orbitals [93] which induces globally ferromagnetic super-exchange interactions between the Mn ions.

Based on reports of a noncentrosymmetric space group ($C2$, see [94]), BMO has been conjectured to be ferroelectric, and thus multiferroic. Later neutron diffraction experiments however indicated a centrosymmetric structure [95], ruling out ferroelectricity in bulk BMO. We note however that first principles calculations [96] have predicted a ferroelectric ground state for compressively strained films and that indications of ferroelectricity have been provided in thin films [97,98]. $BiMnO_3$ [99] and $La_{0.1}Bi_{0.9}MnO_3$ [100] ultrathin films have also been shown to be ferroelectric at room temperature. To date, there are no clear indications that $BiMnO_3$ and related phases are magnetoelectric, aside from magnetocapacitance measurements showing a peak at the ferromagnetic $T_C$ [101].

Orthorhombic rare-earth manganites such as $TbMnO_3$ are so-called type II multiferroics, in which ferroelectricity arises as a consequence of non-collinear spin ordering that breaks inversion symmetry. Multiferroicity in this compound was first discovered by Kimura et al [44], and the existence of an incommensurate spiral spin order was clarified by Kenzelmann et al [102]. Arima et al later confirmed the same spin order in (Tb, Dy)$MnO_3$ compounds [103]. The mechanism leading to the onset of ferroelectricity in the presence of spiral spin order was elucidated through the spin-current model [104], see cf **Figure 8.** a. Experimentally, these compounds become ferroelectric below about 30 K and their polarization is small, in the 0.1 µC/cm² range. However, because the ferroelectric character arises from the spin ordering, they display substantial magnetoelectric coupling. Early on, it was shown that the application of a magnetic field has a strong influence on the ferroelectric properties, notably on the amplitude and direction of the polarization also leading to large magnetocapacitance effects [105], cf **Figure 8.**

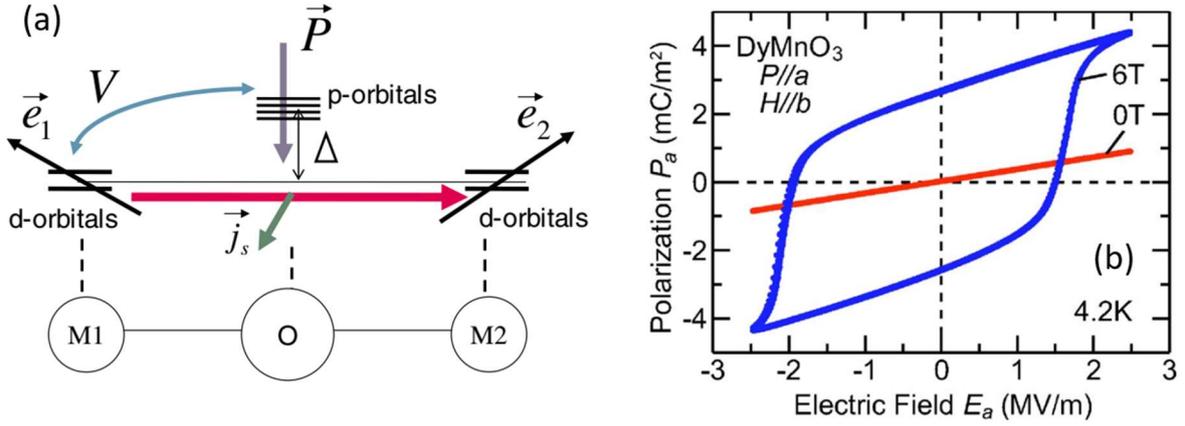

*Figure 8. (a) Spin-current model. Two transition metal ions M1 and M2 are separated by an O ion. M1 and M2 carry non-collinear spin moments $\vec{e_1}$ and $\vec{e_2}$. In this situation, a spin current arises and is expressed as $\vec{J_s} \propto \vec{e_1} \times \vec{e_2}$, with the direction of the $\vec{J_s}$ vector corresponding to the spin polarization. The electric polarization is then given by $\vec{P} \propto \vec{e_{12}} \times \vec{J_s}$ where $\vec{e_{12}}$ is the unit vector connecting M1 and M2. This mechanism is analogous to the inverse Dzyaloshinskii-Moriya interaction, cf. [106]. (b P–E curves obtained at magnetic fields of 0 T (red) and 6 T (blue) for a $DyMnO_3$ crystal, illustrating the magnetic field control of ferroelectric in this compound [107].*

In perovskite manganites, when the size of the A site rare-earth cation is further reduced beyond that of Dy, or A is Y or Sc, the hexagonal structure becomes more stable than the orthorhombic structure. Hexagonal manganites are also multiferroic with a very high ferroelectric $T_C$, around 1000 K, and they are antiferromagnetic with a Néel temperature typically lower than 100 K [108]. Coupling between the two orders was first detected as an anomaly in the dielectric constant at the Néel point for $YMnO_3$ [109]. Dielectric anomalies at magnetic phase transitions were later found in other compounds of the series [110,111]. In general, hexagonal manganites with a magnetic ion at the A site have very complex phase diagram [112] – as for instance $HoMnO_3$ [113] – with spin reorientation temperatures where the dielectric constant shows a pronounced peak [110] and the polarization a kink [114]. The application of a magnetic field allows tuning the system into various magnetic states that have different dielectric properties. So far, this magnetoelectric coupling has not been harnessed to control magnetism by electric field.

### 2.1.1.3. Ferrites

Besides $BiFeO_3$, several other Fe-containing oxides have been explored as possible multiferroics with a sizeable magnetoelectric coupling. Fe-based compounds often have larger magnetic moments and high magnetic transition temperatures, which is appealing for applications.

Fe-based perovskites, i.e., orthoferrites, are directly related to $BiFeO_3$ but lack the lone pair provided by Bi ions that are responsible for the robust ferroelectricity in that compound. Nevertheless, $GdFeO_3$ and $DyFeO_3$ have been shown to be ferroelectric at very low temperatures. The mechanism is of course different from that at play in $BiFeO_3$; here, ferroelectricity is improper and believed to be driven by magnetic order through exchange striction below the ordering temperature of the rare-earth ion, around 3 K [115,116]. While polarization was shown to strongly depend on magnetic field, only a moderate change of magnetization was induced by electric field [115]. Recently, non-stoichiometric $YFeO_3$ was reported to display ferroelectricity at room temperature, qualifying it as multiferroic [117]. It will be interesting to see if this behavior can be reproduced in other systems and if magnetoelectric coupling is present in this new phase.

When the A cation size is small, $AFeO_3$ compounds may be stabilized in a hexagonal structure, resembling that of hexagonal manganites that are ferroelectric. Hexagonal $AFeO_3$ compounds have thus been predicted to be ferroelectric and to display magnetoelectric coupling [118]. Their Néel temperature is around 100 K, that is much lower than in their orthorhombic cousins [119]. Various reports indeed indicate a ferroelectric response at room temperature [120,121]. Electric control of magnetism has been elusive so far, with this family of compounds.

A promising, yet complex, family of ferrites for the electrical control of magnetism is hexaferrites. These compounds have (very) large unit cells with many magnetic sites and can be grouped into six sub-families coined M, W, Y, Z, X and U-type hexaferrites. Their structure is built from blocks labelled R, S and T (R block: $[(Ba,Sr)Fe_6O_{11}]^{2-}$ ; S block or spinel block: $Me^{2+}[Fe_4O_8]$ ; T block: $[(Ba,Sr)_2Fe_8O_{14})]$) ; Me is a divalent metal ion, for instance $Zn^{2+}$ or $Co^{2+}$) [122]. The most well-known is the M-type structure, magnetoplumbite that is built from alternating S and R blocks. While most hexaferrites are ferrimagnetic, some – and in particular Y-type compounds – display non-collinear magnetic order. What is quite unique compared to other non-collinear systems is that in some hexaferrites this order exists at and above room temperature.

The magnetic moments within hexaferrites can be viewed as being organized into two types of stacks with large or low moment. The stacks are then coupled together by super-exchange in a fashion that is sensitive to the concentration of Ba or Sr ions, that tunes the Fe-O-Fe bond angles at the interface between blocks. This results in non-collinear order, such as a proper screw for Y-type ferrites. When a magnetic field is then applied perpendicular to the hexagonal axis, the materials undergo magnetic phase transitions to, e.g., conical structures that cause the appearance of a spontaneous polarization [123]. In most compounds, the finite conductivity impedes the observation of such a magnetoelectric coupling at room temperature, but it has been realized in some Z-type and U-type ferrites [124,125].

Electric-field control of magnetization has been demonstrated in some of these ferrites. In a Co-based Z-type compound, Chun et al reported a change of the magnetization of about 0.6 µB/f.u. over 2 MV/m at room temperature [126]. In these experiments, the field dependence comprised a linear and a quadratic terms but later, working with a Zn-based Y-type compound Chai et al reported magnetization switching between about -2 and +2 µB/f.u. in a field of ±2 MV/m, albeit at 15 K [127], see **Figure 9.** A similar effect up to 250 K was reported subsequently in a related system [128] and even at room temperature, with however a reduced amplitude [129].

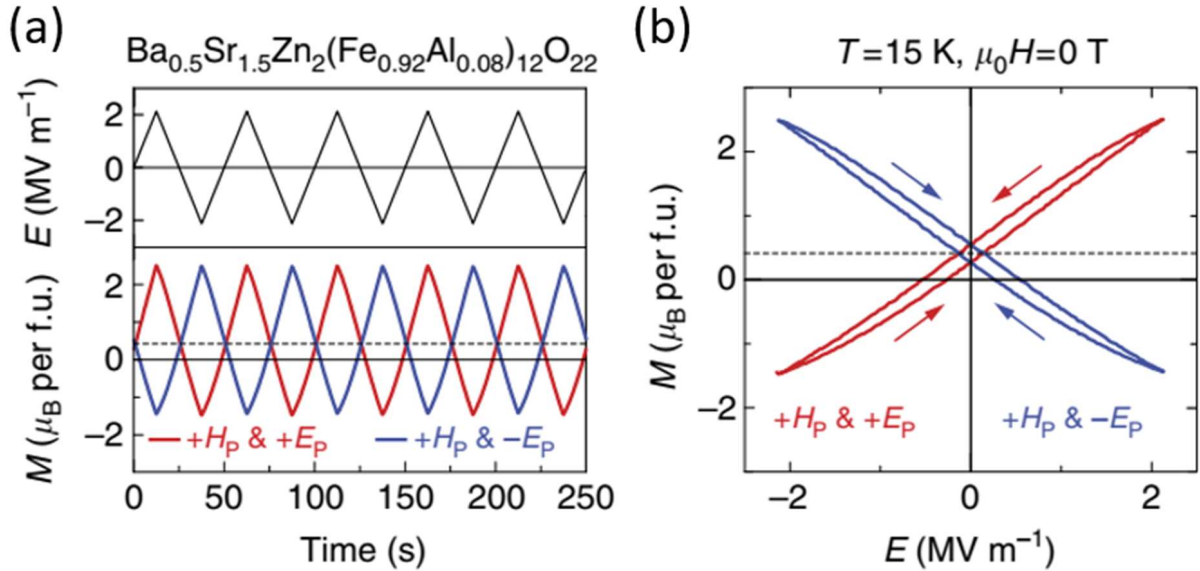

*Figure 9. Electric field modulation of magnetization in a Y-type hexaferrite. (a) Periodic modulations of M along the [100] crystallographic direction at zero magnetic field under repeating triangular waves of E applied parallel to [120], after preparing the system by cooling it to the measurement temperature in electric field and magnetic field (magnetoelectric annealing, see original paper for details).* [127]. *(b) Corresponding magnetization vs electric field loops illustrating the reversal. The red and blue traces in (a) and (b) correspond to opposite direction of the applied electric field during the magnetoelectric annealing procedure* [127].

### 2.1.1.4. Other systems including organics

In contrast to the heavily studied inorganic multiferroics, organic multiferroics have been less explored [130]. Organic materials provide an equally broad palette of materials design building blocks, but face similar challenges as do their inorganic counterparts. Inducing a magnetic state, especially at room temperature requires strong exchange interactions, thus invariably necessitating the introduction of transition metal ions into an organic framework. One could envision a multiferroic tree, as depicted in Erreur ! Source du renvoi introuvable. for the inorganic systems. Before converging into possible multiferroic systems, it is perhaps appropriate to discuss the possible origins of ferroelectricity and magnetism separately in these compounds.

Ferroelectricity in organic materials has been extensively studied [131], with the PVDF (polyvinylidene difluoride) and DOBAMBC (P-(n-(decyloxybenzylidene)-p-amino-(2-methylbutyl)) systems receiving considerable scientific attention. Ferroelectric liquid crystals have also been inveestigated [132,133]. Recent developments in molecular ferroelectrics, such as diisopropyl ammonium bromide (and related compounds) are showing a lot of promise with spontaneous polarization almost equal to the model system, barium titanate [134]. The robustness of the ferroelectric order parameter through charge, permittivity and piezoelectric measurements is a strong positive sign. Further work on the switching dynamics in such order-disorder ferroelectrics would be very welcome. Equally important, pathways to introduce magnetism into such materials would be quite rewarding. Organic charge-transfer based ferroelectrics, such as tetrathiafulvalene-p-chloranil (TTF–CA) [135] are another possible class of ferroelectrics, but with a much lower spontaneous polarization; large polarization values have been reported, but the experimental measurements likely require further validation. Another class of organic ferroelectrics, the metal–organic frameworks (MOFs), such as [NH4]-M(HCOO)$_3$ and [(CH3)2NH2]M(HCOO)3 M = Zn,Mn, Fe, Co and Ni, have shown promising spontaneous polarization due to their order–disorder transition, which however occurs well below room temperature [136]. Given the large body of research into metal organic framework compounds for a wide range of possible applications, such organics hold promise for future study.

Coming to organic multiferroics, the challenges of obtaining magnetic and ferroelectric order are almost exactly the same as in their inorganic counterparts, namely the contradictions in the requirements for these two order parameters to co-exist. One example is tetrathiafulvalene-p-bromanil (TTF–BA), which derives it ferroelectric order from a spin-Peierls-like instability (spin-lattice interaction), albeit at a low temperature of 53 K [137,138]. This is accompanied by the emergence of a relatively small polarization, quite like the emergence of ferroelectricity in the magnetic manganites. Another organic multiferroic of spin-driven polarization is the crystalline thiophene-$C_{60}$ charge-transfer complex [139]. By utilizing the supramolecular assembly strategy to build electron donor thiophene and acceptor $C_{60}$ co-crystals, room temperature magnetism and spontaneous polarization were observed [140] There have been a few demonstrations of multiferroic behavior (once again with a low ferroelectric $T_C$) in MOF's that contain 3d transition metal species.  Organic charge-transfer salts, such as κ-(BEDT-TTF)2Cu[N(CN)2]-Cl, exhibit the converse behavior, i.e., a charge ordering induced magnetism, typically at temperatures ~25 K [141]. Thus, organic multiferroics provide a unique set of chemical frameworks to explore spin-charge coupling, but the challenges for potential translation to devices remain in terms of the ordering temperatures or the strength of the individual order parameter. In this sense, the large room temperature polarization of the diisopropyl ammonium bromide seems promising for further research to make them magnetic. More broadly, organic ferroelectrics/ferromagnets and multiferroics seems to be a topic that is rich for an even deeper and more comprehensive investigation, using the fundamental materials design principles outlined in   **Erreur ! Source du renvoi introuvable.**. It is particularly noteworthy that organics typically do not require the high process temperatures that are characteristic of inorganics such as the oxides, and thus should be more amenable to integration efforts, once the right materials system is discovered.

### 2.1.2. Multiferroic heterostructures

#### 2.1.2.1. *BiFeO₃-based heterostructures*

Thin-film synthesis of $BiFeO_3$ (and other multiferroics) has been a very fruitful pathway to study the materials physics of magnetoelectric coupling as well as pointing the way to possible applications. The perovskite symmetry and lattice parameters (pseudocubic lattice parameter of 3.96 Å) close to a large number of oxide-based substrates means that epitaxial synthesis is possible and has indeed been widely demonstrated [142]. Films with thicknesses down to just a few unit cells and as large as a few microns have been synthesized by physical-vapor deposition (*e.g.*, pulsed laser deposition [47,142,143], sputtering [144], molecular beam epitaxy [145]), chemical-vapor deposition [146], and chemical-solution deposition. Many studies have used conducting perovskite electrodes (such as $SrRuO_3$, $La_{1-x}Sr_xMnO_3$, $La_{1-x}Sr_xCoO_3$) as bottom electrodes to both template the perovskite phase as well as provide a bottom contact for electrical measurements. These synthesis studies have led the way to enable a wide range of materials physics studies.

A particularly important aspect is the stability of the polar state as the thickness is scaled down. Such size effects have been extensively studied in classical ferroelectrics [147] and are characterized by a suppression of the order parameter as the thickness is scaled down.  Similar studies have been undertaken in the case of the BiFeO$_3$ system [148–151], albeit in an incomplete sense. Several studies have shown that the polar order parameter is reduced, but still maintained. The ferroelectric switching process in BiFeO$_3$ is believed to be limited by nucleation and growth of reverse domains [152–154] broadly captured by the Kay-Dunn model [155], in which the coercive field scales as film thickness $d^{-2/3}$. Consequently, progressively larger reductions in film thickness are needed to reduce the coercive voltage as it is pushed to smaller values. In BiFeO$_3$, lanthanum substitution has been shown [156] to reduce the switching energy by reducing the polarization [151], although to an insufficient extent to date. Pushing BiFeO$_3$ close to a phase boundary between ferroelectric and antiferroelectric states or identifying materials without the octahedral rotations of BiFeO$_3$ could be an alternative pathway to smaller coercive fields.

The antiferromagnetic order has also been shown to exist at room temperature in films that are as thin as 4 nm (10 unit cells). What has not been shown is the coupling between the two order parameters at such length scales, and more importantly, electric field manipulation of this coupling. Thus, a deeper, quantitative understanding of the stability of the individual order parameters, the coupling between them as well as E-field manipulation of this coupling at a thickness less than ~10 nm would be of significant interest. This is captured in **Figure 10**

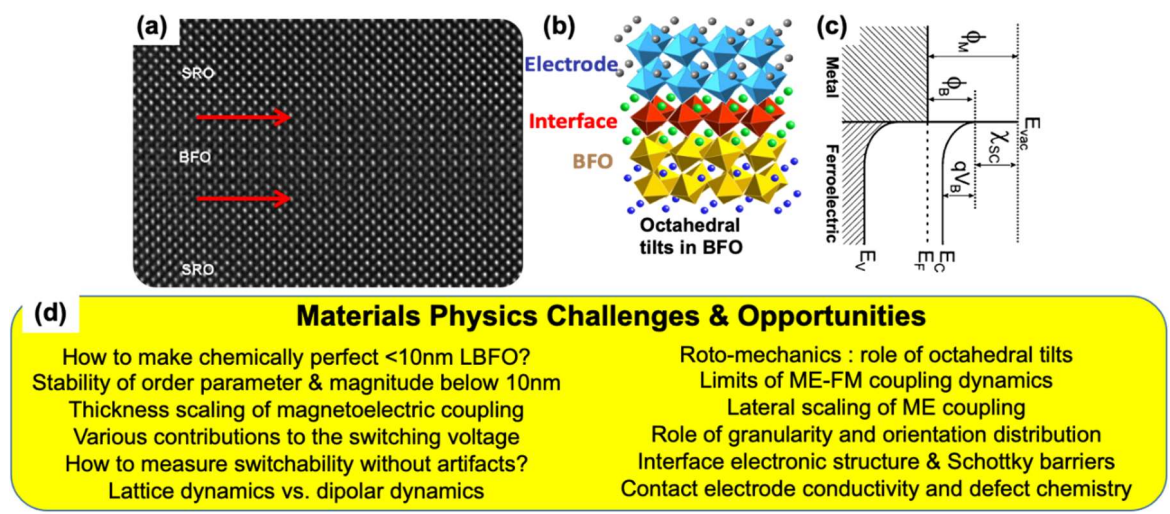

*Figure 10. (a) is an atomic resolution image of a 6 unit cell thick BiFeO$_3$ (BFO) layer sandwiched between epitaxial SrRuO$_3$ (SRO) top and bottom electrodes as a representative of sub-10 nm thick multiferroics as a model system. (b) is the corresponding crystal model showing the octahedral tilts (in both the SRO and BiFeO$_3$ layers). (c) schematically depicts how the formation of a Schottky barrier at the contact metal-BiFeO$_3$ interface can lead to potential drops; (d) lists materials physics challenges and opportunities for multiferroic heterostructures.*

### 2.1.2.1.1. BiFeO$_3$/La$_{0.7}$Sr$_{0.3}$MnO$_3$

Perhaps the most significant breakthrough in the past few years is the demonstration that the magnetization direction in conventional ferromagnets (*e.g.*, Co$_{1-x}$Fe$_x$) can be rotated by 180° with an electric field [157] when it is exchange coupled to BiFeO$_3$ [67,158]. The extension to all-oxide La$_{0.7}$Sr$_{0.3}$MnO$_3$/BiFeO$_3$ interfaces [149] (**Figure 11**), with chemically abrupt *A*-site termination [159], allowed for electric-field control of exchange bias coupling at temperatures below 100 K [160]. Exchange bias refers to the horizontal shift of the magnetization vs field loop of a ferromagnetic layer, due to the exchange coupling to an adjacent antiferromagnetic layer.

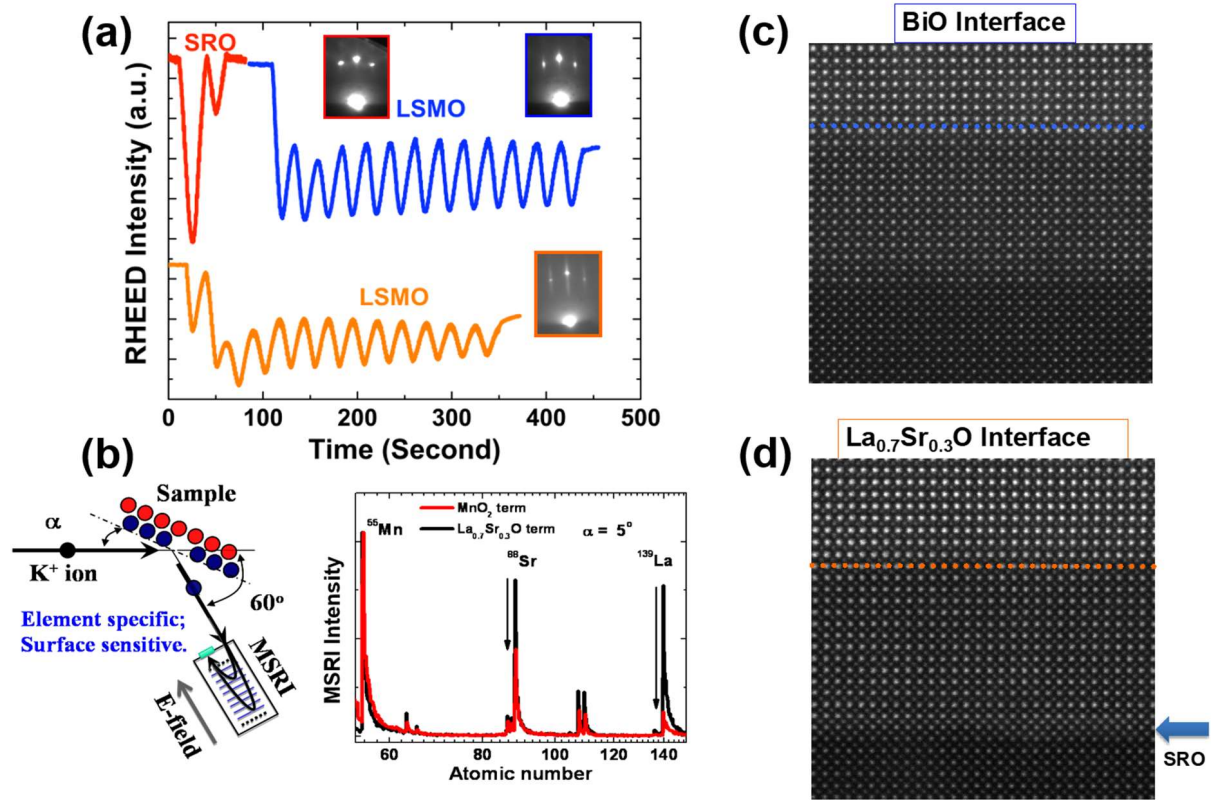

*Figure 11. Synthesis of model systems. This figure illustrates epitaxial synthesis as a pathway to create model systems at the scale of a single unit cell. (a) is a reflection high-energy electron diffraction (RHEED) pattern of the growth of the $La_{0.7}Sr_{0.3}MnO_3$ bottom electrode on a $TiO_2$- terminated $SrTiO_3$ substrate; insertion of 2-unit cells of $SrRuO_3$ leads to a conversion of the termination from B-site to A-site; (b) Time of flight- ion scattering and recoil spectroscopy (TOF-ISARS) of the two types of substrate surfaces. The spectra are normalized to the Mn peak and it is clear that the La content (black) for one of them is much higher than that of the other; (c,d) are atomic resolution STEM images of the two types of interfaces showing that atomically sharp interfaces can be obtained [159].*

Earlier work on the same system has shown the ability to reversibly switch between two exchange-biased states with the same polarity (unipolar modulation) without the need for additional magnetic or electric fields in a multiferroic field effect device [160], but eventually the ability to reversibly switch between these two states with opposite polarity (bipolar modulation) was demonstrated as well (**Figure 12**). The key was modifying the direction of the magnetization in the $La_{0.7}Sr_{0.3}MnO_3$ with respect to the current in the device channel. A reversible shift of the polarity of exchange bias through the zero applied magnetic field axis was thus achieved with no magnetic or electric-field cooling and no additional electric or magnetic bias fields – in essence, full direct electric field control of exchange bias. This also helped clarify the mechanism underlying the change in exchange bias coupling.

An important open problem is the development of oxide ferro- or ferri- magnets with high $T_c$, a significant remanent moment and strong exchange coupling and ohmic contacts with $BiFeO_3$ or other multiferroics; spinels or double perovskites are promising candidates in this regard [161,162]. In a complementary direction, the antiferromagnetic domain orientation in magnetoelectric $Cr_2O_3$, which can be controlled by an electric field, has been shown to affect the exchange-bias coupling to a ferromagnetic overlayer [163] opening a pathway to electric-field switchable exchange-bias devices.

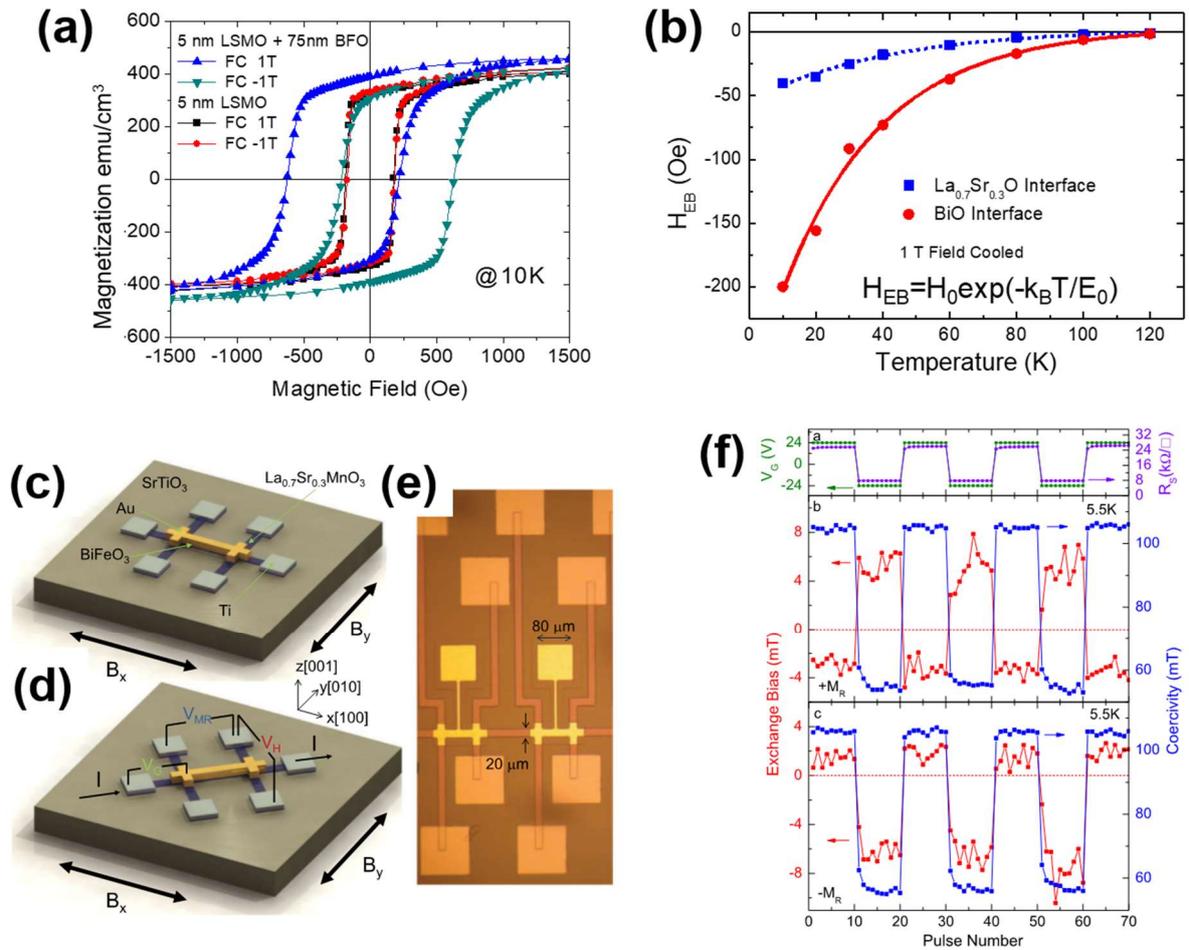

*Figure 12. Electric field manipulation of interfacial magnetic coupling in epitaxial heterostructures. (a) 1 Tesla, field cooled magnetic hysteresis loops at 10 K showing a strong exchange bias of 200 Oe for a $La_{0.7}Sr_{0.3}MnO_3$ (5 nm) / $BiFeO_3$ (75 nm) heterostructure; (b) the magnitude of the exchange bias field as a function of temperature and interface termination (La/Sr-O vs. Bi-O interfaces) [164]. (c,d) are device layouts for magnetoelectric measurements, with the corresponding SEM image shown in (e) ; (f) shows the bipolar voltage profile and the corresponding exchange bias and coercivity showing full electric field switching of the exchange bias [165].*

2.1.2.1.2. $BiFeO_3$/ferromagnetic metals

Metallic ferromagnets, such as the well-studied CoFe system provide a good starting point to explore electric field control of ferromagnetism. Although chemically very different from the oxides, metallic ferromagnets generally have higher Tc's and thus a greater likelihood of strong exchange coupling. The push for ultra-low power logic-memory devices builds from observations of the potential of magnetoelectric control using multiferroics – the key being the ability to control magnetism with electric field at room temperature [166] using a spin-valve device (**Figure 13**a) to demonstrate such a coupling [32]. For example, magnetoelectric switching of a magnetoresistive element was recently shown to operate at or below 200 mV, with a pathway to get down to 100 mV [167]. Reducing the thickness is an obvious pathway to get to such low voltages. A combination of structural manipulation via lanthanum substitution and thickness scaling in multiferroic $BiFeO_3$ has helped to scale the switching energy density to ≈10 µJ.cm$^{-2}$ and provides a template to achieve attojoule-class nonvolatile memories. Using La-$BiFeO_3$, it was possible to show that the switching voltage of the giant magnetoresistance (GMR) response can be progressively reduced from ≈1 V to 500 mV by decreasing the film thickness down to 20 nm (**Figure 13**a). Electric-field control of the magnetization direction in the bottom $Co_{0.9}Fe_{0.1}$ layer was shown in measurements both in a magnetic field of 100 Oe as well as in the remanent state (*i.e.*, zero magnetic field) (**Figure 13**b,c). The low-voltage magnetoelectric switching in multiferroic $Bi_{0.85}La_{0.15}FeO_3$ was further probed by X-ray magnetic circular dichroism XMCD-PEEM imaging at the Co $L_3$ edge via studies (inset, **Figure 13**d,e) where application of +/-500 mV revealed contrast changes consistent with reversal of the in-plane magnetization.

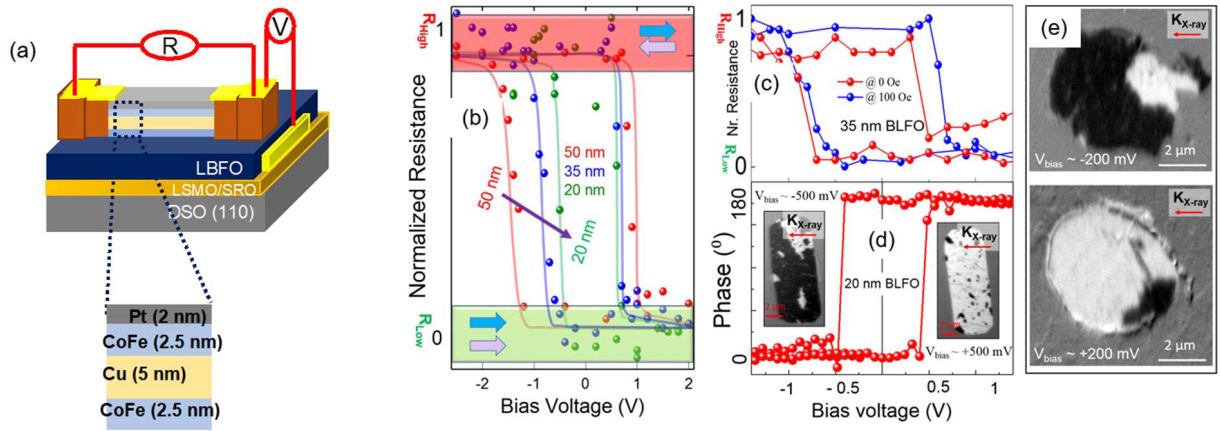

*Figure 13. E-field control of magnetism at room temperature. (a) Schematic of the magnetoelectric test structure comprised of the multiferroic La-BiFeO$_3$ layer which is in contact with a CoFe-Cu-CoFe spin valve which is used as a read out element; (b) shows the normalized resistance (Nr. Resistance) of the spin valve as a function of applied voltage to the La-BiFeO$_3$ layer; (c) is a plot of the normalized resistance vs. electric field at zero magnetic field and at 100Oe, showing no significant difference, thus illustrating that the switching of the spin valve us due to the electric field; (d) piezoelectric hysteresis loop for the 20nm La-BiFeO$_3$ layer showing the full switching at ~500 mV; the insets show XMCD-PEEM images of a Co layer that is in contact with the La-BiFeO$_3$. The contrast reversal illustrates a change in the magnetization direction due to the applied voltage of 500 mV; (e) XMCD-PEEM image of a CoFe-10nm La-BiFeO$_3$ test structure that has been switched by -200 mV (dark) and +200mV (bright) contrast, showing that the magnetization direction has been mostly switched.* [167,168].

## 2.2. Strain-driven control of magnetism using ferroelectrics and piezoelectrics in multilayers

### 2.2.1. Piezoelectric/ferromagnet

Another way to control magnetism with an electric field is to combine piezoelectric materials and magnetic materials in thin film heterostructures. The simplest geometry is to grow a magnetic thin film on top of a ferroelectric (or a relaxor) substrate with large piezoelectric coefficients (a relaxor is a ferroelectric with large electrostriction and piezoelectric coefficient [169]). Pertsev predicted that giant magnetoelectric susceptibility may be achieved in such geometry as a result of the strain-driven spin reorientation in the ferromagnetic thin film [170]. Nickel is often chosen as the magnetic thin film due to its sizeable magnetostriction at room temperature ($T_C$>>300 K). Modifications of the remnant magnetization, magnetic anisotropy or even magnetization direction of a Ni thin film induced by the electric field applied onto its ferroelectric/piezoelectric substrate were reported [171–174]. This is illustrated in **Figure 14**a-c in which the magnetic easy axis of the Ni layer reversibly rotates by 90° (along the in-plane x- or y-axis) depending on the sign of the voltage applied to x-axis of the $Pb(Zr_xTi_{1-x})O_3$ substrate [173]. The electric-field strain-induced modifications of magnetization or magnetic anisotropy were extended to other artificial multiferroics including Fe or $La_{0.7}Sr_{0.3}MnO_3$ on $BaTiO_3$, $Co_{40}Fe_{40}B_{20}$, or $La_{0.7}(Ca,Sr)_{0.3}MnO_3$ on $Pb(Mg_{1/3}Nb_{2/3})_{0.7}Ti_{0.3}O_3$, or $Ga_{1-x}Mn_xAs$ on $Pb(Zr_xTi_{1-x})O_3$, FeGaB on $Pb(Zn_{1/3}Nb_{2/3})O_3$-$PbTiO_3$ [175–180].

In ferroelectrics in which polarization is associated with a strong deformation of the lattice (such as $BaTiO_3$), the application of an electric field can result in a modification of ferroelastic domains and modify the average strain on the adjacent magnetic layer. Combining optical imaging techniques, Lahtinen et al. demonstrated a full imprint of the ferroelastic domains of a $BaTiO_3$ substrate on the magnetic domains of a CoFe thin film grown on top [181]. Furthermore, they were able to electrically control the magnetic domain patterns of CoFe by the voltage applied through the $BaTiO_3$ substrate (**Figure 14**d-e) [182].

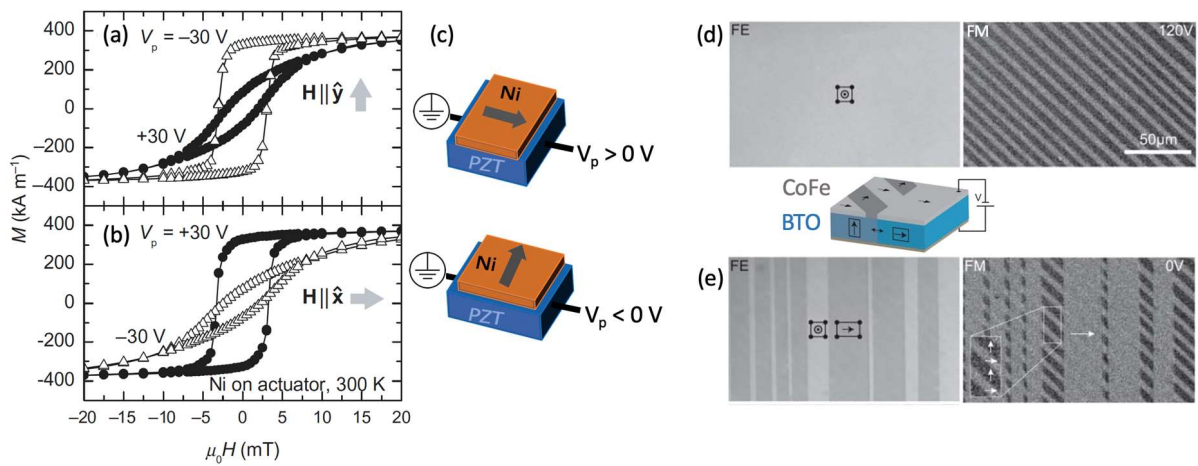

*Figure 14. Piezoelectric control of the magnetic anisotropy. Magnetization vs. magnetic field aligned along (a) y and (b) x axis in Ni thin films on $Pb(Zr_xTi_{1-x})O_3$-based actuators under +30 V and -30 V (along the x axis). (c) Sketch showing that the magnetic anisotropy of the Ni thin film rotates by 90° depending on the sign of the voltage applied to the piezoelectric actuator (a-c from [173]). Ferroelectric domain (left, birefringent contrast) and magnetic domains (right, magneto-optic Kerr contrast) for a CoFe thin film on a $BaTiO_3$ substrate (a) under a vertical voltage of 120 V and (b) when the voltage is turned off. As sketched in between (d) and (e), the voltage changes the population of ferroelastic domains in $BaTiO_3$ and consequently the local strain and magnetic anisotropy. (a-c from [182]).*

### 2.2.2. FeRh-based structures

In parallel to these efforts to control the orientation of magnetization with an electric field, attempts have been made to achieve an electrical control of the magnetic order. For this approach, archetypical magnetic material is FeRh with the CsCl-type structure, which displays a first-order metamagnetic phase transition from a low temperature antiferromagnetic phase to a high temperature ferromagnetic phase, slightly above room temperature (350-370 K) [183]. This first-order magnetic phase transition is accompanied by sharp changes in the volume and resistivity. FeRh thus displays strong coupling between lattice, magnetization and electronic properties. Motivated by the volume change at the ferromagnetic to antiferromagnetic transition in $Fe_{1-x}Rh_x$ [184,185], an electric field was used to drive the reciprocal effect, a ferromagnet-to-antiferromagnet transition induced by a structural deformation. This makes this system promising for the electric-field control of magnetism and resistivity via piezoelectric effects.

Cherifi et al. grew 20-nm-thick epitaxial thin films of FeRh using rf sputtering on BaTiO$_3$ single crystals. Applying a voltage to the BaTiO$_3$ crystal and changing the proportion of c- and a- ferroelastic domains, they were able to modulate the average epitaxial strain and trigger a giant change of magnetization at 385 K (**Figure 15**a) [186]. These results were supported by ab initio calculations as well as XMCD-PEEM images, which demonstrate that turning off the electric field leads to a transition from an antiferromagnetic state (pure c-domains) to a ferromagnetic one (a-domains) (**Figure 15**a-b) [187]. The strain-driven magnetic transition results in a 260% change of the coercive field for FeRh thin films grown on (1-x)Pb(Mg$_{1/3}$Nb$_{2/3}$)O$_3$-xPbTiO$_3$ (PMN-PT) [188]. Interestingly, the electric-field induced phase transition in FeRh/PMN-PT further enables to modulate the spin dynamics of FeRh with a 120% modulation of the magnetic damping (**Figure 15**c-d), resulting from the modification of the relative fraction of the antiferromagnetic/ferromagnetic phases [189].

Since the resistivities of the two magnetic phases of FeRh differ, the magnetic transition is accompanied by a ~25% change in film resistivity. Using FeRh thin films on PMN-PT, Lee et al. demonstrated a giant electroresistance of 8% using the piezoelectric strain modulations at 368 K [190]. This electroresistance is attributed to a variation of the antiferromagnetic to ferromagnetic phase proportions. Later on, similar observations were made on FeRh/BaTiO$_3$ with an electroresistance of 22% at 376 K (**Figure 15**d-f) [191]. Magnetic force microscopy (MFM) investigations under electric field revealed a full magnetic transition in the film (**Figure 15**g). This electric readout of the first-order phase transition opens possibilities for non-volatile magnetic memories in a simple architecture. For more details on the electric-field control of magnetic and resistive properties in FeRh, the reader is invited to look at following reviews [192,193].

Open challenges with this approach include reducing the optimal working temperature from around 100°C to room temperature, tuning the chemical composition to optimize the strengths of the exchange interactions, achieving complete conversion between the ferromagnetic and antiferromagnetic phases and reducing the required applied voltages. Other promising systems are the Mn-Pt intermetallics and half-doped perovskite manganites such as La$_{0.5}$Sr$_{0.5}$MnO$_3$, in which an electric-field-driven charge-ordered antiferromagnetic insulator to ferromagnetic metal transition could be possible [194], although then the Curie temperature is below 300 K.

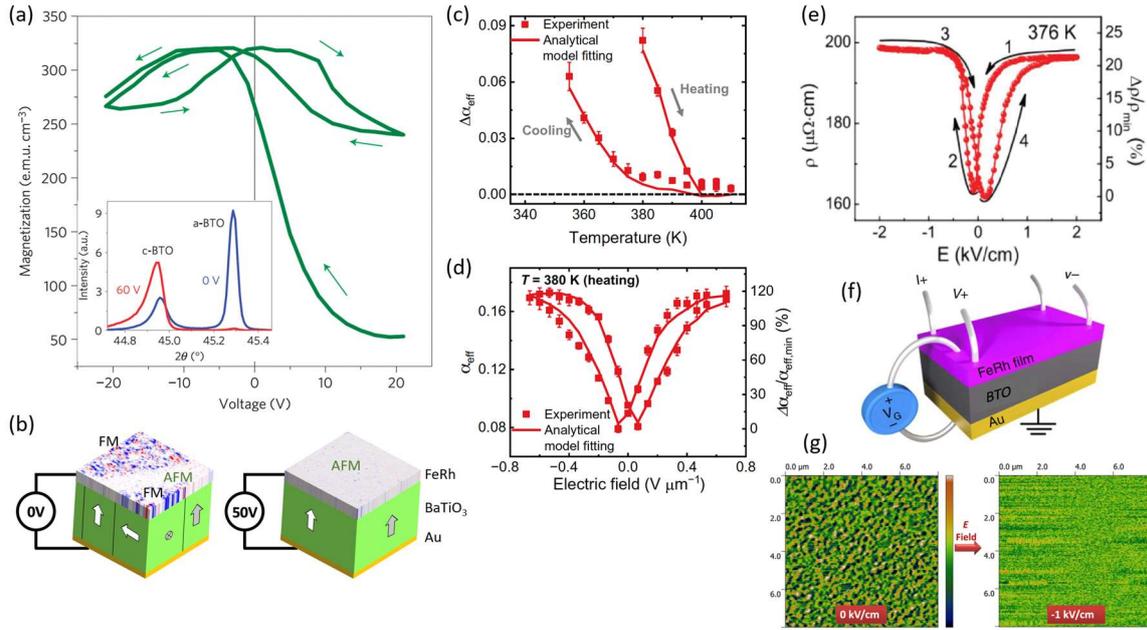

Figure 15. (a) Variation of the magnetization with voltage in FeRh grown on BaTiO$_3$ single crystals at 385 K. The inset shows the X-ray diffraction pattern of the (002) and (200)/(020) reflections of BaTiO$_3$ as a function of voltage at 390 K. Under 60 V, the BaTiO$_3$ is purely c-domains while it consists of a mixed a- and c-domains at 0 V. (b) Sketch of the electric-field induced magnetic phase transition at 385 K with the XMCD-PEEM image overlayed. (a-b from [186]. (c) Temperature dependence of the magnetic damping in FeRh thin films grown on PMN-PT. (d) Electric-field modulation of the damping at 380 K. (c-d from [189]. (e) Large electroresistance of FeRh thin films on BaTiO$_3$ substrates at 376 K. (f) Principle of the experiment. (g) MCD phase images collected at 376 K at zero and -1 kV/cm electric field. (e-g from [190,191].

### 2.2.3. LuFeO$_3$/LuFe$_2$O$_4$

As mentioned in **Erreur ! Source du renvoi introuvable.**, there is a lot of potential in designing magnetoelectrics at the atomic scale using epitaxial superlattices. The original work of Mundy et al [45] on LuFeO$_3$-LuFe$_2$O$_4$ superlattices showed that the epitaxial pathway to magnetoelectric coupling is indeed possible. LuFeO$_3$ belongs to the class of ferroelectrics, termed as improper ferroelectrics, in which the fundamental order parameter is a structural distortion; this distortion coupled to a polar mode (Spaldin et al., Nature Materials) leading to a spontaneous polarization of 3-5 μC/cm$^2$ along the c-axis of the hexagonal structure (cf Section 2.1.1.2). Using the power of epitaxy, the authors prepared atomically perfect superlattices combining LuFeO$_3$ with its sister compound, LuFe$_2$O$_4$ (which is ferrimagnetic with a T$_N$ of ~240 K). The magnetic state in the LuFe$_2$O$_4$ layer has been switched with an electric field [45], with the coupling most likely mediated through the lattice.

Fan and co-workers [195] revealed the microscopic details of the coupling across the ferroelectric (LuFeO$_3$) ferrimagnet (LuFe$_2$O$_4$) interface. A key issue with LuFe$_2$O$_4$ is that the Curie temperature is lower than room temperature (~240 K in the bulk; ~280 K in epitaxial superlattices [45]). Thus, it would be desirable to replace this with other structurally and chemically compatible ferrites. Research in this regard is underway, using CoFe$_2$O$_4$ as the replacement for LuFe$_2$O$_4$.

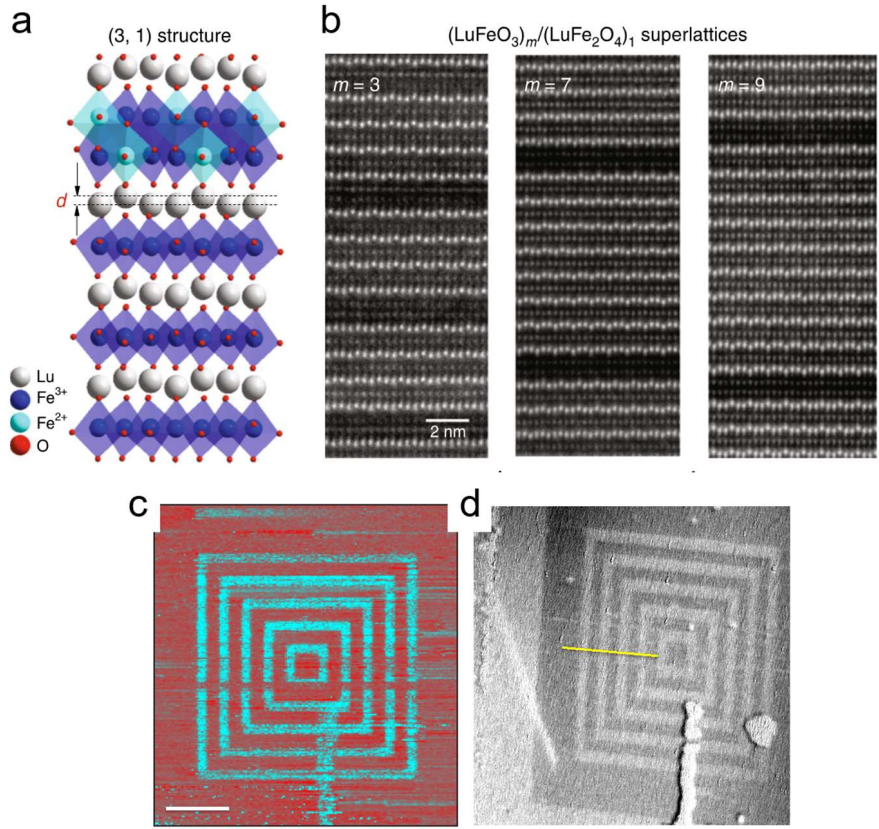

*Figure 16. Epitaxial magnetoelectric superlattices from the improper ferroelectric LuFeO$_3$. (a) a schematic of the crystal structure of LuFeO$_3$/LuFe$_2$O$_4$ superlattice; (b) atomic resolution images of superlattices with various LuFeO$_3$/LuFe$_2$O$_4$ stacking sequences and the corresponding ferrimagnetic T$_C$ of the LuFe$_2$O$_4$; (c) A piezoforce microscopy image of a LuFeO$_3$/LuFe$_2$O$_4$ superlattice showing the box-in-a-box switching of the ferroelectric polarization ; (d) the corresponding XMCD-PEEM image at the Fe-edge showing the switching of the magnetization state [45].*

### 2.3. Electric-field effects in magnetic semiconductors, oxides and metal ultrathin films

Since magnetism is usually intimately linked to the electronic structure and carrier density of materials, accumulating or depleting charges into a magnet may influence its transition temperature, magnetization, anisotropy and even its magnetic order. Charge accumulation/depletion can be achieved using a dielectric or a ferroelectric, in which case the amount of added/removed charge is typically higher (in the $10^{13}$-$10^{14}$ cm$^{-2}$, depending on the ferroelectric polarization value, vs $10^{11}$-$10^{13}$ cm$^{-2}$ with a dielectric, depending on its dielectric constant and on the electric field applied) and remanent. This provides a means to electrically control magnetism in a nonvolatile fashion. Another possibility to accumulate or deplete charge is to use an ionic liquid. When a voltage is applied, a huge electric field of the order of 10 MV/cm is generated at interface between the liquid and the magnetic film due to the formation of an electric double layer. Ionic liquid gating can lead to charge density accumulation up to ~$10^{15}$ cm$^{-2}$.

While the elastic interaction harnessed in strain-driven magnetoelectrics can extend over several hundreds of nanometers, the field effect operates over distances of the order of the Thomas-Fermi screening length ($\lambda_{TF}$), which is a few angstroms in metals and a few nanometers in semiconductors.

In magnetic materials, it has been argued that changes in the magnetic properties may be perceived over distances set by the exchange interaction length which is usually larger than $\lambda_{TF}$ and can approach 10 nm [196].

Several mechanisms occur to electronically drive changes in the magnetic properties. A first one corresponds to electrostatic doping (that is charge accumulation and depletion in a conductor at the interface with a dielectric or a ferroelectric [197]) of the interfacial region in the ferromagnet: if the magnetic properties are strongly doping-dependent, as in carrier-mediated ferromagnets such as (Ga,Mn)As or mixed-valence manganites, charge accumulation or depletion will lead to changes in the magnetic response. A second mechanism is related to the spin-dependent screening in the ferromagnetic of the interface-bound charges of the ferroelectric. In ferromagnetic metals, due to the different density of states for spin up and spin down electrons at the Fermi level, the screening is spin dependent. This spin-dependent screening leads to changes in the surface magnetization and surface magnetocrystalline anisotropy [198]. A third contribution is due to changes in the electronic bonding at the interface between the ferroelectric and the ferromagnet (electronic reconstruction). The displacements of atoms in the ferroelectric due to the polarization reversal influence the overlap between the orbital of the ferroelectric and ferromagnet materials at the interface [199]. This leads to charge redistribution which affects the magnetization, anisotropy and spin polarization at the interface. Related to it, magnetic reconstruction may occur upon accumulating or depleting charges. This mechanism is particularly appealing in materials such as manganites possessing very rich phase diagrams, with magnetic competing phases as function of carrier doping.

In the following we cover these effects for three family of materials namely magnetic semiconductors, magnetic oxides and transition metals. The most spectacular effects have been seen in the former two families, albeit mostly at low temperature due to the low $T_C$ of these compounds. Using ionic liquids, large modulations have also been seen at room temperature with ultrathin transition metal films.

### 2.3.1. Magnetic semiconductors

Charge-driven magnetoelectric coupling was first explored more than twenty years ago in carrier-mediated ferromagnets such as diluted magnetic semiconductors (DMS) [200]. Experimentally, the first demonstration of an electric control of the magnetic state in these systems was in (In,Mn)As thin film in a field-effect transistor geometry using a polyimide layer as the dielectric [201]. The authors measured the anomalous Hall effect of the ferromagnet as a function of the applied gate voltage and could thus detect a modulation of the Curie temperature of about 2 K upon applying a voltage of ±125 V, cf **Figure 17.** A similar but larger effect was later observed using standard magnetometry in (Ga,Mn)As using $HfO_2$ as the dielectric [202]. Importantly, the data can be well explained by simulations using the p-d Zener model, responsible for ferromagnetism in DMS [203]. Similar effects were subsequently reported in other types of DMS, see e.g. [204,205]. Not only the Curie temperature has been modulated electrically in these systems, but also the magnetic anisotropy [206] and magnetic domain wall motion [207]. A non-volatile electric field transition from a ferromagnetic state (accumulation) to a paramagnetic (depletion) one was demonstrated a few years later by replacing the dielectric gate by a ferroelectric one [208].

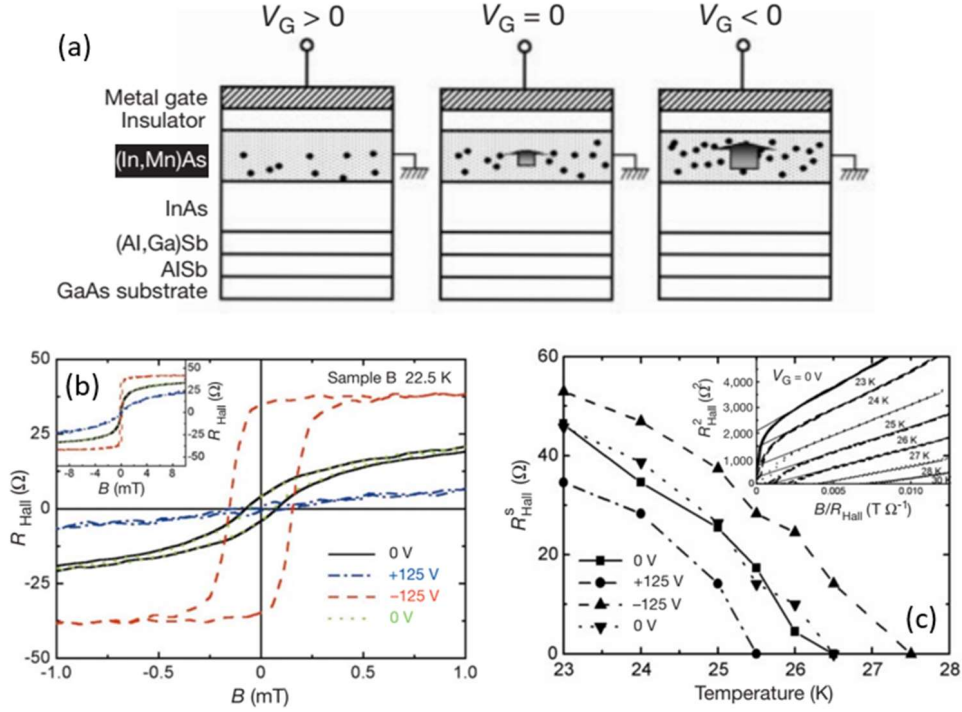

*Figure 17. (a) Field-effect control of the hole-induced ferromagnetism in magnetic semiconductor (In,Mn)As field-effect transistors. The gate voltage $V_G$ applied through the insulator controls the hole concentration in the magnetic semiconductor channel (filled circles). Negative $V_G$ increases hole concentration, resulting in enhancement of the ferromagnetic interaction among magnetic Mn ions, whereas positive $V_G$ has an opposite effect. The arrow schematically shows the magnitude of the Mn magnetization. (b) Hall effect for different gate voltages. When holes are partially depleted from the channel ($V_G$=+125 V), a paramagnetic response is observed (blue dash-dotted line), whereas a clear hysteresis at low fields (<0.7 mT) appears as holes are accumulated in the channel ($V_G$=-125 V, red dashed line). Two Hall curves measured at $V_G$=0 V before and after application of -125 V (black solid line and green dotted line, respectively) are virtually identical (i.e. the effect is volatile). Inset, the same curves shown at higher magnetic fields. (c) Temperature dependence of spontaneous Hall resistance $R^S_{Hall}$ under three different gate biases. $R^S_{Hall}$ proportional to the spontaneous magnetization $M_S$ indicates ±1 K modulation of $T_C$ upon application of $V_G$=±125 V. $T_C$ is determined using Arrott plots (shown in inset) [201].*

### 2.3.2. Oxide heterostructures

Because they crystallize in the same perovskite structures are reference ferroelectrics (BaTiO$_3$, Pb(Zr,Ti)O$_3$, etc), magnetic perovskite oxides can be combined with them into epitaxial heterostructures, to achieve an electrical control of magnetic properties. Being typical carrier mediated ferromagnets, manganites (La$_{1-x}$Sr$_x$MnO$_3$) soon appeared as natural candidates for magnetoelectric effects. For example, Kanki et al. [209] evidenced electric-field-induced modifications in the magnetic moment amplitude of a 10 nm La$_{0.85}$Ba$_{0.15}$MnO$_3$ channel by XMCD experiments close to the metal-insulator transition temperature, using Pb(Zr,Ti)O$_3$ as the ferroelectric gate oxide. This modulation was ascribed to changes induced in the carrier density in the channel depending on the remanent ferroelectric polarization direction in the Pb(Zr,Ti)O$_3$ ferroelectric gate as revealed by the resistance dependence. Lu et al. observed a 10% modulation of the magnetization upon polarization reversal in La$_{0.67}$Sr$_{0.33}$MnO$_3$ (10nm)/BTO bilayers grown on SrTiO$_3$(001) substrates [210]. The large change in magnetization, inversely proportional to the La$_{0.67}$Sr$_{0.33}$MnO$_3$ thickness was ascribed to the carrier modulation and to the shift in the metal-insulator transition near room temperature.

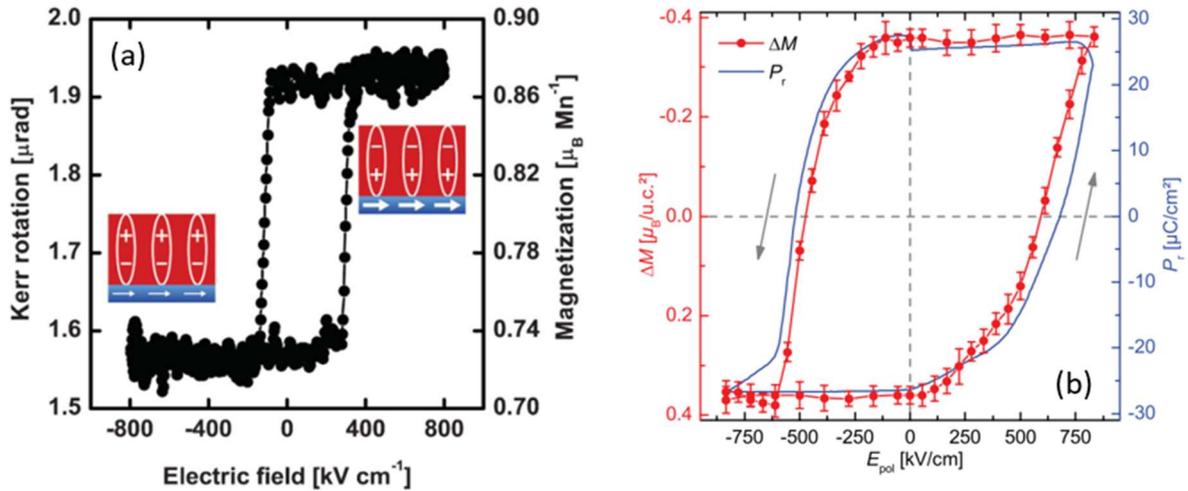

*Figure 18. (a) Magnetoelectric hysteresis curve at 100 K showing the magnetic response of the Pb(Zr,Ti)O$_3$/La$_{0.7}$Sr$_{0.3}$MnO$_3$ system as a function of the applied electric field. The two magnetization values correspond to modulation of the magnetization of the La$_{0.7}$Sr$_{0.3}$MnO$_3$ layer. Insets represent the magnetic and electric states of the La$_{0.7}$Sr$_{0.3}$MnO$_3$ and Pb(Zr,Ti)O$_3$ layers, respectively. The size of the arrows indicates qualitatively the magnetization amplitude [211]. (b) Comparison of the electric field dependence of the remanent ferroelectric polarization P$_r$ and of the magnetic modulation per unit cell area ΔM measured in a Pb(Zr,Ti)O$_3$/La$_{0.7}$Sr$_{0.3}$MnO$_3$ bilayer. Both curves measured consecutively at 50 K and 100 Oe [212].*

Interestingly, an electrically induced magnetic transition was identified in La$_{0.8}$Sr$_{0.2}$MnO$_3$ (4 nm)/Pb(Zr,Ti)O$_3$ bilayers [211]. Important modifications in the Curie temperature and magnetization amplitude at 100K probed by Kerr magnetometry were reported in this system, see **Figure 18.** Additional experiments using X-ray absorption near edge spectroscopy revealed the charge-induced change by polarization switching in the valence state of Mn atoms (0.1 electrons per Mn atom) in the La$_{0.8}$Sr$_{0.2}$MnO$_3$ layer [213]. From combined spectroscopic, magnetic, and electric characterizations of this system, Vaz et al. concluded that the magnetic spin configuration of the La$_{0.8}$Sr$_{0.2}$MnO$_3$ at the Pb(Zr,Ti)O$_3$ interface changes from ferromagnetic in the depletion state to A-type antiferromagnetic in the accumulation state (increase of hole doping) and that this interface-charge-driven ME coupling is at the origin of the effect [214]. In the accumulated state, the interface layer consists of strongly depopulated, antibonding 3$d$ $e_g$ 3$z^2$–$r^2$ states, resulting in a weakening of the double-exchange interaction at these orbitals. An antiferromagnetic coupling to the adjacent layers ensures that the 3$d$ $e_g$ $x^2$–$y^2$ orbitals are energetically privileged, favoring the super exchange interaction and a transition from a ferromagnetic state to an antiferromagnetic one consistent with theoretical predictions for related systems [215]. Ma et al. also reported a change by one order of magnitude in the in-plane and out-of-plane magnetizations at La$_{0.67}$Sr$_{0.33}$MnO$_3$/Pb(Zr,Ti)O$_3$ interfaces due to the appearance of an antiferromagnetic spin alignment induced by hole doping [216].

Perhaps the most spectacular electric-field modulation of magnetism in La$_{0.7}$Sr$_{0.3}$MnO$_3$/Pb(Zr,Ti)O$_3$ bilayers is from Leufke et al, see **Figure 18.** [212]. The excellent correspondence of the polarization vs E and magnetization vs E loops indicates a purely electrostatic doping as the origin of the effect, with negligible contribution from piezoelectricity and/or electrochemistry (see later). The authors analyzed in detail the dependence of the effect on the poling voltage and on temperature to conclude that phase separation between antiferromagnetic and ferromagnetic regions, a common feature of mixed-valence manganites [217], played a significant role in the observed effects.

In heterostructures combining a ferroelectric such as Pb(Zr,Ti)$O_3$ and a ferromagnet like La$_{0.7}$Sr$_{0.3}$MnO$_3$, the influence of the electric field on magnetism may arise from both field effect and strain-driven effects, due to the piezoelectric nature of the ferroelectric. Several studies have evidenced the coexistence of both mechanisms and separated them. Typically, strain-driven effect has an even dependence on electric field while charge-driven ones are odd. Since strain effects can extend over large thicknesses into the magnetic film while charge-driven effect are purely interfacial, studying magnetization vs electric field loops as a function of thickness typically yields a crossover between both types of behavior [218]. Huang et al have evidenced this phenomenon clearly, also concluding on the influence of orbital reconstruction effects in the low thickness limit [219,220].

Gating of manganites with ionic liquids has also been attempted, leading to striking results. As always with electric double layer systems, but perhaps even more importantly with oxides in which oxygen diffusion can be strong, in such experiments electrostatic effects may be accompanied by electrochemistry (that is, ion migration between the electrolyte and the channel material), and both contributions are notoriously hard to separate [221,222]. Dhoot et al reported a resistance change approaching 100% and modulations of the metal-insulator transition temperature (corresponding to $T_C$ in these compounds) by over 30 K [223]. Even larger modulations were later found by others in other manganites e.g. [224–226]. The results of Molinari et al [221] correspond to an actual measurement of magnetization under the influence of ionic liquid gating. Working just above room temperature and just below the $T_C$ of an La$_{0.7}$Sr$_{0.3}$MnO$_3$ film these authors are able to modulate magnetization reversibly over tens of cycles with just ±200 mV [221].

### 2.3.3. Transition metal and alloys

In order to achieve effects at room temperature and in materials more compatible with applications, the electric-field effect has been explored on ferromagnets based on transition metals and their alloys. The first report of voltage-controlled magnetism in transition metals was by Weisheit et al who observed a modulation of the coercive field of FePt ultrathin film by about 5% at room temperature [227], cf **Figure 19**a-b. A couple of years later, the first results on the voltage control of magnetic anisotropy (VCMA) in an all-solid-state system were reported in Fe/MgO [228] and CoFeB/MgO [229], see **Figure 19**c-d. The electric field was applied across a polyimide layer and a ZrO$_2$ layer, respectively. The mechanism underlying the observed VCMA was investigated theoretically and proposed to be related to changes in the hybridization between O 2p states and different Fe 3d orbitals [230,231]. VCMA was used to induce magnetization reversal and thus to switch a MTJ between parallel and antiparallel states. The application of a short voltage pulse induces the precession of the magnetization which, if the pulse is properly timed, reverses.

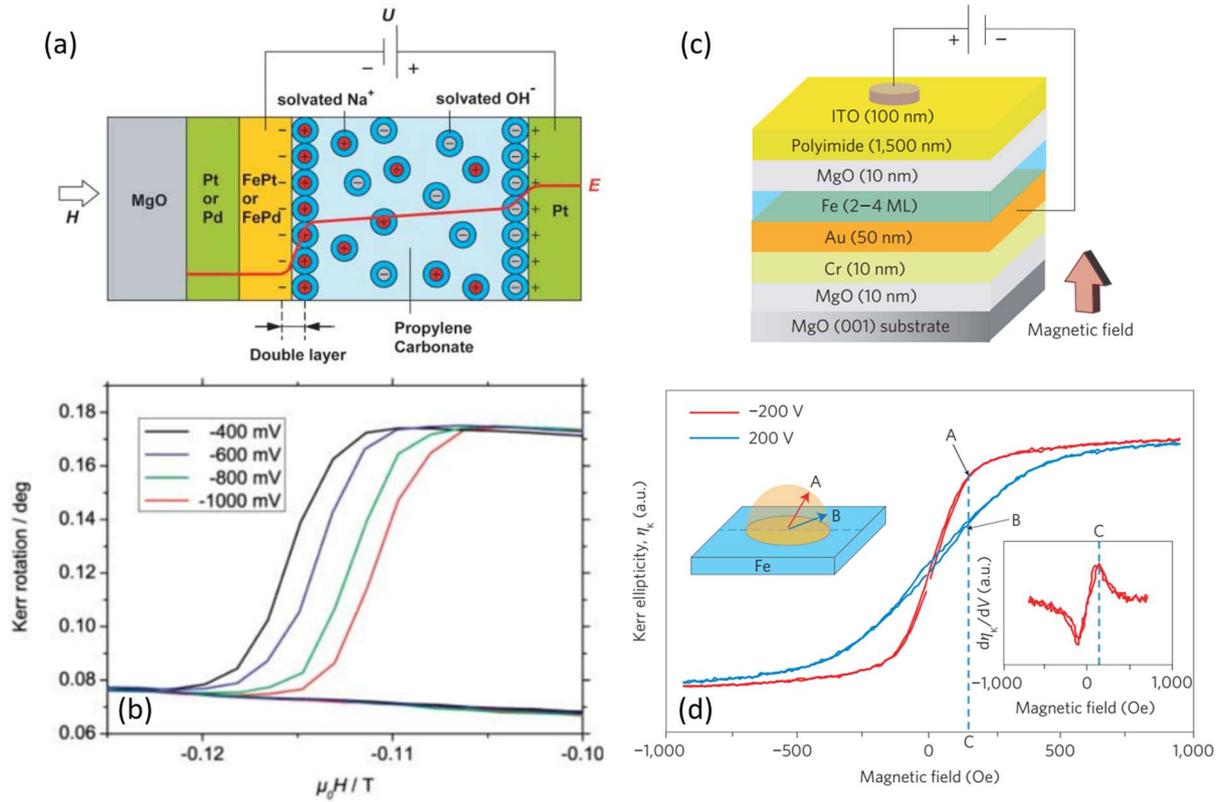

*Figure 19. (a) Fig. 1. Schematic of an electrolytic cell containing the FePt or FePd film within an applied magnetic field H. The potential profile E due to the applied potential U is indicated by the red line. The potential drop at the Pt electrode side is much lower (as compared to that of the sample surface) as a result of the Pt electrode's large surface area. (b) Magnetization switching of the 2-nm-thick FePt film for different U values between the film and the Pt counter electrode.* [227] *(c) Schematic of the sample used for a voltage-induced magnetic anisotropy change. (d) Magneto-optical Kerr ellipticity $\eta_k$ for different applied voltages as a function of applied field. The thickness of the Fe film was 0.48 nm. A significant change in the hysteresis curve indicated a large change in perpendicular anisotropy following application of the bias voltage. The right inset shows the voltage modulation response of the Kerr ellipticity, $d\eta_k/dV$. The left inset illustrates the magnetization direction at points A and B in the hysteresis curves* [228].

Accumulating and depleting charge into a ferromagnet is also expected to yield a modulation of its Curie temperature, which was realized by Chiba et al in 0.4 nm Co films using $HfO_2$ as the gate dielectric [232]. Upon applying ±10 V, these authors were able to shift $T_C$ by about 12 K, resulting in the electrical switching between ferromagnetism and paramagnetism around 320 K.

Parallel to these pioneering results, the possibility to use ferroelectricity to control the magnetism of transition metal layers was explored. Research in this direction has been mainly through first-principles calculations, in particular for the $BaTiO_3$/Fe system [233–235]. In particular, ferroelectric switching was predicted to influence the magnetic moment at the interface and the spin polarization near the Fermi energy, which will be exploited in so-called multiferroic tunnel junctions [236,237] (see Section 5.1.2). Using XMCD at the Co $L_{3,2}$ edge, Heidler et al observed a hysteretic dependence of the Co magnetic moment as a function of electric field in Co/PMN-PT [238]. The data suggested a combination of strain- and charge-induced effects. Mardana et al combined a Co ultrathin film with a ferroelectric polymer, P(VDF-FrFE), to achieve a non-volatile electrical control of magnetic coercivity [239]. Subsequent studies reported a hysteretic dependence of coercivity with electric field in CoFeB/BaTiO$_3$ [240] and Fe/BaTiO$_3$ [241] and of the anisotropy field in CoFe/(Ba, Sr)TiO$_3$ [242].

The properties of ferromagnetic domains can also be tuned by charge accumulation/depletion. Domain wall velocity was found to strongly depend on electric field in Co ultrathin films [243]. Using a meshed gate electrode, Ando et al were able to achieve magnetic domain writing by electrical gating [244]. The fact that such charge accumulation/depletion effects require ultrathin films is particularly appealing to control specific spin textures occurring at such low thickness when the ferromagnet is effectively sandwiched between different layers leading to inversion symmetry breaking and unleashing Dzyaloshinskii-Moriya interaction (DMI). Schott et al have exploited this possibility to turn magnetic skyrmion bubbles on and off with an electric field [245].

Just as for the manganites, the most spectacular effects have been obtained using ionic liquid gating. As displayed in **Figure 20**a-b a shift in T$_C$ by about 100 K was observed upon applying ±2V in ultrathin Co films [246].

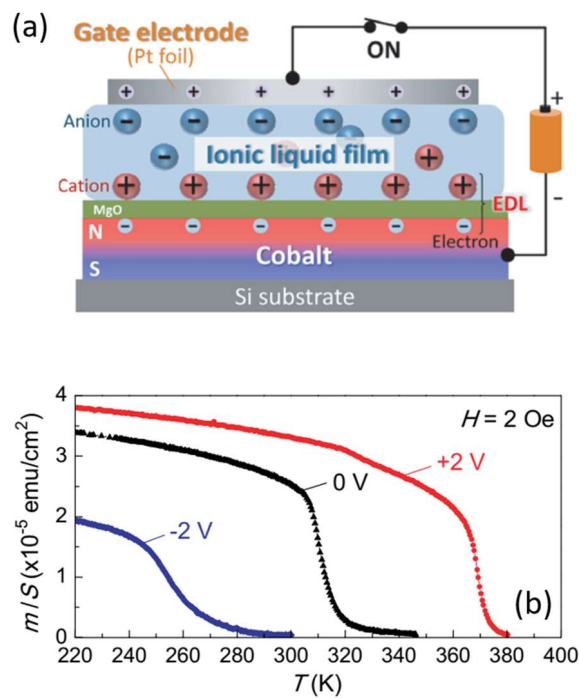

*Figure 20. (a) Sketch of the device for the modulation of the magnetic properties of a Co film. (b) Temperature dependence of the magnetization at H=2 Oe under a gate voltage V$_G$=-2 V, 0 V, and +2 V [246].*

Before moving to the next section on 2D magnets, we briefly assess the advantages and inconvenients of the approaches for electric-field control of magnetism we have just discussed, namely exchange-based magnetoelectric coupling (in single phase materials or in heterostructures involving a room-temperature multiferroic such as BiFeO$_3$), strain-induced control of magnetization and electric field effect. All three approaches have evidenced a response at room temperature, although for the first one the choice of materials is very limited (to BiFeO$_3$ and some hexaferrites with complex unit cells that have not yet been grown as thin films). It is however the most straightforward approach to achieve a 180 degree switching of magnetization. This may also be achieved using strain-based magnetoelectric coupling, but through complex writing protocols, and through field effect [247], although this remains to be shown. As a result, the most promising strategy so far still relies on the use of BiFeO$_3$, although the deterministic nature of the switching is a major issue [248]. This emphasizes the need for both new materials, perhaps in the 2D family (see below), and for further imaginative schemes for strain and field-effect based approaches.

### 2.4. 2D magnets

Before the discovery of intrinsic magnetism in different two-dimensional (2D) materials in 2017, such possibility was disregarded based on the Mermin-Wagner theorem [249], which was formulated for the case of isotropic Heisenberg model with finite-range interactions. However, the presence of uniaxial anisotropy (such as magnetocrystalline anisotropy caused by spin-orbit coupling) allows the stabilization of magnetic order in 2D [250], a possibility which was experimentally confirmed in different van der Waals (vdW) materials.

The first experimental demonstration of 2D magnetism was reported in Cr$_2$Ge$_2$Te$_6$ vdW semiconductors down to the bilayer limit with unprecedented control of the Curie temperature ($T_C$) with low applied magnetic fields [251]. Another breakthrough experiment demonstrated intrinsic 2D magnetism down to the monolayer limit in insulating exfoliated CrI$_3$ [252]. Interestingly, these vdW materials showed layer-dependent magnetism due to behavior alternating between ferromagnetic and antiferromagnetic states as number of layer increases. The third exfoliated material reported to show long range magnetic order in 2017 was metallic Fe$_3$GeTe$_2$, with a higher $T_C$ than the other two materials [253–255]. Some transition metal dichalcogenides (TMD), e.g. VSe$_2$ [256] and MnSe$_2$ [257], have also been reported to be magnetic in some of their crystallographic phases. Ising-type magnetic ordering has also been demonstrated in phosphorous-based insulating antiferromagnets, e.g. in FePS$_3$ [258].

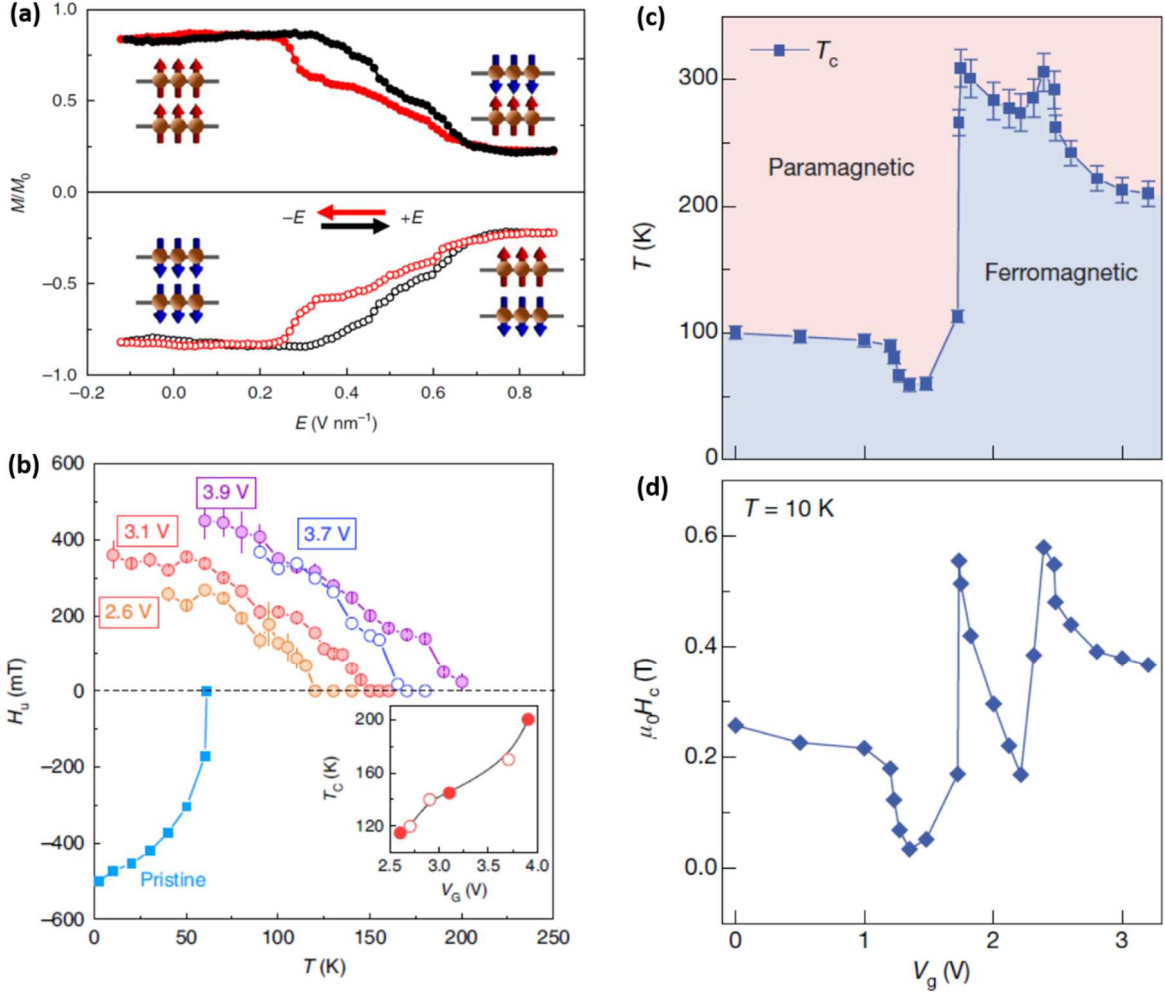

*Figure 21.* (a) Top: Normalized magnetization measured by magnetic circular dichroism (MCD) as a function of the applied electric field (trace and retrace) at 4 K and fixed magnetic field (+0.44 T for top panel and -0.44 T for bottom panel), showing the electrical switching of the magnetic order in bilayer $CrI_3$. The insets represent the corresponding magnetic states [259]. (b) Uniaxial magnetic anisotropy field ($H_u = H_s^\perp - H_s^\parallel$) of multilayer $Cr_2Ge_2Te_6$ as a function of temperature at different gate voltages and in the pristine case. Inset: the dependence of $T_C$ on gate voltage [260]. (c) $T_C$ of a trilayer $Fe_3GeTe_2$ as a function of gate voltage [255]. (d) $H_C$ of a trilayer $Fe_3GeTe_2$ as a function of gate voltage at 10 K [255].

These materials form part of more general families of 2D vdW structures. Such a large number of atomically thin vdW magnets show a wide variety of electrical and magnetic properties, ranging from ferromagnetic semiconductors or metals to antiferromagnetic insulators. Due to their 2D character, they are much more sensitive to external stimuli, in particular electric field, allowing efficient control of their magnetic properties. They can be naturally stacked with a wide range of vdW materials, forming heterostructures with almost ideal interfaces. The electrical control of magnetism in a 2D magnet can occur via different mechanisms, such as linear magnetoelectric coupling or electrostatic doping.

The former mechanism requires the material to break simultaneously time-reversal symmetry and inversion symmetry, a condition fulfilled by bilayer $CrI_3$ in the antiferromagnetic ground state, but not by the ferromagnetic phase or by the monolayer $CrI_3$, in which inversion symmetry is present. Jiang et al. (Jiang et al. 2018a) measured the magnetoelectric response with magnetic circular dichroism (MCD) and using a dual gate structure to apply an electric field in order to take out the effect of doping. Interestingly, the magnetoelectric coupling was maximum around the spin-flip transition that occurs at ~0.5 T. This made it possible to switch electrically bilayer CrI$_3$ between the antiferromagnetic and ferromagnetic states at a constant magnetic field (close to the spin-flip transition, see **Figure 21**a).

The control of magnetism is also possible via electrostatic doping in 2D magnets. This mechanism has the benefit that it does not require the specific symmetry of the linear magnetoelectric coupling and, in addition to bilayer CrI$_3$ (Jiang et al. 2018b; B. Huang et al. 2018), it is also present in monolayer CrI$_3$ (Jiang et al. 2018b) and in Cr$_2$Ge$_2$Te$_6$ [260,263]. In the case of monolayer CrI$_3$ (Jiang et al. 2018b), saturation magnetization ($M_S$), coercive field ($H_C$) and $T_C$ increase (decrease) with hole (electron) doping. In bilayer CrI$_3$, electron doping (~2.5×10$^{13}$ cm$^{-2}$) reduces the spin-flip transition almost to zero magnetic field (Jiang et al. 2018b). Although this should enable electrical switching of magnetization at zero field, a magnetic field near the spin-flip transition is required for a fully reversible switch (Jiang et al. 2018b; B. Huang et al. 2018). Electrostatic doping using ionic liquid gating has also been reported in multilayer Cr$_2$Ge$_2$Te$_6$ [260,263]. Wang et al. [263] used magneto-optical Kerr effect (MOKE) measurements to report that saturation field ($H_S$) decreases and $M_S$ increases as a function of doping levels (both electron and hole), while $H_C$ and $T_C$ are insensitive to doping. This performance was tentatively attributed to a moment rebalance of the spin-polarized band structure while tuning its Fermi level. On the contrary, Verzhbitskiy *et al.* [260] showed a shift of $T_C$ from ~61 K to up to 200 K when an electron doping of ~4×10$^{14}$ cm$^{-2}$ was applied, using magnetoresistance measurements. Additionally, the magnetic anisotropy was dramatically changed, moving from perpendicular to in-plane (see **Figure 21**b). The authors attribute the occurrence of such effect to a double-exchange mechanism that is mediated by free carriers, which dominates over the super-exchange mechanism of the original insulating state.

A voltage control of magnetism with a completely different origin has been reported in multilayer CrI$_3$. In this material, memristive switching is observed when a large enough voltage is applied, where the two resistive states are coupled to the magnetic phases [264]. The origin of the effect is a thermally induced mechanism when current flows across CrI$_3$.

Voltage control of magnetism has also been reported in Fe$_3$GeTe$_2$ which, unlike the previous 2D magnets mentioned in this subsection, is metallic. Deng et al. [255] applied ionic gating to bring $T_C$ from ~100 K up to ~300 K in trilayer Fe$_3$GeTe$_2$ (see **Figure 21**c), a very remarkable observation as to date no pristine 2D magnet is ferromagnetic at room temperature. As plotted in **Figure 21**d, $H_C$ roughly follows the variation of $T_C$ with voltage gate. The large electron doping induced by the ionic gate (~10$^{14}$ cm$^{-2}$ per layer) causes a substantial shift of the electronic bands of Fe$_3$GeTe$_2$. The large variation in the density of states (DOS) at the Fermi level leads to appreciable modulation in the ferromagnetism, in agreement with the Stoner model for itinerant electrons [255,265]. Finally, metallic ferromagnet Fe$_5$GeTe$_2$ (F5GT) has been electron doped with protonic gating, which can induce a transition to an antiferromagnetic phase at 2 K [266].

### 2.5. Electric-field control of magnetic skyrmions

Magnetic skyrmions are two dimensional topological solitonic spin textures that can be stabilized in chiral magnets thanks to the Dzyaloshinskii-Moriya interaction (DMI), anisotropic interactions existing in the absence of inversion symmetry, either in non-centrosymmetric lattices [267,268] or when the breaking of inversion symmetry is due to defects or interfaces [269–271]. Section 3.5 is complementary to the present one; it describes skyrmions in more detail and discusses how they can be manipulated by electrical currents. Skyrmions have some similarities with magnetic bubbles, which were used to store data in a non-volatile memory, popular in the 1970s and 1980s [272], before being replaced by more advanced technologies such as hard disk drives and flash memories. However, skyrmion devices have the potential to offer much higher data storage densities than bubble memory, due to the smaller size of skyrmions and their stability given by the topological protection. Another difference is the way that the data is manipulated: while skyrmion devices use spintronic techniques based on charge currents, bubble memory used magnetic fields to move the bubbles, which did not favor downscaling.

Over the last decade, magnetic skyrmions were observed in a wide range of materials and heterostructures including metallic MnSi [273,274] or FeGe [275], but also insulating $Cu_2OSeO_3$ [276]. In insulating skyrmion lattice compounds, the chiral lattice gives rise to a magnetoelectric (ME) coupling between electric and magnetic orders, opening a path for electric-field control of magnetic skyrmions, with potentially no Joule-heating dissipation. In single crystal $Cu_2OSeO_3$, it was demonstrated that electric field can induce a rotation of the skyrmion lattice via this ME coupling [277]. These giant skyrmion lattice rotations (spanning in a range of 25°) operate via skyrmion distortion, as supported by calculations. However, this skyrmion lattice is restricted to a narrow temperature (54-58 K) and magnetic field region in $Cu_2OSeO_3$. The electric-field control of the skyrmion phase pocket was revealed combining magnetic susceptibility and microwave spectroscopy (**Figure 22**a) [278] and further confirmed using neutron scattering [279]. Thus, the metastable skyrmion lattice can be created and erased isothermally under electric fields and in a non-volatile manner [278,280]. Using real-space methods such as Lorentz transmission electron microscopy, a skyrmion lattice could be reversibly written and erased under electric field pulses from a helical spin background in transistor devices based on single crystal $Cu_2OSeO_3$ (**Figure 22**b-c) [281].

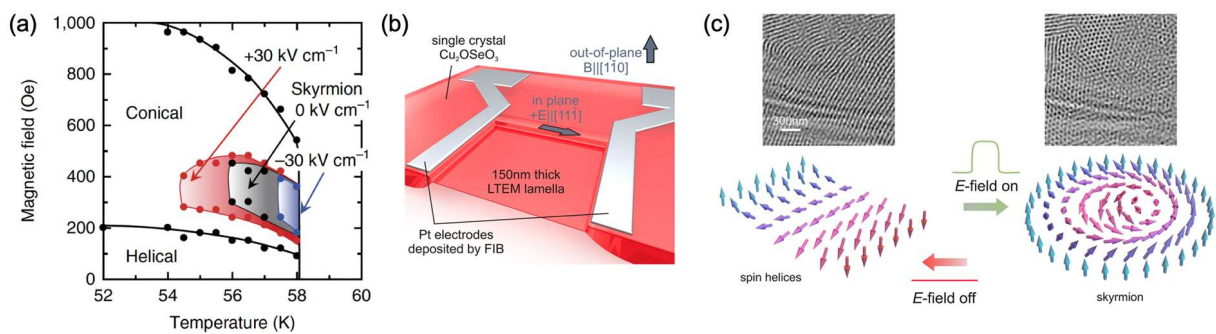

*Figure 22. (a) Electric-field control of the skyrmion phase pocket in single crystal $Cu_2OSeO_3$. Electric and magnetic fields are parallel to the [111] direction of the crystal (from [278]). (b) Schematic of the single crystal $Cu_2OSeO_3$ sample configuration using patterned Pt electrodes to apply in-plane electric fields of 3.6 V/μm. (c) Reversible electric field transition between the helical spin state and the skyrmion lattice visualized by for Lorentz transmission electron microscopy (T = 24.7 K under an out-of-plane magnetic field of 254 Oe). (b-c from [281]).*

Attempts have also been made to stabilize skyrmions in oxide heterostructures (see e.g. [282,283]) and to control them by electric field. We mention the results from Wang et al who reported the observation of skyrmion bubbles (see 2.5 for more on skyrmions and their electric-field control) in $SrRuO_3/BaTiO_3$ bilayers with a skyrmion density and associated topological Hall effect tunable by ferroelectric polarization [284], cf **Figure 23**. We note however that reports of skyrmions in SrRuO₃ heterostructures and the interpretation of the topological Hall effect are still under intense debate, cf e.g. [285,286].

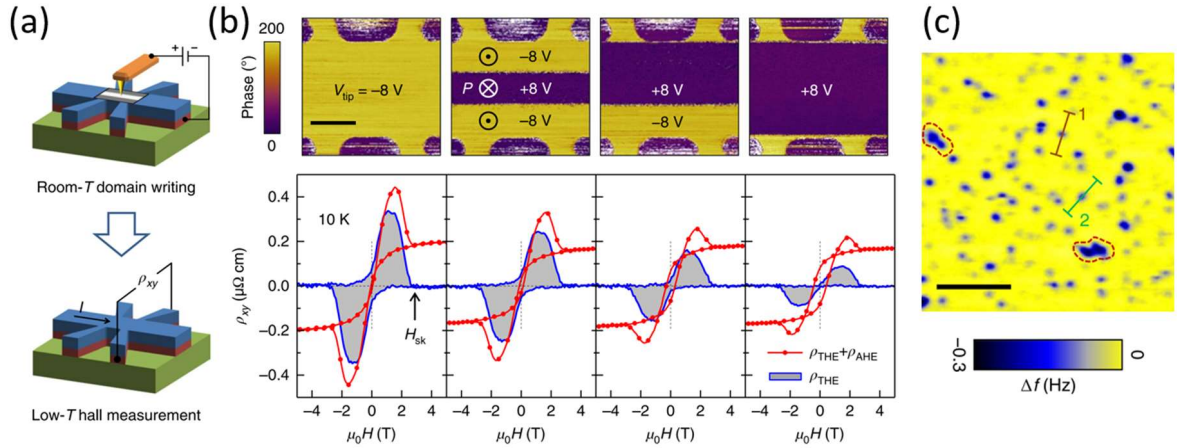

*Figure 23. (a) Schematic diagram of the experimental set-up for ferroelectric domain switching using an antiferromagnetic tip and to perform Hall measurements. (b) Piezoresponse force microscopy phase images (top panels), Hall and extracted topological Hall curves (bottom panels) of a SrRuO₃/BaTiO₃ sample for different ferroelectric poling states. The scale bar corresponds to 10 μm. (c) Difference in MFM contrast between images taken at two different magnetic fields. See original paper for details. [284]*

Novel 2D multiferroic materials were predicted in Co intercalated MoS₂ dichalcogenides, with degenerate DMIs in the two ferroelectric states. The chirality of the skyrmions stabilized in such 2D multiferroics can therefore be reversed by electric fields thanks to the ME coupling [287]. Combining a bilayer van der Waals of WTe₂/CrCl₃ with a 2D ferroelectric CuInP₂S₆, the electric-field writing and deletion of Néel-type skyrmions was predicted, where an interfacial ME coupling involving polarization-induced electronic reconstruction gives rise to a non-volatile control of the DMI [288].

While in single phase chiral magnets the skyrmion phase is limited to low temperature, asymmetric multilayer stacks of heavy metals and ferromagnetic layers can give rise to room-temperature skyrmions [289–291], stabilized by interfacial DMIs [292,293]. In multiferroic heterostructures consisting of such asymmetric [Pt/Co/Ta]₅ multilayers and a ferroelectric Pb(Mg$_{1/3}$Nb$_{2/3}$)$_{0.7}$Ti$_{0.3}$O₃ (PMN-PT) layer, the strain-mediated electric-field control of skyrmions was recently demonstrated [294]. The observations of electric-field-induced creation, deformation, and annihilation of the skyrmions were corroborated by strain-induced variations of both the magnetic anisotropy and the interfacial DMI. Electromechanical and micromagnetic simulations revealed that applying a voltage between two lateral electrodes in such multiferroic heterostructures can give rise to a transverse strain gradient, because of the non-uniform electric-field profile in the piezoelectric material. Owing to the magnetoelastic coupling, this strain gradient can be used to compensate the skyrmion Hall angle and propagate more efficiently skyrmions under spin transfer torque [295].

Writing and deleting individual skyrmions with an electric-field was originally demonstrated at low temperature (7.8 K) using spin-polarized scanning tunneling microscopy on an ultrathin Fe layer on Ir(111) [296]. The main mechanism involved was a change of the magnetic exchange interaction with the electric field, leading to either a ferromagnetic ground state (positive electric field) or a skyrmion state (negative electric field). Using a CoO/Co/Pt trilayer, in which large interfacial DMIs were reported and with a Co thickness close to the ferromagnetic-paramagnetic transition at room temperature, micron-size skyrmion bubbles could be reversibly written and erased using an electric-field [245]. These modifications were interpreted by a modulation of the magnetization and anisotropy under electric-field, possibly via changes in the electron density of state of the ultrathin Co layer. In Ta/FeCoB/TaO$_x$ trilayers, a 130% variation of the DMI under voltage could be detected using Brillouin light spectroscopy and magneto-optic Kerr microscopy [297]. These results and the correlated size variations of the skyrmion bubbles were explained by the large sensitivity of the FeCoB/TaO$_x$ Rashba DMI to the electric field. The electric-field creation and directional motion of chiral domain walls and skyrmion bubbles could be achieved in SiO$_2$/Pt/CoNi/Pt/CoNi/Pt multilayer with thickness gradient and interfacial DMI [298]. The SiO$_2$/Pt interface provides a large electric-field induced magnetic anisotropy change due to the electric quadrupole induction. Recently, a femtosecond pulse electric field was predicted to generate a DMI in single ultrathin metallic thin films [299]. This mechanism allows the coherent nucleation of skyrmions, as well as other exotic topological defects (antiskyrmions, target skyrmions, etc) by modifying the properties of the ultrafast electric field pulse.

To make a brief aside, polar skyrmions and other possible topological objects (polar vortices, center domains, merons, etc) are now gathering a lot of interest among the ferroelectric community [300], as these objects would be smaller than their magnetic counterparts and naturally controlled by an electric field [301–303]. Indeed, polar skyrmions were recently observed in PbTiO$_3$/SrTiO$_3$ superlattices at room temperature [304,305]. This field is still in its infancy and the complex competition between depolarizing fields, strain and electric field gradients is currently investigated. Stabilizing polar chirality in domain walls and bubbles is a prerequisite [69,306,307], while the underlying mechanisms for such polar chirality are not clearly identified. Recently, the electric analog of the DMI was proposed [308], opening an avenue for the design of topological objects in ferroelectrics and multiferroics.

### 2.6. Dynamics

The dynamics of the antiferromagnetic and ferroelectric states and the coupling between them can be probed in either time domain or frequency domain-based measurements. While the fundamental physics of magnons, electromagnons and ferroelectromagnons are best studied in frequency domain measurements, from a more practical perspective, especially in digital electronics, time-domain measurements are more valuable. The emergence of antiferromagnetic spintronics provides another impetus to both aspects. There have been some key reviews of the high frequency dynamics of multiferroics in recent years [309,310] and we refer the interested reader to these reviews for further details. While there have been a large number of papers published on the physics of the polarization switching process in ferroelectrics over sixty years [311,312], true time-domain studies are still evolving. In capacitive elements such as a ferroelectric or multiferroic capacitor, the time domain dynamics of switching of the order parameter is invariably convoluted with the circuit level parameters (and parasitics), which then obfuscate the intrinsic time dynamics. Thus care is needed to probe the dynamics in such capacitive elements by reducing, resistive losses, as well as circuit level capacitive parasitics.

### 2.6.1. Magnonics

In magnonics, spin waves form the fundamental excitation [313,314]. This field has experienced a re-emergence over the last decade as exciting discoveries have yielded a breadth of interesting new physics as well as the potential for low power computing [315] such as magnon logic [313] antiferromagnetic spin wave field-effect transistors and all-magnon transistors based on magnon-magnon scattering with resonant excitation. There are several ways to create magnons [316], and spin transport via magnon currents have already been reported in a variety of systems [317]. Though resonant excitations are typically used to study spin waves [318], magnon currents can be excited incoherently by a thermal gradient through the spin Seebeck effect (SSE) [319] or the spin accumulation mechanism (SAM) [320] from the spin Hall effect (SHE) to probe non-local spin transport. Previous research has demonstrated non-local spin transport [321] in insulating ferrimagnets [322] and antiferromagnets [323], thermally excited spin-transport over exceptionally long distances [324], and non-volatile magnetic field control [320].

### 2.6.2. Electric control of magnons; ferroelectromagnons

Early work in the fifties and sixties [325] provided the fundamental backbone for the study of coupled spin/charge waves, termed as electromagnons (or more precisely ferroelectromagnons) [326,327]. In simple terms, ferroelectromagnons are the coupling between spin waves and charge waves. A good example is the case of antiferromagnetic spin waves in the prototypical rare earth ferrites [318] such as $DyFeO_3$. Such antiferromagnetic resonances are typically in the 300-350 GHz range, as a direct consequence of the large antiferromagnetic anisotropy field compared to ferromagnets. Replacing Dy with Bi to create $BiFeO_3$ leads to ferroelectromagnons in the 600-800 GHz range. There have been a few studies of such ferroelectromagnons, particularly using Raman and optical probes [63,66,70–72]. **Figure 24** presents Raman experiments from [72] evidencing magnon modes of the cycloidal spin order of a $BiFeO_3$ crystal. Series of modes (cyclon and extra-cyclon) are present due to zone folding. The energy of the modes can be strongly modulated by electric fields and in a hysteretic fashion.

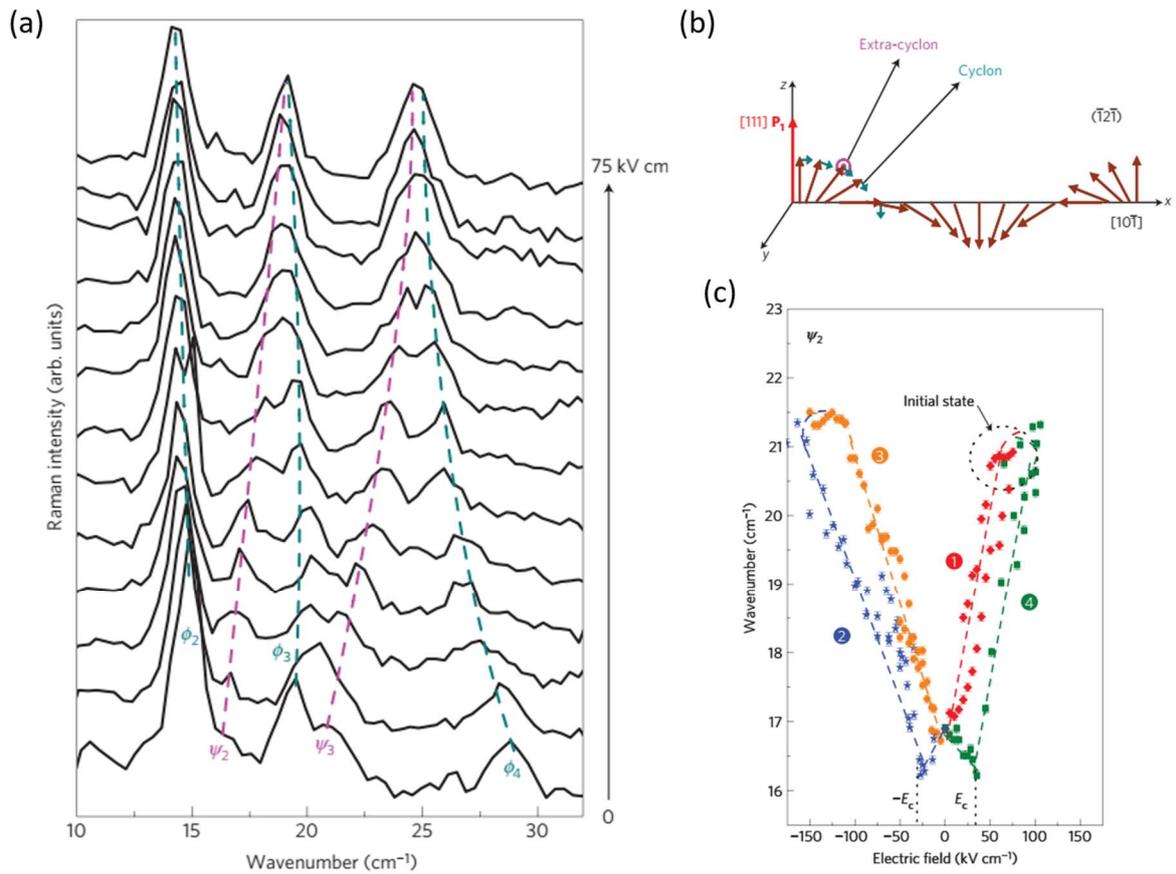

*Figure 24. (a) Raman spectra showing magnon modes (cyclon: $\phi_2$, $\phi_3$, $\phi_4$; extra-cyclon: $\psi_2$, $\psi_3$) in a BiFeO$_3$ single crystal for increasing electric fields. (b) Sketch of the magnon modes in the cycloidal order of BiFeO$_3$. (c) Electric-field dependence of the energy of the $\psi_2$ showing a strong and hysteretic modulation. From [72].*

BiFeO$_3$ provides a good model system to harness the electric-field control of magnons. The ferroelectric and antiferromagnetic domain structures in BiFeO$_3$ exhibit a one-to-one correspondence [68] and deterministic control of magnetic order via manipulation of the ferroelectric state (with applied electric fields) has already been demonstrated [64,65]. The transport of magnons in BiFeO$_3$ in a non-local geometry is shown schematically in **Figure 25**. The devices consist of a metal with a large spin orbit coupling, such as (Pt), deposited on the magnet. One strip functions as injector and the other as detector. When a charge current *I* is sent through the injector, the spin Hall effect [328] generates a transverse spin current (see Section 3.1.3). A spin accumulation then builds up at the Pt/magnet interface. When its spin orientation is parallel (antiparallel) to the average magnetization *M*, magnons are annihilated (excited), resulting in a non-equilibrium magnon population in the magnet. The non-equilibrium magnons diffuse in the magnet, giving a magnon current propagating from injector to detector. At the detector, the reciprocal process occurs: magnons interact at the interface, flipping the spins of electrons and creating a spin imbalance in the Pt [315]. Owing to the inverse spin Hall effect, the induced spin current is converted into charge current, which under open-circuit conditions generates a voltage *V*. **Figure 25**c demonstrates a novel manifestation of magnetoelectric coupling in BiFeO$_3$ to manipulate magnon current [329]. Non-volatile, hysteretic, bistable states of magnon current were observed with an applied electric field, indicating that the electric field induced switching results in changes to the magnon spin polarization pointing across the channel. Thus, in principle, one should be able to sense the magnetic state of the multiferroic by this approach. However, to facilitate magnonic elements operating with a linear response at room temperature, the ideal signal pathway would be: input electronic charge signal → electron spins → magnons → electron spins → output charge signal. This will require exploring both thermal magnons via the spin Seebeck effect and the isothermal spin accumulation mechanism.

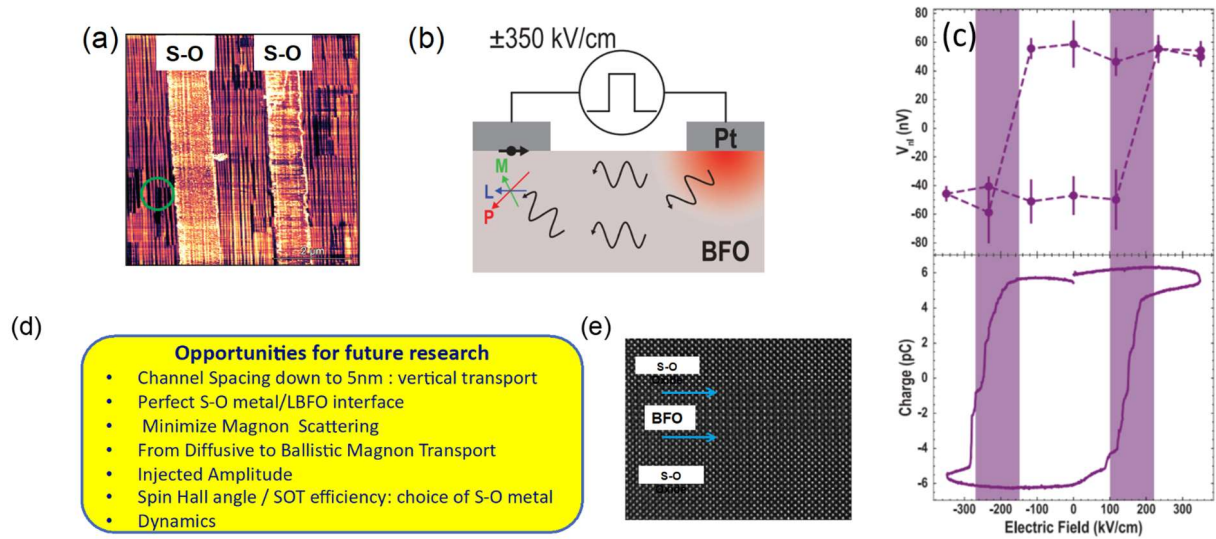

*Figure 25. (a) Piezoforce microscopy (PFM) image of a 100 nm thick BiFeO$_3$ layer on a DyScO$_3$ substrate illustrating the typical 71-degree stripe domains; the two broad stripes notated as S-O are the metal layers (typically a metal with strong spin-orbit coupling such as Pt) that are used to probe the inverse spin Hall and spin Seebeck responses due to the propogation of magnons in the BiFeO$_3$ layer as illustrated in (b); an electric field applied between these two metal strips enables switching of the ferroelectric polarization state of the BiFeO$_3$; (c) the top panel shows the non-local spin Seebeck voltage as a function of dc electric field applied to the BiFeO$_3$ while the lower panel shows the corresponding ferroelectric switching; (d) is a panel that summarizes some areas of research, specifically a focus on ballistic spin/magnon transport in epitaxial heterostructures such as the one shown in (e)* [329].

While much remains to be understood on the fundamentals of magnon transport and its electric field manipulation, the results of these studies point to a rich frontier of spin dynamics in such multiferroics. Of equal importance is the potential for such approaches to lead to larger inverse spin Hall voltages, perhaps through a thorough search for possible candidate materials (such as topological insulators, heavy transition metal-based complex oxides with exotic electronic band structure, such as SrIrO$_3$). Particularly, the fact that the antiferromagnetic state of the multiferroic can be directly read-out using the inverse spin Hall effect means that a ferromagnetic layer to sense the antiferromagnetic state is not required. This should also help in eliminating the effects of interfacial degradation between the ferromagnet and the multiferroic oxide.

We expect that dynamical effects in multiferroics will increase in importance over the next years, driven by new experimental capabilities such as ultrafast X-ray sources (for example, Linac Coherent Light Source (LCLS) at Stanford Linear Accelerator Center, Stanford University), and that the fundamental limits on the dynamics of spin-charge-lattice coupling phenomena will be experimentally established. Theoretical proposals of dynamical multiferroic phenomena, in which a time-dependent polarization induces a magnetization in the reciprocal manner from that in which spin spirals induce polarization [330] should be validated by careful experiments. At the same time, more work on antiferromagnetic resonance in multiferroics is required; while many studies were carried out in the 1960s and 1970s [318] on conventional antiferromagnets, such measurements with modern multiferroics, which typically have higher resonance frequencies has been scarce. The recent surge in antiferromagnetic spintronics should be a welcome boost to such studies [73,331]. In a similar vein, there appears to be a great opportunity for fundamental and applied studies of nonlocal measurements of spin transport and its electric field manipulation [315,329]. We expect such approaches to be of significant scientific and technological interest in the next few years, especially if pathways to enhance the magnitude of the nonlocal spin Hall voltage are discovered.

In addition to static modulations of the exchange bias at $BiFeO_3$/$La_{0.7}Sr_{0.3}MnO_3$ interfaces, one might as well expect potential modulations of the spin dynamics of $La_{0.7}Sr_{0.3}MnO_3$ by the multiferroic. Merbouche et al. demonstrated that the transmission of spin waves across a two-micron channel of $La_{0.7}Sr_{0.3}MnO_3$ can be modulated by the domain structure of the adjacent $BiFeO_3$ layer [332]. The 13-nm-thick $La_{0.7}Sr_{0.3}MnO_3$ thin film was optimized on $NdGaO_3$(001) in order to obtain low Gilbert damping values of the order of $6 \times 10^{-3}$ [333]. The spin waves were probed in the Damon–Eshbach configuration by means of propagative spin wave spectroscopy [332] (**Figure 26**a). Using PFM, the out-of-plane polarization of $BiFeO_3$ was electrically controlled in order to define a magnonic crystal structure (**Figure 26**b). While the homogeneous up and down states show rather similar transmission properties, the periodically-poled pattern give rise to a gap in the spin wave transmission at 3.54 GHz with more than 20 dB rejection (**Figure 26**c-d). This constitutes the first example of a non-volatile electric-field induced reconfigurable magnonic crystal based on $BiFeO_3$/ferromagnetic metal systems. Indeed, the whole field of antiferromagnetic spintronics/magnonics and electric field driven magnonics is worthy of a significantly deeper investigation, again within the perspective of low energy manipulation of magnons as the principal carriers of information.

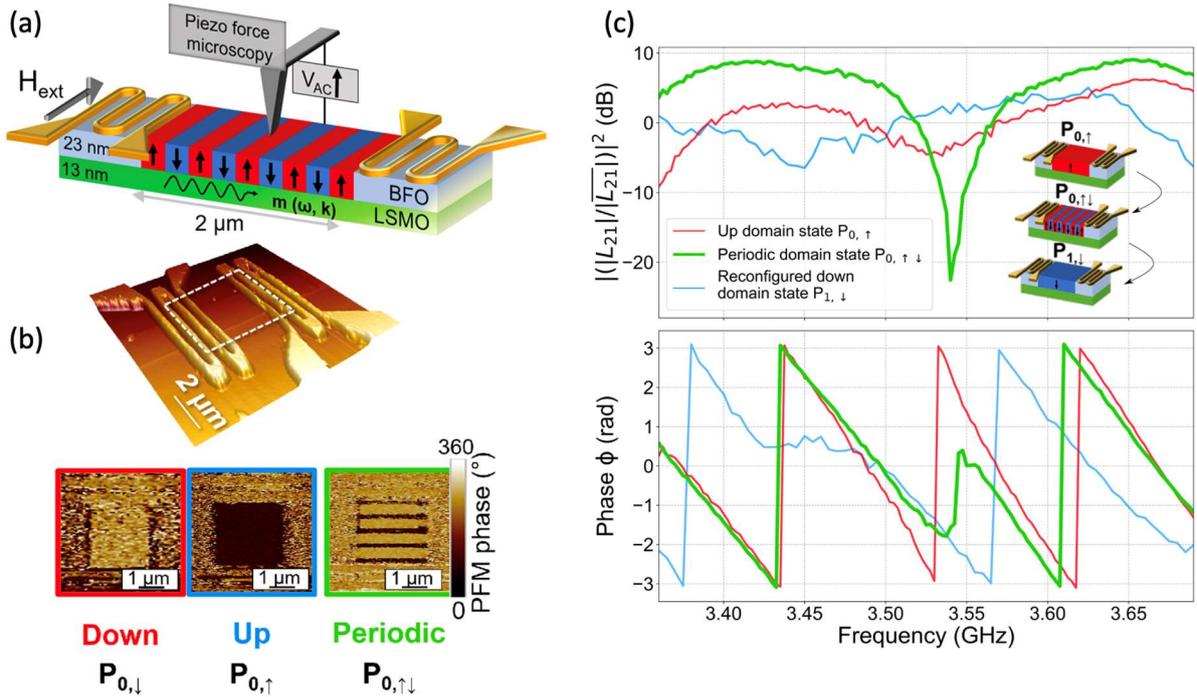

*Figure 26. Voltage-controlled reconfigurable magnonic crystal based on BiFeO3/La0.7Sr0.3MnO3. (a) Sketch of the setup in which spin waves are injected by an antenna and collected by the other with a 2-micron gap in between. The ferroelectric domains are read and controlled by PFM. (b) 3D view of the actual device and PFM phase images of the gap in three different polarization configurations: down (red), up (blue) and periodic (green) with a period of 500 nm. (c) Frequency dependence of the inductance (top) of the phase (bottom) showing a 20 dB rejection at 3.54 GHz for the periodically-poled configuration (green) as well as an accident in the phase. From [332].*

### 2.6.3. Ultrafast measurements of time domain dynamics

Despite all of the prior work, switching a ferroelectric state (as well as a multiferroic state) with a voltage as small as 100 mV remains a challenge and a research opportunity. Work so far with the La-BiFeO$_3$ system points to the possibility of switching time scales below 100 ps, if the measurement circuit is fast enough. Since the electric field scales with the dimensions of the ferroelectric, progression towards switching voltages of 100 mV automatically require that either the switching field is very low or that the switching behavior scales well with thickness. Therefore, it is critical to understand ferroelectric switching behavior in the ultrathin limit (< 20 nm). Quantitative studies of the switching dynamics at such a thickness and at time scales of hundreds of ps are still lacking and should be a fruitful area of research especially on the experimental side. What are the limits to the switching speed of ferroelectrics / multiferroics? There have been speculations that one limit could be the acoustic phonon mode (i.e., the velocity of sound in the material) since the switching of the polar state clearly involves the time-dependent deformation of the lattice at least in such perovskite-based ferroelectrics. For nominal values of the velocity of sound in such oxides (a few km/s), this would suggest switching time of the order of a few tens of picoseconds. Thus, the role of lattice dynamics during the dipolar switching event needs considerable further work. This is also true of ferroelectrics: the strong coupling between the spontaneous dipole at the lattice, immediately suggests that the dipolar switching dynamics in a thin film attached to a substrate will be strongly convoluted by the lattice dynamics. Recent *ab initio* and experimental studies of the switching dynamics of BiFeO$_3$ [153,334] indeed point to such a difference, which can be probed by studies of free-standing films compared to a film tethered to a substrate [335]. Of course, such a substrate clamping effect on the lattice dynamics can be mitigated by reducing the lateral dimensions of the magnetoelectric element, such that it is essentially unclamped [336]. Measuring at such time scales requires very fast electronics (for example, pulse generators with rise times smaller than a few tens of ps and oscilloscopes that can capture the switching transients at commensurate speeds); thus it is not surprising that there have been only a few measurements of the polarization switching dynamics approaching such time scales [337]. This is true for both ferroelectrics and multiferroics [338] and as we go forward into this exciting field of electric field controlled magnetic devices, such studies are critically needed.

## 3. Control of magnetism by current-induced torque

The main tool for the control of magnetism by current is the spin-transfer mechanism introduced by Slonczewski and Berger (Slonczewski 1996; Berger 1996), that is the transfer of the spin angular momentum and associated magnetization carried by a spin-polarized current (a spin current) to the magnetization of the magnet. This topic has been reviewed exhaustively, see for instance [342] (for the case where the spin current is generated by a magnet) or [341] (for the case where the spin current is generated by a system with spin-orbit coupling). In here, we will focus on the main experimental results, highlighting the potential applications of the current-induced torques.

We first describe the different types of spin currents and the different ways used to produce them, as summarized in **Figure 27**.

## 3.1. Spin currents

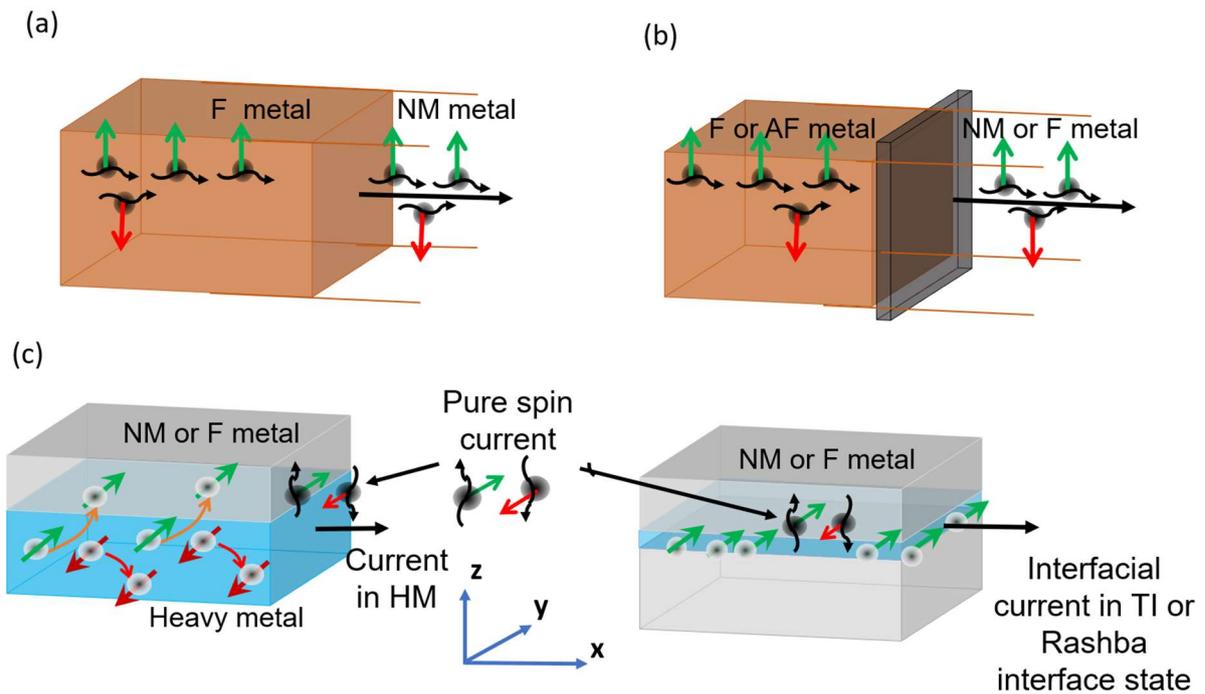

*Figure 27. Spin currents: (a-b) Spin-polarized currents flowing inside a magnetized material (a) and tunneling from this material (b). At the interface with a nonmagnetic material, the spin polarization extends with an exponential decrease in the range of the spin diffusion length. (c) For current along x, emission along z of a pure spin current into a magnetic or nonmagnetic layer by SHE in a heavy metal (HM) (left) and by diffusion from an Edelstein polarization in the surface/interface states of a topological insulator, Dirac semimetal or Rashba 2DEG (right).*

### 3.1.1. Spin-polarized current in a magnetic (ferromagnetic, ferrimagnetic) conducting material

The first way to produce a spin current is simply the exploitation of the two-current conduction (Mott and Fowler 1936; A. Fert and Campbell 1968) in a magnetic (ferro- or ferrimagnetic) material with different currents carried by the electrons having their spin parallel or opposite to the magnetization (spin down and spin up), as represented in **Figure 27**a. We will call this type of current a spin-polarized current. At the interface of the magnetic material with a nonmagnetic conductor and for both directions of the current, the spin polarization extends with an exponential decrease into the nonmagnetic material at a distance from the interface which is called spin diffusion length ($\lambda_{sf}$) [344–346].

### 3.1.2. Spin-polarized current tunneling from a magnetic material

A current tunneling from a ferromagnetic or ferrimagnetic material into another material is also spin polarized, as represented in **Figure 27**b, which is exploited in the tunnel magnetoresistance (TMR) of the MTJs [347–351]. In the approximation of Julliere model [347], the spin polarization of the current tunneling from a magnetic material into, for example, a nonmagnetic material reflects simply the spin polarization of density of states at the Fermi level in the magnetic material. However, in the real situation, the spin polarization of the spin current can also depend on the filtering of different types of wave functions by the material of the tunnel barrier [350–355]. Actually, such filtering effects have been exploited to obtain very high spin polarizations of the tunneling current and large TMRs [356,357].

### 3.1.3. Conversion between charge and spin currents by the Spin Hall Effect (SHE) and Spin Anomalous Hall Effect (SAHE); pure spin currents.

The spin Hall effect (SHE) of a nonmagnetic material, for example a heavy metal (HM) with large spin-orbit coupling (SOC), is related to the SOC-induced deflection of the electrons of opposite spins in opposite directions [328,358–362]. In the example of **Figure 27**c with a charge current along $\hat{x}$, the electrons with spins along $\hat{y}$ ($-\hat{y}$) are deflected upward (downward) along $\hat{z}$. This leads to what is called a pure spin current and can be described as the combination of opposite flows of electrons with opposite spins. In isotropic materials, the SHE is characterized by the spin Hall angle $\theta_{SHE}$. Quantitatively, in an infinite material and for spin current emission along $+\hat{z}$ generated by a charge current in the $x-y$ plane, a charge current density $J_c$ flowing in the direction of the unit vector $\hat{j}$ emits along $\hat{z}$ a spin current density $J_s$ polarized along

$$\hat{\sigma} = \pm(\hat{j} \times \hat{z}) \quad \text{Eq. 1}$$

i.e., $\pm\hat{y}$ for $\hat{j}$ along $\hat{x}$ in **Figure 27**, ± depending on the sign of $\theta_{SHE}$. If the charge and spin current densities are defined as the respective flows of positive charges -e and unit spins, $J_s$ and $J_c$ are related by $J_s = \theta_{SHE} J_c$. Typical values of $\theta_{SHE}$ are, for example, 0.06 for Pt, 0.15 for Ta or 0.3 for W [362–365].

In an isolated layer, the SHE leads to an accumulation of opposite spin at opposite interfaces. With a conducting layer covering the layer of a heavy metal with SHE, as represented in the left panel of **Figure 27**c, the accumulation of spin along $+\hat{y}$ (in the figure) diffuses into the top layer, the charge neutrality condition leads to an attraction of spin $-\hat{y}$ and this situation is described as an injection of a pure spin current density $J_s$ into the neighbor material. The amplitude of the injected spin current depends on the transparency of the interface and also on the possibility of large enough spin absorption (i.e., short enough $\lambda_{sf}$) to limit the spin accumulation in the neighbor material and the resulting repulsion of the injected spins (one says, to prevent reflection of the spin current). In the best conditions i.e., transparent interface and large enough absorption of the injected spins, the injected spin current keeps approximately its value $\theta_{SHE} J_c$ in the heavy metal.

Spin currents are also generated by current in ferromagnetic or ferrimagnetic materials. Until recently, it was supposed that, due to exchange interactions being much stronger than spin-orbit interactions, the transverse component of a SOC-induced spin current was completely dephased by exchange-induced precessions and its spin polarization aligned with the magnetization. What remains is the so-called spin anomalous Hall effect (SAHE) with a spin current polarized along the magnetization direction $\hat{m}$ [366,367]. In an infinite material and for the spin current along $\hat{z}$ generated by a charge current in $x-y$ plane, a charge current density $J_c$ flowing in the direction of the unit vector $\hat{j}$ emits along $\hat{z}$ a spin current density $J_s$ polarized along $\hat{m}$ with

$$\boldsymbol{J_s} = \boldsymbol{\theta_{ASHE}} [(\hat{j} \times \hat{m}) \cdot \hat{z}] \boldsymbol{J_c} \quad \text{Eq. 2}$$

where $\theta_{SAHE}$ is the spin anomalous Hall angle.

However, more recent theoretical works by Amin *et al.* (V. P. Amin et al. 2019; V. P. Amin, Haney, and Stiles 2020) or Kim *et al.* [370] have shown that the alignment of the SOC-induced spin current with the magnetization direction is incomplete in most magnetic materials. This gives rise to the coexistence of SAHE-type and SHE-type spin currents. This coexistence has clearly been shown in the experiments of Das *et al.* [371] and was also found in other recent works [372–374]. In particular, the experiments of Céspedes-Berrocal *et al.* have shown that, for GdFeCo ferrimagnetic alloys, the 5d character of the Gd electrons leads to particularly large currents of SHE and SAHE symmetries coexisting with respective spin Hall angles $\theta_{SHE} \approx 0.16$ and $\theta_{SAHE} \approx 0.6$ [375].

The generation of a pure spin current from a charge current by SHE or SAHE can be described as a conversion of a charge current into a spin current. Inversely, in another type of experiment, a spin current injected into a material (say, HM) can be converted into a charge current in the heavy metal by the so-called inverse spin Hall effect (ISHE), as expected from Onsager reciprocity [376]. Typical examples with the ISHE of Pt can be found in the literature [362,363,376–378].

### 3.1.4. Conversion between charge and spin current by spin-orbit coupling in surface or interface states.

Charge currents flowing in or scattered by surface/interface states can generate spin currents [379]. Here we will only describe the generation of spin currents by the Edelstein Effect (EE) in topological surface states or Rashba states [341,380–383].

**Figure 28**a displays the classical image of the Dirac cone of topological 2D states at the surface or interface of 3D topological insulators or Dirac semimetals [384–386]. The corresponding Fermi contour is shown in **Figure 28**b and is characterized by the locking between spin and momentum represented on the figure. In a similar way, the Rashba interaction generated by spin-orbit coupling and inversion symmetry breaking at surfaces or interfaces [374,387] leads to the type of dispersion surfaces shown in **Figure 28**c, which gives the two Fermi contours with different radii and opposite spin-momentum locking shown in **Figure 28**d. As represented in **Figure 28**e, a current flowing in a topological surface/interface state generates an overpopulation of spin oriented in a transverse direction with respect to the current and a depletion of the opposite spins. This is the Edelstein spin polarization induced by current in the surface states [380]. If the topological 2D states are at an interface with a conducting material, the spin accumulation diffuses through the interface and a pure spin current density $J_s$ with polarization perpendicular to the 2D charge current is injected into the adjacent material [381,382]. For a current flowing in a Rashba two dimensional electron gas (2DEG), a similar mechanism, with a partial compensation of the opposite contributions from the two Fermi contours, leads also to a similar production of spin current, see [341,368,388].

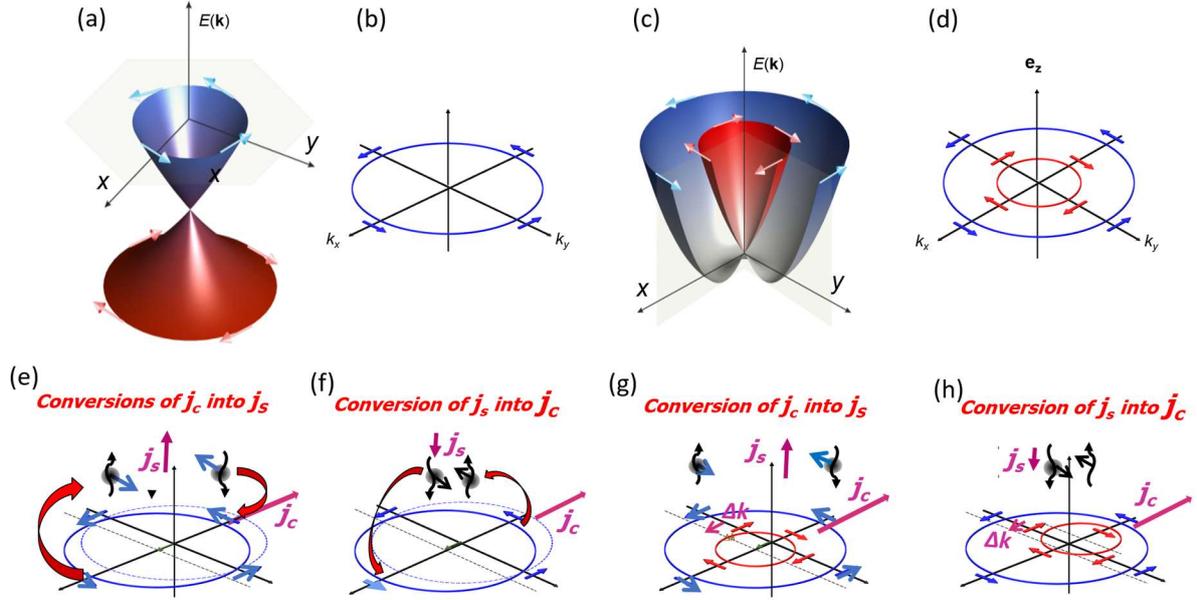

*Figure 28. (a) Sketch of the electronic energy dispersion surfaces in the surface states of a topological insulator (Dirac cone). (b) Fermi contour at constant energy illustrating the spin-momentum locking: at any **k** position on the contour, the spin is perpendicular to **k**. (c) Electronic dispersion surfaces of a Rashba system. (d) In contrast to the case of topological insulators, here the systems comprises two Fermi contours; on each the spin is locking perpendicular to **k** both for the spins curl clockwise one contour and counterclockwise for the other contour. (e) Charge to spin conversion with a topological insulator: the application of a charge current $j_c$ along -x causes a shift of the Fermi contour and generates an extra population of states with spin along y. This generated spin density can then diffuses vertically as a spin current $j_s$. (f) Spin to charge conversion with a topological insulator: spins oriented along y injected into the topological insulator populate states with momentum along x (which is accompanied by the ejection of spins oriented along -y from states with momentum along -x), causing an overall shift of the Fermi contour and thus the generation of charge current along - x. (g) Charge to spin conversion in a Rashba system. The situation is similar to that in (e) except that spin densities with opposite spin polarizations are generated by the injected charge current for the inner and outer contours. They however do not compensate, yielding the generation of a finite spin density that may diffuse vertically as a spin current. (h) Spin to charge conversion in a Rashba system. Again the situation is similar to that in (f) but here the injection of spins causes shifts of the Fermi contours in opposite direction, albeit without a full compensation, which results in the generation of a finite charge current.*

In both situations of topological insulators and Rashba interfaces, the conversion of a 2D charge current into a 3D pure spin current can be characterized by the parameter $q_{ICS}$ (in m$^{-1}$) introduced for topological insulators surfaces by [381] and relating the 3D spin current density $J_s$ (in A/m²) to the 2D charge current density $J_c$ (in A/m)

$$J_s^{3D} = q_{ICS} J_c^{2D} \quad \text{Eq. 3}$$

with experimental results corresponding to values of $q_{ICS}$ in the nm$^{-1}$ range [381,389].

The reverse conversion, by the inverse Edelstein effect (IEE), can be understood from **Figure 28**f and h: the injection of a pure spin current into topological or Rashba 2D states leads to an overpopulation of occupied states on one side of the Fermi contour and to a depletion on the other side, that is to a charge current flowing in the 2D states. In other words, there is a conversion between an injected 3D spin current and a 2D charge current in the 2DEG at the surface or interface. For Rashba Fermi contours, there is only a partial compensation between the two contours and the same type of spin-to-charge conversion exists. In both cases, the conversion of a 3D spin current into a 2D charge current by the IEE is characterized by a length, $\lambda_{IEE}$, with values in the nm range or exceeding 10 nm [388,390–398].

It is interesting to compare the spin currents generated by the EE and those produced by the SHE of a heavy metal [399]. For SHE, in the optimal conditions with transparent enough interfaces, the transferred spin current density $J_s^{3D}$ is simply related to the charge current $J_c^{3D}$ in the SHE layer by the expression [400,401]:

$$J_s^{3D} = \theta_{SHE}[1 - sech(t/\lambda_{sf})]J_c^{3D} \quad \text{Eq. 4}$$

where $t$ and $\lambda_{sf}$ are the thickness and the spin diffusion length of a heavy metal. Expressing the current in the heavy metal in terms of a 2D charge density $J_c^{2D} = tJ_c^{3D}$, one finds from Eq. 4 that the maximum value of the ratio $J_s^{3D}/J_c^{2D}$ (to be compared to $q_{ICS}$ in Eq. 3) is obtained for $t \cong 1.5\lambda_{sf}$ and is expressed by $q_{SHE} = 0.38\,(\theta_{SHE}/\lambda_{sf})$. With typical values of $\theta_{SHE}$ and $\lambda_{sf}$ in the respective ranges of 10 % and a few nm, one finds values of $q_{SHE}$ smaller than $10^{-1}$ nm$^{-1}$, more than one order of magnitude below that of the $q_{ICS}$ of the EE in 2DEGs [399]. Larger spin currents are thus expected from EE at surface or interface 2DEGs than from SHE at 3D layers, in agreement with the experimental results on switching by spin-orbit torque (SOT) discussed below.

For the opposite conversion from spin to charge, comparisons between experimental values of the conversion coefficient $\lambda_{IEE}$ for various topological insulators or Rashba surface/interface states and the effective conversion coefficient $\lambda_{SHE} = \theta_{SHE}\lambda_{sf}$ of heavy metals see **Table 2** in (Rojas-Sánchez and Fert 2019). The coefficient $\lambda_{IEE}$ of topological insulator (TI) or Rashba surface/interface states can be larger than the effective $\lambda_{SHE}$ of heavy metals by one or two orders of magnitude.

### 3.1.5. Spin currents in insulating materials

In insulating magnetic materials, spin currents can be carried by magnons [313,315,402,403]. Such spin currents carried by magnons in a magnetic insulator layer can be electrically generated by a spin current carried by conduction electrons in a metallic layer via the spin accumulation at the interface. The conversion between metallic spin current and magnon spin current is controlled by the interfacial spin mixing conductance [404,405]. Typical examples are the direct and inverse conversions between conduction electron spin currents in heavy metal and magnon spin currents in $Y_3Fe_5O_{12}$ (YIG) based magnetic insulators [405,406].

### 3.2. Spin transfer, spin-transfer torques (STTs) and magnetization switching by STT

The concept of spin transfer and spin-transfer torque (STT), introduced by Slonczewski and Berger [339,340], is illustrated schematically in **Figure 29**a for the typical case of 3*d* ferromagnetic metals with ferromagnetic layers F1 and F2 separated by a nonmagnetic layer, either a tunnel barrier as MgO or a nonmagnetic metal as Cu. A spin-polarized current is prepared by F1 to obtain, in the spacer layer, a spin polarization obliquely oriented with respect to the vertical magnetization of the second magnetic layer F2 (the spin polarization in the spacer layer is not simply the polarization of the current inside F1 and, in general, is intermediate between the polarizations of F1 and F2). When this current enters F2, the exchange interactions with the local spins induce precessions of the transverse component of the injected spins around the magnetization axis of F2 and the dephasing of these precessions by the distribution of the exchange interactions makes that the global transverse polarization disappears. As the exchange interaction is spin conserving, this dephasing corresponds to an absorption of the transverse component of spin current. The absorption is complete after penetration beyond the so-called spin dephasing length, in general of the order of one or a few nm (or incompletely absorbed if the thickness of the magnetic layer is smaller than the dephasing length). In the first situation of a thick enough layer, if also the spin-lattice relaxation by spin-orbit coupling can be neglected, the total transverse spin component lost by the current is transferred to the total spin of F2. This can be also described as a STT acting on F2 and given by the following expression as a function of the unit vectors $\hat{m}$ along the magnetization $m$ of the magnetic layer and $\hat{\sigma}$ along the spin polarization of the injected current:

$$\boldsymbol{T_{STT}} = \boldsymbol{\tau_{DL}} \left[\hat{\boldsymbol{m}} \times (\hat{\boldsymbol{m}} \times \hat{\boldsymbol{\sigma}})\right] + \boldsymbol{\tau_{FL}} (\hat{\boldsymbol{m}} \times \hat{\boldsymbol{\sigma}}) \quad \text{Eq. 5}$$

The first and main term, the damping-like (DL) torque, is a direct consequence of the spin-transfer mechanism and the coefficient $\tau_{DL} = \frac{\hbar}{2e} J_s^{abs}$ for the torque by spin area can be directly related to the density of absorbed spin current, $J_s^{abs}$. **Figure 29**b shows that, for $m$ precessing around its equilibrium direction, the damping-like torque is in the same direction as the damping torque of the Landau‑Lifshitz‑Gilbert (LLG) equation and acts to reduce or enhance the damping. For theoretical expressions of the damping-like torque with different types of injectors and as a function the interfacial coefficient called spin-mixing conductance, we refer to [407] or [408]. The field-like (FL) torque is a corrective term, generally much smaller, related to the exchange field generated by the injected spin polarization [409] and to the imaginary part of the spin-mixing conductance [408].

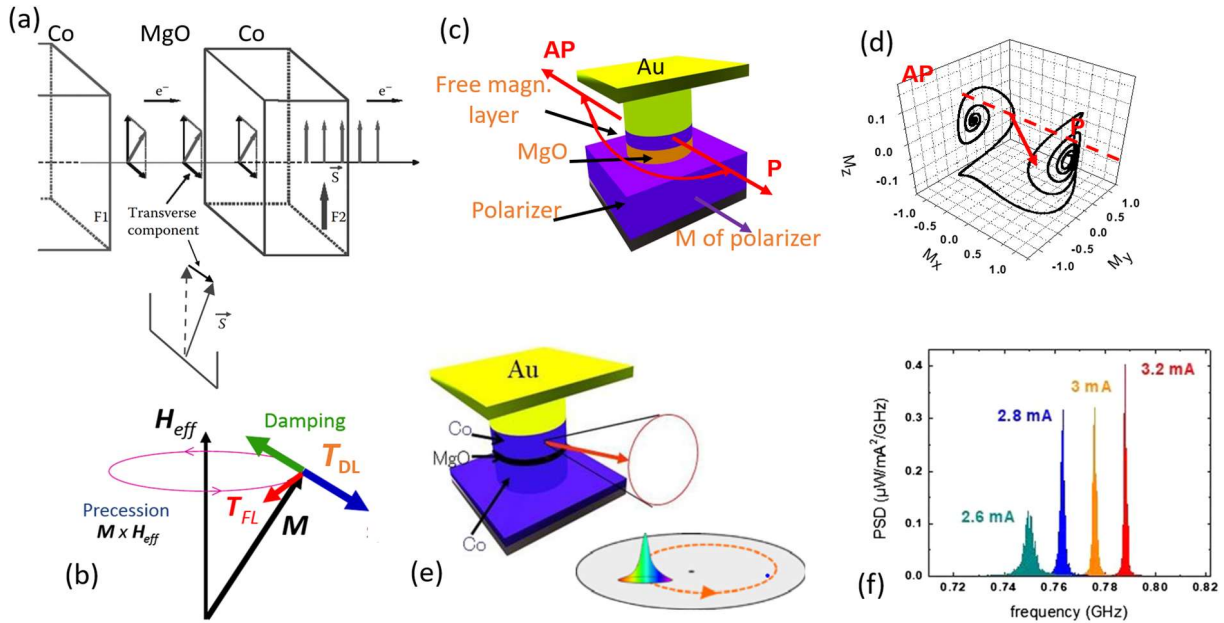

*Figure 29. (a) Concept of spin-transfer torque: A spin-polarized current (prepared by a magnetic material F1) is injected through a nonmagnetic layer (tunnel barrier or metal) into the magnetic material F2. Inside F2, exchange-induced precessions dephase the transverse components of injected spins and lead to a transfer of the transverse component of the injected spin current into F2 or, equivalently, to a torque on its magnetization. (b) Schematic of the damping-like and field-like torques on a magnetization M departing from its equilibrium orientation along Heff and precessing around Heff in the situation in which the damping-like torque is opposite to the LLG damping torque and enlarges the precessions. (c, d) Switching by STT: Schematic of a nanopillar with two ferromagnetic CoFeB layers (polarizer and free*

*layer) separated by a MgO layer in c. Macrospin simulations of the precession and switching of the free layer from P to AP by the STT induced by the injection of a vertical spin current from the polarizer into the free layer in (d). P and AP stand for parallel and antiparallel, respectively. (d) Magnetization dynamics for a device of the type in (c) in the regime in which the STT generates a steady state gyration of the magnetization in the free layer (or a gyration of a magnetic vortex in the free layer, see inset with vortex core and its trajectory shown as a dashed line). (f) Experimental example of microwave power emission generated by vortex gyration. (a-e) Adapted from [343]. (f) From [410].*

The first experimental evidences were obtained using either point contacts or pillar-shaped devices (**Figure 29**c) in which the STT created by the spin-polarized current emitted by the thick reference magnetic layer (polarizer) can switch the magnetization of the thin free magnetic layer between the parallel (P) and antiparallel (AP) orientations of the two layers [411–413]. A macrospin simulation of the progressively extended precessions and switching of the magnetization of the free layer is also shown in **Figure 29**c. A small switching current is obtained when the coefficient α characterizing the damping torque in the LLG equation and the energy barrier between the P and AP states are small. An early experimental example of device switched by STT is displayed in the right of **Figure 29**c. In a second type of regime in the same device, the STT can be used to generate magnetic excitations in the free layer, steady state precession of the magnetization or gyrations of a magnetic vortex, which leads to ac voltage via TMR or GMR and microwave power emission (**Figure 29**d).

As far as applications are concerned, the appearance of STT has boosted the development of the MRAMs which are called STT-MRAM for those using STT. Since their first demonstrations in the mid-2000 [414], the STT-MRAMs have been frequently described as a potential universal memory having arguments to compete with all the main types of electronic memories. During the last years, several major companies started a massive production of STT-MRAM [415,416]. The SOT-MRAMs, based on the spin-orbit torque (SOT) discussed in the next subsection are promising to take over with, in particular, great progress in term of high speed. The second type of interesting application is the spin-torque nano-oscillator in the microwave technologies.

## 3.3. Spin-orbit torques (SOT) and magnetization switching by SOT.

### 3.3.1. General (metallic magnetic materials)

SOTs are the torques induced by the transfer of spins from a spin current $j_s^{3D}$ generated by spin-orbit coupling [341,417]. Such spin currents can be generated by the SHE of a material of large spin-orbit coupling (such as Pt or Ta), by the SAHE of a ferromagnetic material, or by the EE in topological or Rashba surface/interface states. In the most general case (rotational invariance around the out-of-plane axis), the torque acting on the magnetization of unit vector $\hat{m}$ has the same form as the STT of Eq. 5 and includes a damping-like and field-like torques:

$$\boldsymbol{T_{SOT}} = \boldsymbol{T_{DL}} + \boldsymbol{T_{FL}} = \tau_{DL}[\hat{m} \times (\hat{\sigma} \times \hat{m})] + \tau_{FL}(\hat{\sigma} \times \hat{m}) \quad \text{Eq. 6}$$

where $\hat{\sigma}$ is the unit vector along the polarization of the current injected into the magnetic layer. For both SHE and EE and for a current along $\hat{x}$, $\boldsymbol{\sigma}$ is along $+$ or $-\hat{y}$ depending on the sign of $\theta_{SHE}$ or $q_{ICS}$ and, for SHE, on the direction of emission (+ or -). We show in **Figure 30**a for SHE (in **Figure 30**b for EE) an example of the orientation of the damping-like and field-like torques in the situation with an out-of-plane magnetization.

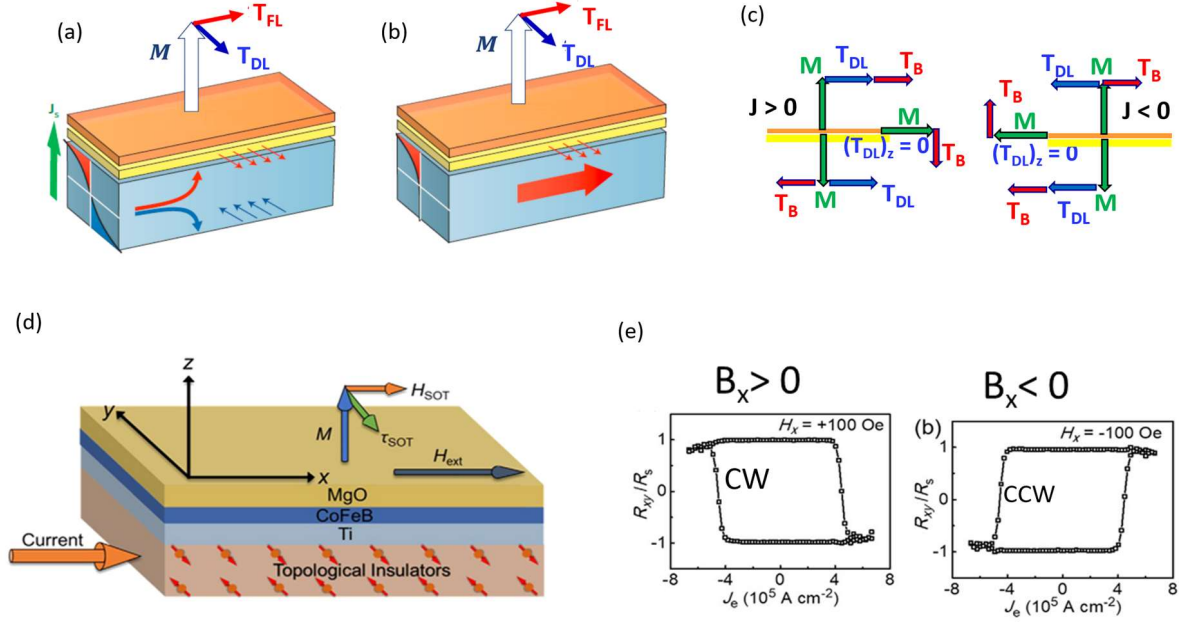

*Figure 30. (a-b) SOTs, $T_{DL}$ and $T_{DL}$, induced by spin currents (polarizations indicated by small arrows) due to SHE in a heavy metal (a) and Edelstein Effect in a Rashba or topological 2DEG (b). (c) Switching in the macrospin limit is used to illustrate the symmetry of the reversal of perpendicular magnetization under the additive actions of damping-like torque $T_{DL}$ and torque $T_B$ induced by an applied field along the current direction. Left: for J > 0, $T_B$ helps $T_{DL}$ to reverse M from up to down, especially at midway, when M is in-plane and $T_{DL}$ = 0 whatever the orientation of M in the plane. Right: same applied field with J < 0 for a reversal from down to up and a clockwise loop. Reversing the applied field leads to a counter-clockwise loop. (d-e) In the device shown in (d), (e) shows the switching loops of the ferromagnetic CoFeB layer magnetization under the conjugated actions of the spin current generated by the Edelstein Effect in the surface state of the topological insulator $(BiSb)_2Te_3$ and an applied field $B_x$ in the current direction. The switching loops, detected by AHE, are clockwise for $B_x > 0$, or counter-clockwise for $B_x < 0$ [418].*

The damping-like torque is generated by the Slonczewski mechanism of transfer of the spin momentum injected into the magnetic material and, as in the STT case, is related to the density of absorbed spin current. When the spin current is injected from a SHE material, the damping-like torque is generally predominant and the field-like torque is a small corrective term due to exchange interactions between **m** and the spin accumulation introduced into the magnetic layer [409]. When the spin source is a Rashba polarization at an interface of the magnetic material itself and directly interacting by exchange with its magnetization, the field-like torque is generally larger but the damping-like torque due to the diffusion of a spin current from the Rashba interfacial polarization can also be large if this spin current is efficiently transferred out of the magnetic layer.

The damping-like and field-like torque (in units of eV/m³) can be expressed as

$$\boldsymbol{T_{DL}} = \frac{\hbar}{2e}\xi^j_{DL}\frac{j_c}{t_F}\hat{\boldsymbol{m}} \times (\hat{\boldsymbol{m}} \times \hat{\boldsymbol{\sigma}}) \quad \text{Eq. 7}$$

$$\boldsymbol{T_{FL}} = \frac{\hbar}{2e}\xi^j_{FL}\frac{j_c}{t_F}(\hat{\boldsymbol{m}} \times \hat{\boldsymbol{\sigma}}) \quad \text{Eq. 8}$$

where $t_F$ is the thickness of the magnetic layer. The coefficients $\xi^j_{DL(FL)}$ express the efficiencies of the conversion of a charge current density $j_c$ into the spin current density $j_s$ transformed into torque. Detailed expressions of $\xi^j_{DL}$ as a function of the conversion coefficients $\theta_{SHE}$ (for SHE) or $q_{ICS}$ (for topological or Rashba 2DEGs), the interfacial transmission coefficients called spin-mixing conductances and the spin diffusion lengths $\lambda_{sf}$ in the different layers can be found in several publications [341,419,420]. In the case of spin emitted by SHE from a heavy metal and generating a torque in another material, the efficiency coefficient $\xi_{DL}^j$ can be seen as an effective spin Hall angle characterizing the finally transferred spin current. Its maximum value for optimal transmission is the intrinsic $\theta_{SHE}$ of the heavy metal. When 2D surface or interface states of a layer generate spin currents from 2D charge currents, an usual simplified picture is that of a layer with only SHE and an uniformly distributed effective $\theta_{SHE}$ taking into account, approximately, both the bulk and surface effects. In this situation, $\xi_{DL}^j$ can be larger than 1, as it is observed with efficient spin emission by Rahba or TI surface/interface states. The expressions are more complex for $\xi_{FL}^j$ as they also depend on the exchange interaction between spin accumulation and magnetization.

Alternatively, the SOT of Eq. 7 and Eq. 8 can be rewritten in term of SOT-induced effective fields $\boldsymbol{B_{DL}}$ and $\boldsymbol{B_{FL}}$ inducing the damping-like and field-like torques on the magnetization:

$$\boldsymbol{T_{DL,FL}} = \boldsymbol{m} \times \boldsymbol{B_{DL,FL}} \qquad \text{Eq. 9}$$

As already pointed out, the above expressions of SOT, Eq. 7 and Eq. 8, are for rotational invariance around the out-of-plane axis, that is for the most frequent situation where the spin source is a polycrystal. A material of lower symmetry for spin source leads to more complex expressions of SOT [421]. An experimental example of the complex symmetry of damping-like torque is given by the SOT generated by $WTe_2$ in which the surface crystal structure has only one mirror symmetry and no two-fold rotational invariance [422]. Another example of low symmetry SOT can be found in Liu *et al.* [423]. The authors show that the symmetry at the $L1_1$-ordered interface of a CuPt/CoPt epitaxial bilayer gives rise to out-of-plane SOT and makes it possible to switch the out-of-plane magnetization of CoPt in zero applied field with a three-fold angular dependence of the switching.

### 3.3.2. Magnetization switching by SOT

The realization of current-induced magnetization switching by SOT is a greatly promising direction for the development of SOT-RAMs and the relay to the STT-RAMs in production today. In particular, the high speed of the switching of layers with perpendicular magnetic anisotropy (PMA) by the damping-like torque is especially appealing. This is the type of switching that we will describe and discuss in the main part of the subsection.

Experimental examples of switching of magnetic layers with PMA by SOT are displayed in **Figure 30**d-f in both situations of spin current induced by SHE in a heavy metal and EE in the surface states of a topological insulator. The schematics in **Figure 30**a-b indicate the spin polarization of the spin currents injected into the top magnetic layer by SHE in heavy metal (a) or EE in 2DEG (b) and the orientation of the SOTs (from Eq. 7 and Eq. 8) acting on a vertical magnetization. The damping-like torque does not break the symmetry between the up and down states and the switching between these states is only possible by adding an applied field along the current direction, as it can be understood in the macrospin model of **Figure 30**c: with a positive current and an applied field $B_x > 0$, the additive actions of the damping-like torque and field-induced torque allow the magnetization to switch from up to down because the field-induced torque is nonzero when the SOT is zero at midway from up and down. The demand for an in-plane field can be justified more generally by symmetry arguments for systems with rotational symmetry around the axis perpendicular to the layers [341]. SOT switching of PMA layers in the presence an applied field is also the usual observation when the switching process is by nucleation and extension of domains with opposite magnetization [424,425].

In the experimental examples of **Figure 30**d-e (TI as source of STT) or **Figure 31**c-f (heavy metal as source of SOT), an in-plane field along the current direction is necessary to switch the magnetization and leads to clockwise or counter-clockwise magnetization cycles depending on the direction of the applied field with respect to the current direction. A similar behavior is also found when the magnetic layer is a magnetic insulator in which the spin current emitted by SHE or EE cannot flow but can be transmitted by magnetic excitations into the insulator. We refer to recent examples with thulium garnet films (Avci et al. 2017; Shao et al. 2018; H. Chen et al. 2020).

Another conclusion can be derived from the comparison between SOT/switching experiments with the SHE in heavy metal and the EE in 2D states of TI, Dirac semimetals or Rashba interfaces. As already discussed [399], the conversion between charge and spin current is generally more efficient by one or two orders of magnitude by using the EE in 2D states than with the SHE of 3D states. This result is confirmed by a direct comparison of the SOT efficiencies and writing powers in experiments of torque and magnetization switching.

In **Table 2**, adapted from [428], we present a selection of experimental results at room temperature on the SOT efficiency coefficient $\xi_{DL}^{j}$ and the writing power $\rho_{WP}$ in different systems (heavy metal, metallic oxides, TI, Dirac semimetals and Rashba interfaces), for the production of spin current. The efficiency is derived from experiments of SOT and switching with different magnetic materials. The writing power, $\rho_{WP} = ((1+s)/\xi_{DL}^{j})^2 \rho_{SOC}$ where $s$ is the ratio of the shunting current to the switching current and $\rho_{SOC}$ is the resistivity of the spin-orbit coupling material, expresses that a total energy $\rho_{WP}(j_c)^2$ is needed for the transfer into the magnet of a flux of spins equal to the flux of electrons in $j_c$ [428,429]. It is an essential element to probe the potential of a SOT material/magnetic material system for devices, for example, SOT-MRAM type (see Section 5).

For the SHE of heavy metal, although all the determinations have not been always obtained in the same conditions, there is a good convergence of the results for a given heavy metal and we present typical data for three heavy metals: Pt, β-W and AuPt. The stronger efficiency of the metallic oxide SrIrO$_3$ probably reflects the combination of SHE in the layer and EE from surface states of SrIrO$_3$, as in other systems with SrTiO$_3$. For TI, there is a huge dispersion of experimental results, due mainly to the difficulty of the separation between the 2D (EE) and 3D (SHE) contributions and to the variety of more or less valid techniques which have been used to derive the SOT. Publishing a huge table of largely dispersed data would not be necessary and we selected only four systems. We have included the very attractive result obtained on Bi$_{0.9}$Sb$_{0.1}$ by Khang et al [389]. This result has drawn a lot of attention. However, it needs to be confirmed by other groups to be realistically promising for applications. In spite of the dispersion of the results, it turns out that, for the efficiency and also low power consumption, the 2D systems (TI, Dirac semimetal) are more performant than the usual heavy metals by two orders of magnitude or more. For applications, other aspects must be accounted for. For example, the advantage of Bi$_{0.9}$Sb$_{0.1}$ in terms of efficiency at low power is compensated by the disadvantage of a preparation by MBE, a non-usual technology in spintronic devices, and the requested in-plane field. In contrast, α-Sn is somewhat below in terms of efficiency and low power but has the advantage of fabrication by sputtering.

| Spin-orbit Material | coupling | Resistivity $\rho_{SOC}$ (×10⁻⁴ Ω·cm) | Current Ratio s | SOT Efficiency $\xi_{DL}^j$ | Writing Power $\left(\dfrac{1+s}{\xi_{DL}^j}\right)^2 \rho_{SOC}$ (×10⁻⁴ Ω·cm) | Reference |
|---|---|---|---|---|---|---|
| Heavy metal | Au$_{25}$Pt$_{75}$ | 0.83 | 0.255 | 0.35 | 10.68 | 429 |
| | Pt | 0.20 | 0.061 | 0.055 | 74.5 | 429 |
| | β-W | 3.0 | 0.923 | 0.33 | 102 | 429 |
| Oxides (metallic) | SrIrO$_3$ | 12 | 1.8 | 1.1 | 31.8 | 430 |
| Topological insulator | Bi$_{0.9}$Sb$_{0.1}$ | 4.0 | 1.2 | 52 | 0.007 | 389 |
| | Bi$_x$Se$_{1-x}$ | 130 | 40 | 18.6 | 632 | 429 |
| | (Bi,Se)$_2$Te$_3$ | 40.20 | 12.37 | 0.4 | 44900 | 429 |
| Topological Dirac semimetal | α-Sn | 0.81 | 0.119 | 6.15 | 0.027 | 428 |
| Rashba 2DEG | LAO/STO | no data | no data | 1.8 | no data | 431 |

*Table 2. Comparison of the SOT efficiencies and writing powers (at room temperature except for LaAlO$_3$/SrTiO$_3$, i.e. LAO/STO) obtained in a selection of spin-orbit coupling materials, heavy metal, metallic oxide, TI, Dirac semimetal, Rashba 2DEG (Table adapted from [428]), with a majority of results from [429]. Compared to heavy metal, the strong efficiency of the metallic oxide SrIrO$_3$ probably expresses the combination of SHE and interfacial EE (its writing power has been estimated for this Table from the transport data on SrIrO$_3$/CoTb in [430]). The very strong efficiency and low energy consumption for Bi$_{0.9}$Sb$_{0.1}$, if confirmed, is very promising for devices. The 2DEG Rashba system LaAlO$_3$/SrTiO$_3$ cannot be characterized by a 3D resistivity and a writing power in terms of 3D resistivity. A general conclusion is that, with respect to heavy metal, $\xi_{DL}^j$ can be larger by two orders of magnitude or more in materials with spin-orbit coupling in surface or interface states (and the writing power can be much smaller, too).*

The requirement of an in-plane applied field to switch an out-of-plane magnetization by SOT in the conditions of Eq. 7 and Eq. 8 (i.e., in the general situation of samples of rotational invariance around the out-of-plane axis) is a disadvantage for devices based on SOT and PMA layers. However, an important advantage of the reversal of PMA layers by SOT is its much faster dynamics in comparison with what can be obtained by SOT with in-plane magnetizations or STT, as we discuss now.

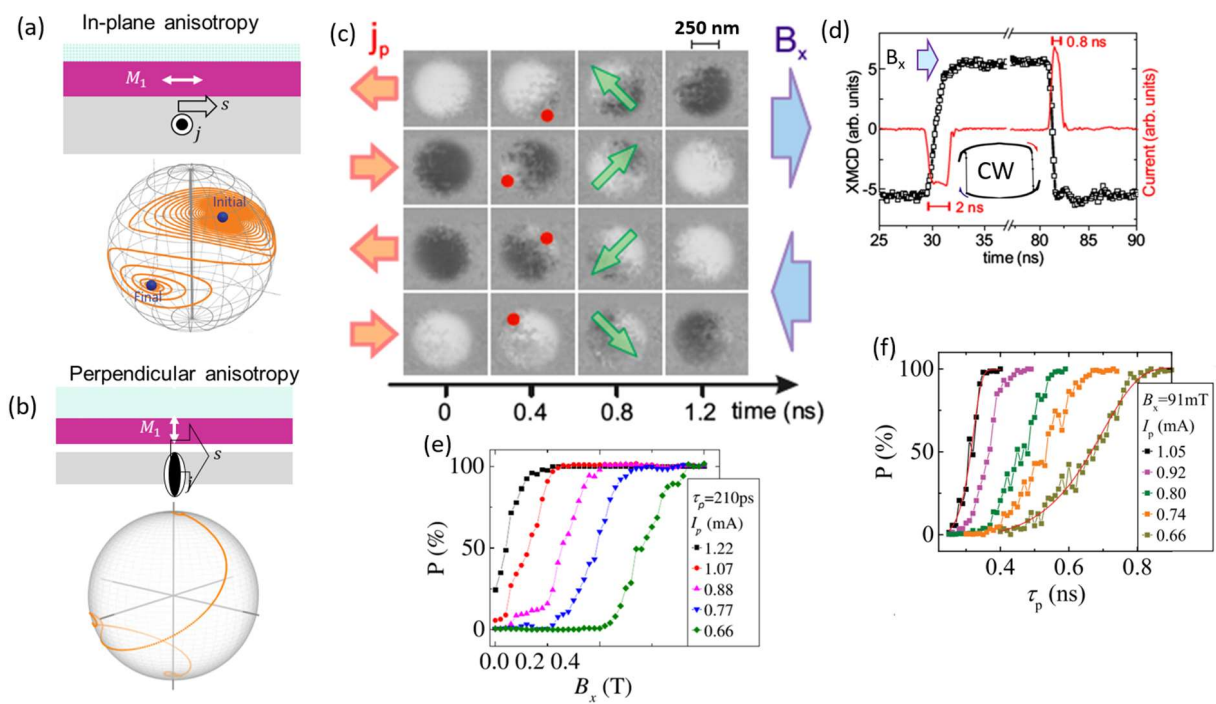

*Figure 31. Symmetry and dynamics of the switching of a magnetic layer with PMA by SOT. (a-b) Macrospin simulations of the switching by SOT of in-plane (a) and out-of-plane (b) magnetizations (courtesy of P. Gambardella). For in-plane magnetization (a), the SOT enlarges progressively precessions of the magnetization around its initial orientation to finally reverse it. The long incubation time (successive precessions) leads to long switching times. For PMA (b), the action of SOT is immediate and can lead to much shorter switching times around or below 1 ns. (c-d) Switching of a ferromagnetic layer with PMA in the process of nucleation and extension of domains. In (c), snapshots of X-ray magnetic dichroism images of a dot of a Pt/Co/MgO with PMA during the reversal of its magnetization by the SOT induced by current pulses in Pt (adapted from [424]). With an applied field along x (- x), the SOT induces a reversal from up (down) to down (up) for positive current and from down (up) to up (down) for negative current, as in the magnetization loops. The nucleation of a reversed domain starts on the edges at a point (see dot) where the combination of the applied field and DMI interaction favors this nucleation. (d) derived from the images in (c), time trace of the average out-of-plane magnetization (squares) during current injection (line). Successive pulse amplitudes of - 3.1 x $10^8$ and + 4.4 x $10^8$ A/cm$^2$ and $B_x$= 0.11 T [341] (e-f) Switching probability P of a square of a Pt(3 nm)/Co(0.6 nm)/AlO$_x$ layer as a function of $B_x$ at different current amplitudes of pulses of 210 ps in (e) and as a function of pulse length at a fixed field of 91 mT and varying current amplitudes in (f) [432].*

In **Figure 31**a-b, we show the macrospin simulations of switching by SOT of in-plane (**Figure 31**a) and out-of-plane (**Figure 31**b) magnetizations. With initial in-plane magnetization, the SOT enlarges progressively precessions of the magnetization around its initial orientation to finally reverse it, as it was also the process for switching by STT in **Figure 29**d. This long incubation time leads to switching times of a few ns or longer. As shown in **Figure 31**b for PMA in the same macrospin picture, the action of SOT is immediate and can lead to very short switching time below 1 ns. Analytic expressions as well as macrospin simulations reproduce not only the short switching times in the ns range but also several other features related to symmetry, such as the requirement of an in-plane field $B_x$ and the dependence of the switching current on the anisotropy field and $B_x$ [433]. For a realistic interpretation of experiments on samples larger than the width of a domain, macrospin models are no longer realistic and it is necessary to consider mechanisms related to the nucleation and extension of domains of opposite out-of-plane magnetizations. However, even in this nucleation-extension regime, the SOT switching of PMA layers is also very short, as it turns out from **Figure 31**c [424] showing snapshots of X-ray magnetic dichroism images of a dot of Pt/Co/MgO with PMA during the reversal of its magnetization by current pulses in Pt by SOT. The nucleation of reversed domain starts on the edges at a point (in red) where the combination of the applied field and the DMI favors this nucleation. In addition, as again expected from symmetry, an applied field along x (- x), the SOT induces a reversal from up (down) to down (up) for positive current and from down (up) to up (down) for negative current. As shown in **Figure 31**d-f, some of the reversals occur in less than 1 ns. **Figure 31**e-f shows that the probability of switching increases with the amplitude of the in-plane field as well as with the amplitude and duration of the current pulses.

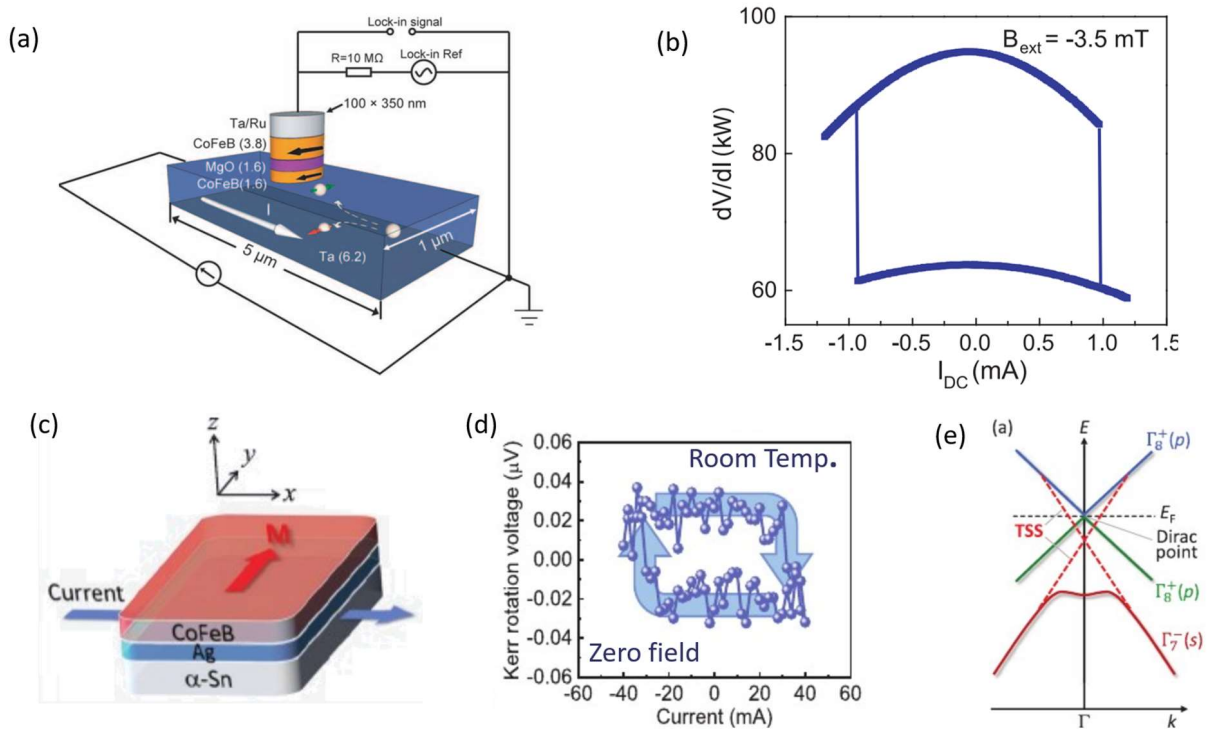

*Figure 32. (a-b) Current-induced switching of in-plane magnetization of a CoFeB layer by SOT generated from SHE in a Ta layer [364], with experimental device in (a) and switching loop at room temperature detected by TMR in a CoFeB/MgO/CoFeB MTJ in (b). (c-e) Current-induced switching of the in-plane magnetization of a CoFeB layer generated by EE in the surface states of the Dirac semimetal α-Sn in a α-Sn/Ag/CoFeB trilayer [428], device in (c), switching loop at zero field and room temperature detected by MOKE in (d) and band structure and Dirac cone of α-Sn in (e).*

Experimental examples of switching of in-plane magnetization by SOT are also displayed in **Figure 32**, with SOT induced either by the SHE of the heavy metal Ta in **Figure 32**a-b or the EE in the topological surface states of the Dirac semimetal α-Sn in **Figure 32**c-d. As pointed out above and illustrated by **Figure 32**a-b in a macrospin picture, the disadvantage of in-plane magnetizations by SOT is a long incubation time during progressively enlarging precessions. The resulting slow dynamics compared to layers with PMA makes the latter the most promising SOT-based devices. However, for some types of applications, the advantage of in-plane magnetism is the possibility of switching by SOT in zero applied field, as illustrated by **Figure 32**c-d for the switching by SOT generated by EE in the interface states of α-Sn.

### 3.3.3. Magnetization switching of single magnetic layers by SOT

Most of the experiments described in the previous paragraph are performed with bilayers including a magnetic layer and a layer with large spin-orbit coupling (heavy metal or material having Rashba or topological surface states). The bilayer structure breaks the inversion symmetry, which is the condition for current-induced torque on a magnetic layer in an heterostructure. Additionally, the spin-orbit coupling of the nonmagnetic layer is used to generate the spin current for the SOT. However, switching by SOT of a single magnetic layer can also be obtained if the magnetic layer itself has a large spin-orbit coupling generating spin currents (for example, the spin-orbit coupling of the 5d band of rare earths or Pt magnetic alloys) and, in addition, no inversion symmetry. The absence of inversion symmetry can be obtained, for example, with a non-centrosymmetric crystal structure [423], by introducing a composition gradient along the out-of-plane axis [375,434–436], or with non-symmetric interfaces [375].

Another example of electrical switching of a single magnetic layer with non-centrosymmetric crystal structure, we have the antiferromagnetic CuMnAs [437] and $Mn_2Au$ [438]. In these cases, the antiferromagnetic order and the particular crystal structure result in staggered SOT in each sublattice, leading to current-induced switching of the Néel vector. Such effect has been shown to be deterministic and multi-level, with potential towards embedded memory-logic applications [439].

We can also cite the pioneering results of Miron et al. on a Co layer between Pt and MgO, as those described in [440]. The perpendicular switching could be ascribed either to the SHE of Pt or to the Rashba effect induced at the interfaces of Co with Pt and MgO. In the second case, it would correspond to a switching of a single Co layer thanks to its asymmetric interfaces with Pt and MgO.

### 3.3.4. Field-free switching by SOT

Since applying an in-plane field $B_x$ to reverse a perpendicular magnetization by SOT is an important disadvantage for the development of applications, several approaches have been developed to solve the problem. The first one is to introduce additional magnetic stripes to provide a dipole field or an exchange-induced effective field [441–443]. An effective $B_x$ can also be created by an in-plane exchange bias field provided by an antiferromagnet [441,444,445].

An interesting solution was also proposed by Wang et al [446] by combining the STT and SOT to achieve the field-free and low power switching of the out-of-plane magnetized free layer of a MTJ.

Finally, field-free switching has also been obtained in single crystal structures by going out of the rotational invariance of the standard polycrystalline structures. Liu et al [423] could achieve field-free switching with $L1_1$-ordered CuPt/CoPt bilayers in which the low-symmetry point group 3m1 generates a SOT depending on the relative orientation of current and crystal axes and leads to field-free switching for some of these orientations. By tuning the composition of the CoPt layer, the same authors are able to achieve self-switching, which is also field free due to the low symmetry at the Co platelet/Pt interfaces present in the CoPt alloy [447].

### 3.3.5. Current-induced magnetization switching of insulating magnetic material

As described in 3.1.5, the injection of a spin current into a magnetic insulator can be achieved by interfacial conversion of a spin current carried by electrons in a metallic layer into a spin current carried by magnons in the magnetic insulator. The resulting torques on the magnetization obey the same symmetry rules as those described for magnetic metals in the preceding subsections. A typical example is the switching of the out-of-plane magnetization of TmIG in W/TmIG bilayers by the spin current initially induced by SHE in W. As for perpendicularly magnetized metallic magnetic layers in 3.1.2, the switching is induced by the combination of SOT and in-plane magnetic field [406].

### 3.4. Current-induced motion of domain walls

The study of the current-induced motion of domain walls (DW) [448,449] (now, accelerated by the proposition of DW-based racetrack memory by Parkin et al [450]), has been an intense field of research in the recent years. An important progress came from the prediction by Thiaville et al that DW of Néel type can be stabilized by DMI and moved at high velocities by a current [451]. Most of the recent studies have been developed on this type of DW.

**Figure 33**a displays a schematic of the DMI interaction at an interface between a magnetic metal and a nonmagnetic heavy metal, $H_{DMI} = (\mathbf{S_1} \times \mathbf{S_2}) \cdot \mathbf{D_{12}}$. In a magnetic layer with PMA, the DMI favors a given direction of rotation when one goes from $\mathbf{S_1}$ to $\mathbf{S_2}$ and leads to the chiral Néel DWs described by Thiaville et al [451] and presented in **Figure 33**b: when one goes from left to right in the figure, the rotation of the spins is counter-clockwise in both DWs and the direction of the central spins are opposite in the up/down and down/up DWs. Thiaville et al [451] showed that such chiral DWs created by DMI can be moved at high velocity by the SOT induced by SHE in the heavy metal layer below or above the magnetic layer. **Figure 33**c displays a typical calculated variation of the velocity as a function of the magnitude of DMI.

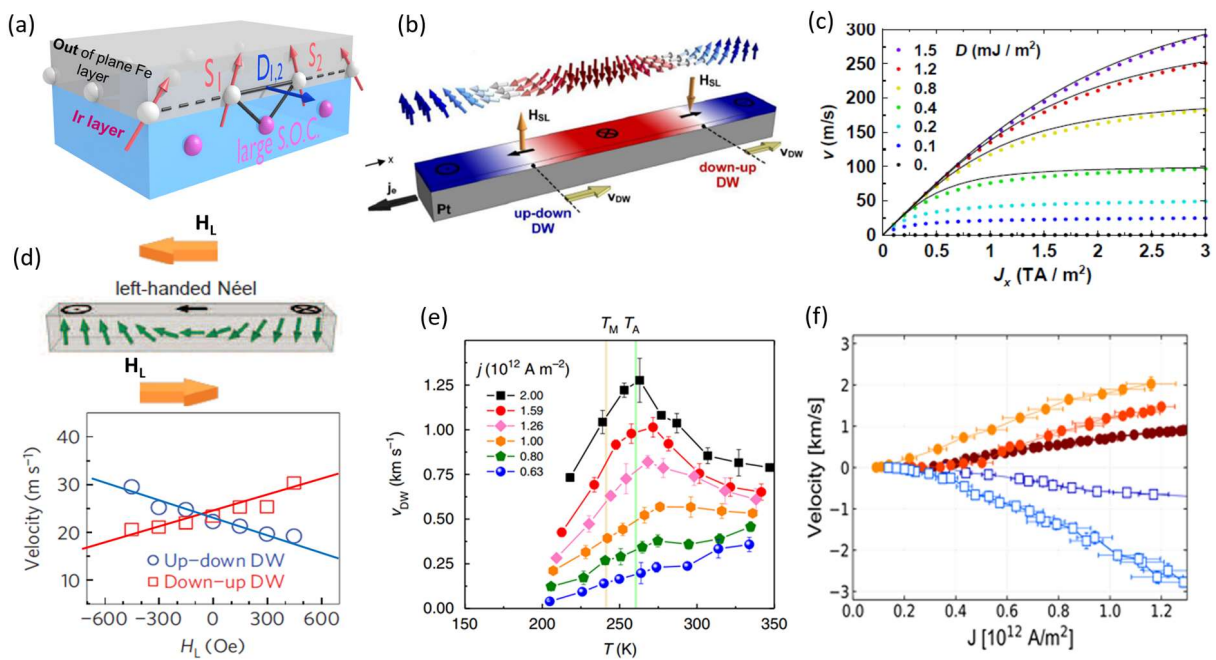

*Figure 33. (a) Illustration of the DMI induced by spin-orbit coupling and breaking of inversion symmetry at the interface between a magnetic layer and a heavy metal. (b) Image of a left-handed chiral DW in Pt/CoFe/MgO. The effective field $H_{SL}$ induced by SHE in Pt moves adjacent up-down and up-down DW in the same direction against electron flow $j_e$. (adapted from [452]. (c) Velocity of chiral DW vs current density for several DMI values, as calculated by Thiaville et al [451]. (d) Top: Schematic showing that an applied in-plane field $H_L$ along the current axis can help the DMI (top arrow) or compete (bottom arrow) with for the formation of a chiral Néel DW (adapted from [452]. Bottom: Dependence of the DW velocity on the sign and magnitude of applied field along the current axis (adapted from [452]). (e) SOT-induced velocity of Néel*

*DWs as a function of temperature in the vicinity of the compensation temperatures $T_M$ (blue vertical line) and $T_A$ (green vertical line) for a ferrimagnetic $Co_{44}Gd_{56}$ layer on Pt [453]. (f) STT-induced velocities, up to about 3 km/s, for Néel DWs as a function of temperature in the vicinity of the compensation temperature of $Mn_{4-x}Ni_xN$ films [454].*

One of the first experimental indications of the influence of spin-orbit coupling effects on the current-induced motion of DW came from the experiments of Miron et al. and Moore et al., who proposed that the current induced DW motion observed on Pt/Co/MgO and not in symmetric Pt/Co/Pt was due to the field-like torque generated by Rashba interactions [455,456]. Then, the 2012 prediction of Thiaville et al. [451] on the conjugated effects of DMI and SOT was clearly confirmed by the experimental papers of Emori et al. and Ryu et al. [452,457]. The role of the DMI was tested by looking at the variation of the velocity when an in-plane magnetic field is applied along the axis of the current. As shown in **Figure 33**d, an in-plane field $H_L$ depending on its orientation along the current axis, helps the DMI in the stabilization of the down-up left-handed Néel DW or competes with it. What is expected is an increase (decrease) of the velocity of the down-up (up-down), as observed in the experimental results in the bottom of **Figure 33**d. Another test was based on the knowledge that Pt and Ta present opposite signs of SHE. Emori et al [452] compared the DW velocities in Pt/CoFe/MgO and Ta/CoFe/MgO and the observed opposite velocities confirmed that the origin of the current-induced motion is the spin current generated by SHE and the resulting torque on the DW.

Most efforts, after 2013, have been devoted to improving the potential of current-induced motion of chiral DW for applications with two main objectives: higher velocities with smaller current and thinner DW width to reduce the bit size in nanodevices. **Figure 33**e shows an example of a remarkable result obtained in $Pt/Gd_{44}Co_{56}/TaO_x$ films for temperatures close to the angular compensation temperature $T_A$ of the ferrimagnet $Gd_{44}Co_{56}$ at which there is a compensation of the angular momenta of the antiferromagnetically aligned Gd and Co [453]. At this temperature and around, the precessional regime of the dynamics is strongly reduced, which gives an immediate motion and high velocities, as shown in **Figure 33**e with velocities exceeding 1 km/s. In addition, as the magnetic compensation temperature $T_M$ is close to $T_A$, the magnetization is small in this temperature range, what reduces the stray field interactions and the width of the DW.

Other directions have been explored to obtain large velocities in absence of SOT by exploiting STT in magnetic materials of strongly spin-polarized conduction and small magnetization, as with $Mn_4N$ grown epitaxially on $SrTiO_3$ and velocities above 1 km/s [458]. Doping $Mn_4N$ with Ni led to velocities close to 3 km/s for a sample of very small magnetization at the vicinity of the magnetic compensation. Because the current spin polarization is related to the spin on the Mn(I) site, the sign of the velocity changes when the global spin of the alloy becomes opposite to the Mn(I) spin at the Ni concentration for compensation, as shown in **Figure 33**f [454].

### 3.5. Current-induced motion of magnetic skyrmions

A magnetic skyrmion is a local whirl of the spin configuration in a magnetic material, a type of topological spin structure already referred to in Section 2.5. As shown in **Figure 34**a for a Néel skyrmion in a magnetic layer with out-of-plane magnetization, the spins inside the skyrmion rotate progressively with a fixed chirality, for example from the up direction at one edge to the down direction in the center and then to up direction again on the other edge. The type of non-trivial topology characterizing the skyrmions was introduced by T.H.R. Skyrme in nuclear physics as topological solitons in the nuclear field [459]. In the case of skyrmions in magnetic materials [460–462], the spin configuration is generally determined by chiral interactions of the DMI type and, consequently, skyrmions can be found in non-centrosymmetric lattices in which they were first observed by neutron scattering [273] or Lorentz microscopy [463]. Later, skyrmions could be found in systems with DMI induced by inversion symmetry breaking at interfaces [271] and were first observed by spin-polarized scanning tunnelling microscopy experiments on Fe monolayers grown on Ir [464]. The non-trivial topology of the spin configuration of skyrmions makes that it cannot be twisted continuously to result in a trivial magnetic configuration. This can be described as a topological protection. To be more precise, the skyrmions can form a skyrmion lattice which is the DMI-induced ground state of the spin system [462–464] or exist as individual skyrmions which can be described as metastable local spin configurations stabilized by their topological protection [465,466].

For the specific property of electrical control of magnetism discussed in our review, the crucial property of the skyrmions is their solitonic nature: they can be electrically moved as particles and this possibility is at the basis of many applications. The first experimental results of motion were obtained for skyrmions in non-centrosymmetric lattice from a combination of neutron scattering and Hall effect measurements [467] and from real-space Lorentz TEM images of skyrmion lattices in FeGe in which the motion of skyrmions is induced by electrical currents or gradients of magnetic field or temperature [468]. The current-induced motion of skyrmions can be described as due to STT [465,469–471] or, alternatively in terms of the emergent electromagnetic field generated by the skyrmion spin texture [467,472]. Most applications that have been proposed are based on the current-induced motion, fusion, or annihilation of such individual skyrmions, the best known being the racetrack memory based on the current-induced motion of trains of individual skyrmions.

The most recently studied systems for application are skyrmions induced by DMI at the interface of a thin enough magnetic layer with a heavy metal (Pt, etc.) or an oxide (MgO, etc.), see **Figure 34**b. As a small skyrmion in a single thin layer can be destabilized by thermal fluctuations at room temperature, a convenient and classical structure is a multilayer as that displayed in **Figure 34**b with additive interfacial DMI for Co between Pt and Ir [465]. A small ferromagnetic interaction between Pt/Co/Ir trilayers couples the skyrmions of successive trilayers, which leads to columnar skyrmions of the type represented in **Figure 34**c. Typical magnetic force microscopy images are displayed in **Figure 34**d.

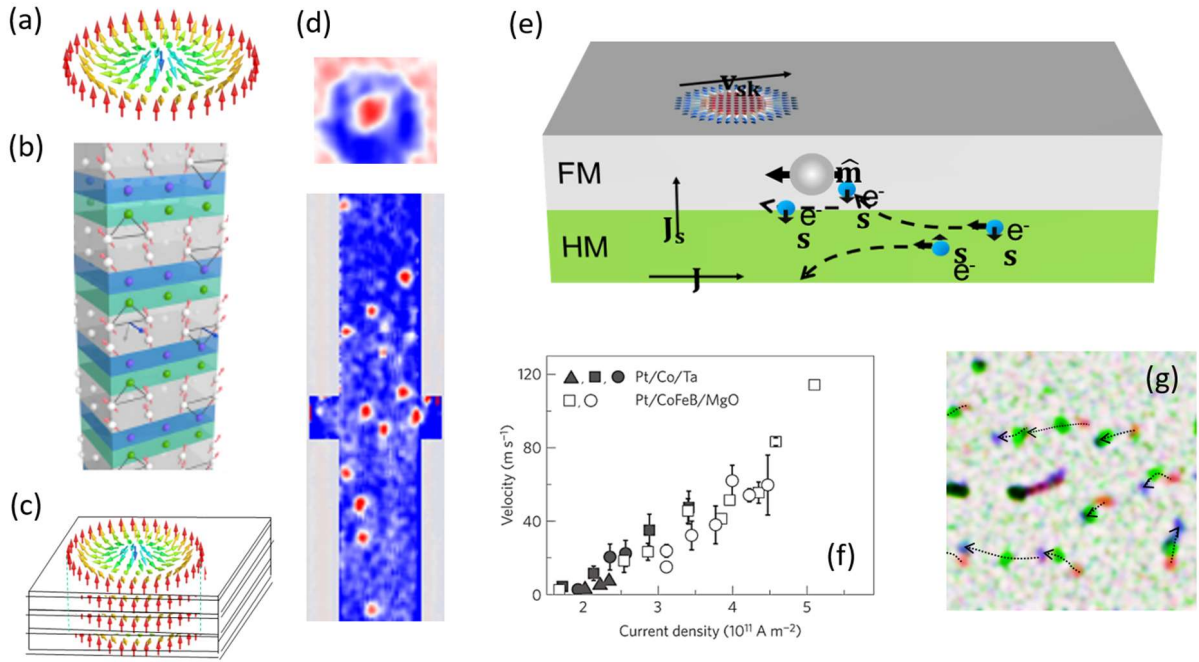

*Figure 34. Current-induced motion of magnetic skyrmions. (a) Spin configuration in a Néel skyrmion. (b) Multilayer with additive DMI at the top and bottom interfaces of the Co layers. (c) Column of coupled skyrmions in a multilayer with interfacial DMIs. (d) Magnetic Force Microscopy images of skyrmions in a multilayer of the type shown in (b) [473]. (e) Motion of skyrmions driven by the SOT induced by SHE in the heavy metal below the magnetic layer. (f) SHE-induced skyrmion velocity as a function of current density in two types of multilayers [289]. (g) Snapshots of the SOT-driven motion of skyrmions in a |Ta10|Pt8|(Co1.4|Ru1.2|Pt0.6)x3|Pt2.4 multilayer induced by $7 \times 10^{11}$ A/m² pulses of 12 ns (courtesy of N. Reyren).*

After the presentation of the current-induced motion of DW in the preceding subsection, the simplest way to understand the current-induced motion of a skyrmion is to consider it as a couple of DWs, up/down from one edge to the center, down/up from the center to the other side. In agreement with what was found for DWs, an efficient way to move the skyrmions is by the SOT generated by SHE in the heavy metals below or above, as represented in **Figure 34**e, or due to EE at the interfaces of the magnetic layer [465,466]. A general feature of the current-induced motion of skyrmions is the coexistence of a longitudinal motion (i.e., along the direction of the current) and a transverse motion (the so-called skyrmion Hall effect) generated by gyrotropic forces related to the topology of the skyrmion. The direction of the longitudinal motion depends both on the chirality of the skyrmion and the spin polarization of the injected current (typically, the motion is in the direction of the charge current for the DMI at the Pt/Co interface and the SHE of Pt). The transverse deflection of the skyrmion, left or right, depends on the spin polarization at the center of the skyrmion. Experimental results on the velocities obtained by SOT are presented in **Figure 34**f. An almost linear variation of the velocity with the current density starts only after a creep regime in which, due to the pinning by defects, the skyrmions do not move or move at only very low velocity, while the skyrmion Hall angle is small. Above a critical current, the velocity increases linearly as expected by theory [474] and in **Figure 34**f, reaches values around 100 m/s. However, with this type of multilayers generally fabricated by sputtering, the scattering and pinning by defects have significant effects even in the quasi-linear regime, which usually leads to the type of non-uniform motions illustrated by **Figure 34**g. A current challenge is obtaining skyrmions in materials with less defects, single crystal layers or 2D van der Waals magnets (see next subsection). Another challenge is the suppression of the transverse motion and promising results have been obtained with antiferromagnetically coupled skyrmions in successive layers [475,476].

### 3.6. Control of magnetism by current-induced torques in 2D magnets.

As for the 3D magnets, the magnetization of 2D magnets can be controlled and manipulated by current-induced torques, STT or SOT. However, the SOT will play a more important role in the case of 2D magnets for the following reason: because the Mermin-Wagner theorem rules out magnetic ordering for isotropic systems of Heisenberg spins [249], magnetically ordered materials exists in 2D only if they can escape from the Mermin-Wagner theorem thanks to large magnetic anisotropies induced by the large spin-orbit interactions of elements such as, for example, Te, I or Bi [250]. In addition, because interfaces play a particularly important role in the properties of 2d materials, the generation of spin current at interfaces by interfacial Rashba interactions and EE can be particularly relevant in heterostructures of 2D magnets (vdW heterostructures).

The first example of magnetization control by SOT shown in **Figure 35**a is the switching of the out-of-plane magnetized 2D ferromagnet $Fe_3GeTe_2$ by the spin current generated by the SHE of Pt deposited on a metallic $Fe_3GeTe_2$ layer [477,478]. As in the switching of PMA of 3D magnets by SOT in subsection 3.3.2, an applied field along the current direction is required to switch the magnetization of $Fe_3GeTe_2$. Clockwise (**Figure 35**b) or counter-clockwise (**Figure 35**c) loops are observed depending on the direction of the applied field. Similar switching of 2D magnets with PMA have been also been obtained with semiconducting $Cr_2Ge_2Te_6$ in combination with Ta or Pt [479–481]. For possible future application to SOT-MRAM devices, it is interesting to compare the current densities and in-plane fields required for SOT switching in 3D and 2D magnetic materials. **Figure 35**d from [480] compares experimental data of some bilayer of 3D or 2D magnetic materials with heavy metals or topological insulators. A smaller switching current density is required for Ta/ $Cr_2Ge_2Te_6$, an order of magnitude below the density for the classical Ta/CoFeB system, the disadvantage of the 3D CoFeB compared to a 2D magnet coming mainly from the useless large current shunting in the metallic CoFeB layer. The current density required for Ta/ $Cr_2Ge_2Te_6$ is even smaller than for a bilayer of Ta and the magnetic insulator TmIG. Concerning the required in-plane field, the values are similar for 2D and 3D magnetic materials. However, it is needless to say that the bottleneck of 2D magnets for applications is still the required low temperature, even if some recent experiments have shown that, in some 2D magnets, the ordering temperature can be raised above room temperature, as it has been already achieved for $Fe_3GeTe_2$ grown on $Bi_2Te_3$ [482] or with electrostatic doping [255].

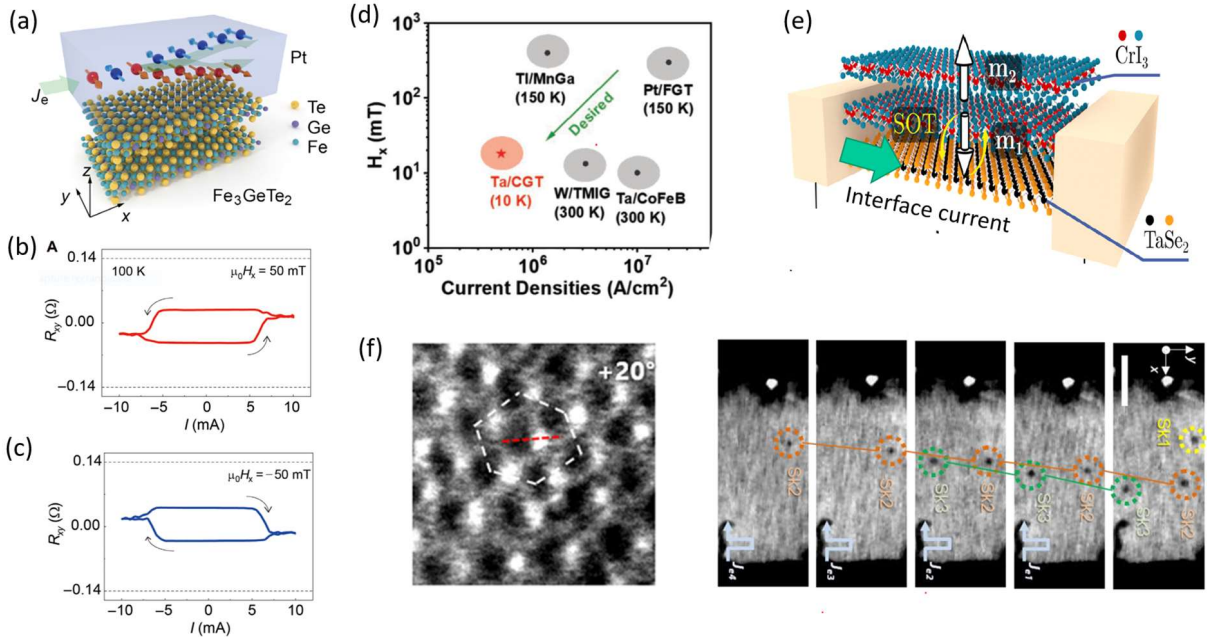

*Figure 35.* Control of magnetism by currents in 2D magnets. (a) Image of a Fe$_3$GeTe$_2$/Pt bilayer [478]. (b-c) SOT switching of the bilayer displayed in (a) in the presence of positive (b) or negative (c) in-plane field along the current direction [478]. (d) Comparison of the current densities and in-plane fields required for SOT switching in devices based on 3D magnets (CoFeB, MnGa, TMIG) and 2D magnets (Fe$_3$GeTe$_2$, Cr$_2$Ge$_2$Te$_6$), with best results for Ta/ Cr$_2$Ge$_2$Te$_6$ [480]. (e) Schematic view of a (CrI$_3$ bilayer/TaSe$_2$) hetero structure in which the SOT induced by the interfacial current can drive the relative orientation of the magnetizations of the two CrI$_3$ layers from parallel to antiparallel [483]. (f) Lorentz microscopy images of skyrmions in a Fe$_3$GeTe$_2$ film with oxidized interfaces and current-induced motion of the skyrmions [484].

In the examples of switching of 2D magnets by SOT just described above, the sources of spin current are 3D heavy metals. Alternatively, both the magnetic layer and the spin source can be 2D materials forming a van der Waals heterostructure with spin currents generated at their interface. Dolui et al [483] have developed a first-principle quantum model for the transport in van der Waals heterostructure (TaSe /CrI$_3$ bilayer) in which, at equilibrium, there is an antiferromagnetic coupling between the two CrI$_3$ layers. They find that a current flowing in the 2DEG at the interface between TaSe and the bottom CrI$_3$ layer (see **Figure 35**e) can switch by SOT the magnetization of this bottom layer to induce a ferromagnetic CrI$_3$ bilayer. An experimental demonstration of a magnetization switching by SOT in an all-van der Waals heterostructure has been recently reported in WTe$_2$/Fe$_3$GeTe$_2$ by two different groups [485,486].

The last point on current-induced magnetization control in 2D magnets is the manipulation of skyrmions. The only example we know is presented in **Figure 35**f and shows the motion of skyrmions in Fe$_3$GeTe$_2$ foils. Magnetic skyrmions in 2D magnets have been observed in several groups [484,487–489], the skyrmions in **Figure 35**f being Néel skyrmions generated by interfacial DMI at the oxidized interfaces of Fe$_3$GeTe$_2$ (in first approximation, interfaces between Fe$_3$GeTe$_2$ and oxidized Fe$_3$GeTe$_2$). The results in **Figure 35**f are promising as the motion seems less affected by defects and more uniform than in the usual sputtered multilayers of magnetic and heavy metals. Many points remain to be understood for skyrmions in 2D magnets as, for example, the exact mechanism inducing the motion, STT or SOT.

# 4. Combined use of electric field and current-induced torques

In Section 3, we have reviewed the control of magnetization by current-induced torques, a field with a large potential for applications in MRAM technology. One of the major drawbacks of using current-induced torques is the energy dissipation associated to the high current densities required for the switching. In this regard, the use of an electric field (voltage) to assist the current-induced torque is of extreme interest to lower the energy consumption of MRAM technology.

The electric field can modulate different ingredients in a current-induced torque system. One of them is the free layer storing the non-volatile information, whose magnetic anisotropy can be controlled with the application of a voltage (VCMA effect, reviewed in Section 2.3.3). Another one is the electric-field control of the spin-charge interconversion, the mechanism at the core of SOTs, which is reviewed in the following Section. Such electric control has been recently shown that it can also be performed through ferroelectricity, reviewed in Section 4.2. Finally, examples where the electric field is used to assist switching in STT and SOT systems will be reviewed in Section 4.3.

## 4.1. Electric field control of spin-charge interconversion

Nowadays, the most widely used way to create spin currents without the use of a ferromagnet is with charge-to-spin current conversion effects in systems with high spin-orbit coupling such as the SHE (see Section 3.1.3) or the EE in interfaces with Rashba coupling and surface states of topological insulators (see section 3.1.4). Conversely, spin currents can be detected with spin-to-charge current conversion from the corresponding inverse effects. Since the conversions fulfill Onsager reciprocity, we will use the term spin-charge interconversion to refer to both the direct and the inverse conversions. In this section, we review the various possibilities for electrical control of such spin-charge interconversion, which can open the path to new functionalities for future energy-efficient electronic devices.

The first observation of SHE controlled by an electric field was reported in GaAs. In this material, the different valleys in the band structure have different spin-orbit coupling properties. Okamoto et al. [490] excited spin-polarized electrons at valley $\Gamma$ by circularly polarized light and applied an electric field to induce an electrical intervalley transition in the conduction band from valley $\Gamma$ to L, which shows larger spin-orbit coupling (**Figure 36**a). The spin Hall angle, determined by the generated transverse voltage ($V_{SH}$) in a GaAs Hall bar (inset in **Figure 36**b), could be tuned from 0.0005 to 0.02 by the electric field (**Figure 36**b).

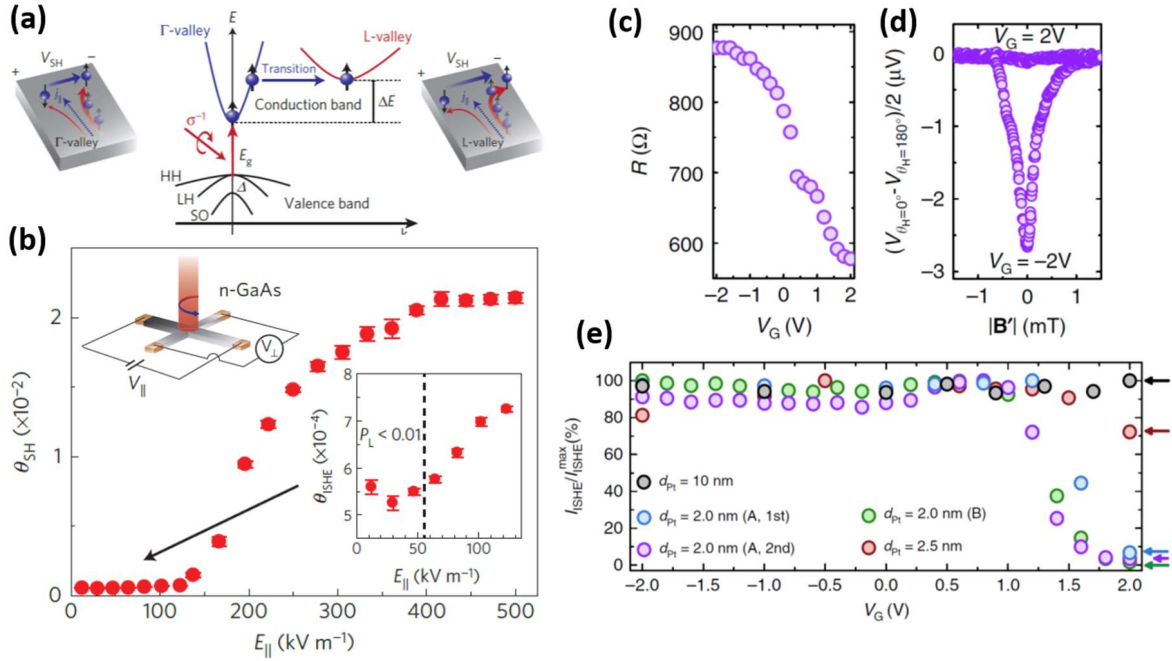

*Figure 36. Electric control of spin-charge interconversion effects in different systems. (a) GaAs band structures and spin-polarized electrons generated by circularly polarized light absorption. A high electric field induces the transition of the spin-polarized electrons from the Γ-valley to the satellite L-valley where part of its p-character provides a larger effective SOC. Sketches at left and right show the optically induced SHE for Γ-valley and L-valley, respectively, being larger in the latter. (b) Electric field dependence of the spin Hall angle in GaAs. Right inset: zoom at low field. Left inset: measurement configuration of the optically induced SHE. (c) Resistance as a function of gate voltage $V_G$ for a 2-nm-thick Pt on YIG. (d) Output voltage detected during spin pumping at the same sample for $V_G$=2 V and -2 V. (e) Normalized spin-to-charge output current as a function of $V_G$ for different Pt thicknesses. (a) and (b) from [490]. (c)-(e) from [491].*

The SHE in a heavy metal has also been tuned by voltage using a ionic liquid gate on ultrathin Pt. Dushenko et al. [491] showed the resistivity of Pt could be tuned by gating (**Figure 36**c). They used the spin pumping technique from an adjacent YIG layer to inject a spin current and measure the transverse charge current to quantify the SHE in Pt (**Figure 36**d). Since the spin Hall angle in Pt depends on its resistivity, a clear gate dependence was observed for the thinnest Pt films (**Figure 36**e). This experiment allowed the authors to reach the dirty regime of the SHE in Pt.

Spin-charge interconversion has been intensively studied in topological insulators, due to the spin-momentum locking present in their Dirac-cone-type surface states. Its efficiency is in principle independent of the Fermi level position within the Dirac cone [492]. Indeed, Wang et al. [493] did not observe any gate dependence of the spin-charge interconversion efficiency in epitaxial $Cr_{0.08}(Bi_xSb_{1-x})_{1.92}Te_3$ thin films measured by spin pumping. However, the carrier density of the surface states is tunable with electric gating [494] and, therefore, the output signal can also be tuned [495]. For instance, Tian et al. [496] observed in $Bi_2Te_2Se$ a modulation of the spin signal measured by spin potentiometry with backgate voltage, due to the gate tunability of its resistance. Voerman et al. [497] also observed a tuning of the spin signal with the backgate voltage when measuring $BiSbTeSe_2$ combined with graphene using a non-local spin valve technique, although the origin remains elusive. In general, one must be aware that experiments involving TIs have the additional complication that bulk is hardly an ideal insulator. Therefore, it can contribute to transport and be a potential source of spin-charge interconversion via SHE.

Graphene is a Dirac semimetal in which the Fermi level can easily tuned with an applied gate, therefore allowing to control its transport properties [498]. While graphene is an outstanding material for long-distance spin transport [499] due to its weak intrinsic spin-orbit coupling and negligible hyperfine interaction, it is not a preferred material for spin-charge interconversion because of the very same reason. Nevertheless, a small but measurable spin-charge interconversion was reported in pristine graphene using spin pumping techniques from an adjacent YIG layer, although the origin of the effect, SHE [500,501] or EE [502], was a source of controversy. Dushenko et al. [501] measured the spin-charge interconversion as a function of gate with a ionic liquid and observed a sign change of the spin-charge interconversion signal when the carrier type was tuned from electrons to holes. Such a sign change with the carrier polarity is a result of symmetry [503]. The small spin-charge interconversion efficiency in graphene can be greatly enhanced by inducing spin-orbit coupling by proximity with a transition metal dichalcogenide (TMD), which gives rise to a spin texture with both an out-of-plane and a helical in-plane component. Theoretical calculations predicted a very large SHE [504] and EE [505,506] in graphene/TMD van der Waals heterostructures, in which both effects can be modulated by tuning the Fermi energy of the system, changing sign with the carrier polarity. While the SHE gives rise to a spin current and spin accumulation with spins pointing out-of-plane when a current is applied in the proximitized graphene (red arrows in **Figure 37**a), the REE generates a non-equilibrium spin density with spins pointing in plane (blue arrow in **Figure 37**a). Using a non-local spin valve technique that allows to distinguish the direction of the generated spins (**Figure 37**a), a large SHE was first experimentally confirmed in graphene/MoS$_2$ [507], followed by the simultaneous observation of SHE and EE in graphene/WS$_2$ [508,509]. In particular, Benitez et al. [509] confirmed experimentally the predicted sign change of the SHE (below ~200 K) and the EE (up to room temperature) with carrier concentration, which is tuned by gating the graphene (**Figure 37**b). A gate dependence of the EE in proximitized graphene has also been reported with WS$_2$ [508], TaS$_2$ [510], (Bi,Sb)$_2$Te$_3$ [511], and MoTe$_2$ [512]. A large variation of the SHE with applied gate has also been observed in graphene/WSe$_2$, with an unprecedented spin-charge interconversion efficiency [513].

A different system of high interest for spin-charge interconversion are 2DEGs that occur at interfaces of oxide heterostructures. A primary example is the 2DEG present in the SrTiO$_3$/LaAlO$_3$ system [514]. By using spin pumping (**Figure 37**c), spin-charge interconversion with in-plane spins originating from the EE was observed in SrTiO$_3$/LaAlO$_3$, showing a large efficiency [392]. A strong gate-tunability associated to the band structure of the 2DEG allows to change the sign of the spin-charge interconversion. A different gate dependence in the same system have been reported at room temperature [515]. A more dramatic modulation of the EE by gate voltage has been subsequently obtained in the 2DEG present in a SrTiO$_3$/AlO$_x$ interface [393], where the spin-charge interconversion efficiency parameter ($\lambda_{IEE}$) changes sign several times with gate voltage (**Figure 37**d). The evolution of this parameter with gate and its large value can be explained by the different contributions of the electronic bands involved, which have different properties from Rashba-like splitting to topological avoided crossings. Spin-charge interconversion with spins out-of-plane originating from SHE has also been observed in the 2DEG at SrTiO$_3$/LaAlO$_3$ interface using a non-local double Hall bar setup [516,517]. The gate control achieved is also attributed to the complex band structure of the 2DEG [517]. An electric field control of charge to spin conversion was also recently reported through unidirectional magnetoresistance measurements in SrTiO$_3$-based 2DEGs [518,519].

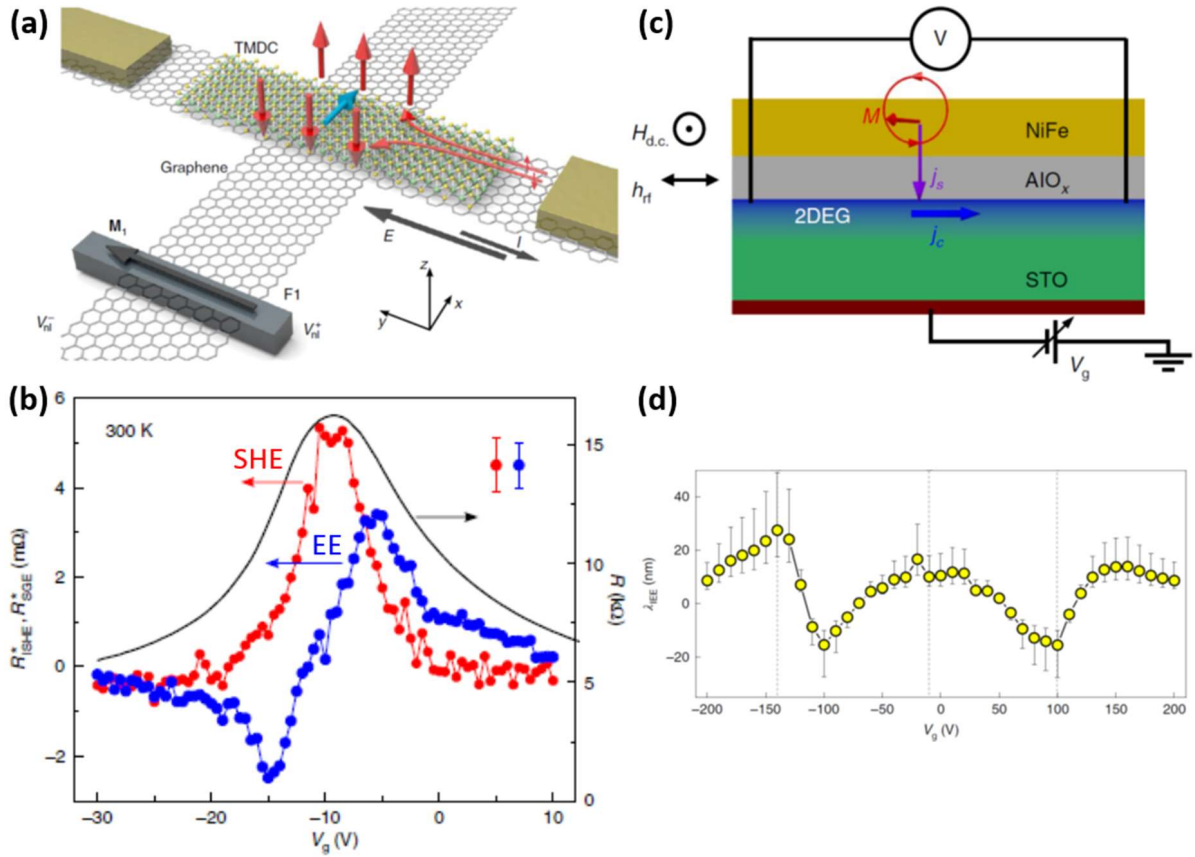

*Figure 37. (a) Sketch of the non-local spin valve concept for spin-charge interconversion measurement in a graphene/TMD van der Waals heterostructure. A current I along the graphene/TMD arm (y-axis) generates a non-equilibrium spin density due to the EE with spins along x (blue arrow) and a spin accumulation with spins out of plane (along z) due to the SHE with opposite orientation at opposite edges of the graphene/TMD arm (red arrows). The induced spins diffuse in graphene towards ferromagnetic electrode F1 and are detected by measuring $V_{nl}^F = V_{nl}^+ - V_{nl}^-$. The EE and SHE contributions to $V_{nl}^F$ are separated via spin precession by applying an external magnetic field along z or x, respectively. (b) Spin-charge interconversion signals for the SHE (red) and the EE (blue) as a function of $V_G$. The sheet resistance of graphene vs $V_G$ is also plotted to show the charge neutrality point. (c) Sketch of the spin pumping experiment to quantify spin-charge interconversion in a SrTiO$_3$/AlO$_x$ 2DEG. (d) Gate dependence of the Edelstein length $\lambda_{IEE}$ of a SrTiO$_3$/AlO$_x$ 2DEG at 15 K. (a) and (b) from [509]. (c) and (d) from [393].*

### 4.2. Ferroelectric control of spin-charge conversion

Being polar materials, ferroelectrics are a natural place to look to engineer Rashba SOC. In addition, their ability to accumulate and deplete charge (depending on the polarization direction) into adjacent materials induces electric fields (over the Thomas-Fermi screening length) whose amplitude and even sign may be switched (cf. **Figure 38**a). If the adjacent material possesses a sizeable SOC, this may generate a region prone to display a Rashba spin-orbit coupling tunable electrically, and in a non-volatile way: in the most simple case, the chirality of the spin contours would be reversed upon switching polarization, as sketched in **Figure 38**b. Injecting a spin current into such system would then lead to the generation of a charge current whose sign will be set by the ferroelectric polarization direction (**Figure 38**c) [520]. The device operating would thus be equivalent to that of a Rashba system combined with a ferromagnet in which magnetization switching would yield a produced charge current of positive or negative sign, with the notable difference that here the sign of the output current is caused by switching a ferroelectric with an electric field rather than by switching a ferromagnet with a magnetic field (or spin-torque). Following Manipatruni et al, this is typically 1000 times more energy efficient [2].

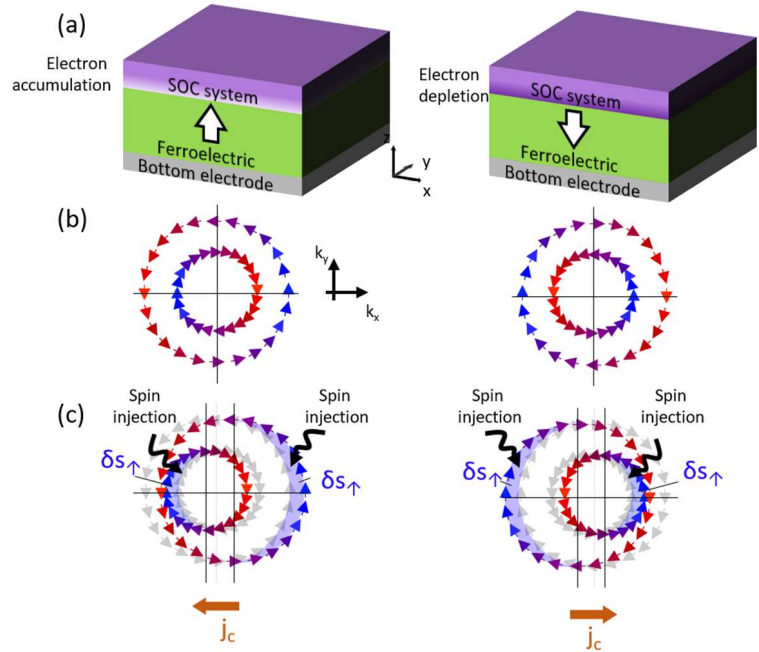

*Figure 38. (a) Sketch of an interfacial ferroelectric Rashba system, in which the ferroelectric accumulates or depletes carrier into an adjacent layer with large spin-orbit coupling (top layer), generating a Rashba state at the interface. (b) Spin contours in the Rashba states. The chirality is reversed upon switching ferroelectric polarization direction. (c) Ferroelectric control of the IEE. From [520].*

Perhaps the first system in which the combination of ferroelectricity and Rashba spin-orbit coupling was considered is GeTe [521]. This compound is the best known member of the family of ferroelectric Rashba semiconductors (FERSC) [522]. GeTe has a ferroelectric $T_C$ of about 700 K in which Ge and Te are displaced along the [111] direction from their ideal rocksalt sites [523]. Its bandgap is only ~0.6 eV [524], which led to difficulties in showing polarization switching, that finally came through piezoresponse force microscopy experiments [525]. GeTe displays a giant Rashba splitting of $\alpha_R$~5 eV.Å owing to several factors, namely the presence of heavy atoms with large SOC, a narrow gap and the same orbital character of the valence and conduction bands. The electronic structure evidencing Rashba-split bands was first reported by Liebmann et al using ARPES and spin-polarized photoemission [526]. Soon afterwards, two papers reported the dependence of the bands spin texture with ferroelectric polarization direction [527,528]. While Rinaldi et al. [527] reported different spin textures for separate samples with up or down ferroelectric polarization tuned by the surface termination, Dil et al. [528] applied an electric field in situ to detect this change.

The ability to control spin textures by ferroelectricity triggered studies on the influence of ferroelectric on spin-charge interconversion. Zhang et al found that the spin Hall conductivity could be strongly tuned by ferroelectricity [529]. Experimentally, Varotto et al made a major advance in the integration of GeTe into spin-orbitronic devices. Not only they provided evidence of ferroelectric switching from electric measurements, but they also showed that the amplitude and sign of spin-charge interconversion efficiency (of amplitude comparable to that of Pt) changed with polarization switching, at room temperature [530], cf. **Figure 39**. This paves the way towards advances devices based on FERSC. We note that the related material SnTe has also been predicted to be a FERSC [531,532].

The low band gap of GeTe leads to the search for more insulating FERSC in the traditional ferroelectric family, perovskite oxides. This includes $BiAlO_3$ [533], $PbTiO_3$ [534,535], $BiInO_3$ [536], strained $KTaO_3$ [537] and strained $SrBiO_3$ [538]. In $SrBiO_3$ in particular, ferroelectric polarization switching was predicted to lead to a reversal of the spin chirality of the Rashba state at the conduction band minimum [538]. Djani et al have however argued that the pseudocubic perovskite oxide family is possibly not the best family to achieve a ferroelectrically tunable Rashba state because in most cases the tunable Rashba state will not be present at the valence band minimum or conduction band maximum but in other bands. They proposed that Aurivilius phases such as $Bi_2WO_6$ are more promising in this respect [539]. Perovskite halides have also been proposed as FERSC [540,541], as well as perovskite nitrides [542]. Outside of the perovskite family, an electrically reversible spin texture has also been proposed for $HfO_2$ [543]. However, to date there have not been experimental demonstration of a Rashba state in most of these compounds, let alone of the possibility to tune it through ferroelectricity.

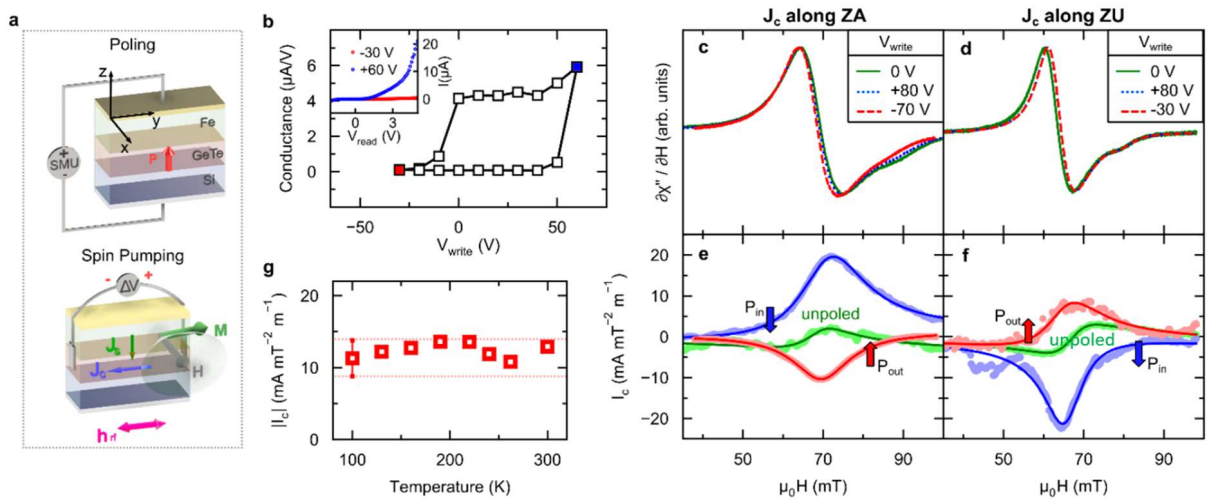

*Figure 39. Ferroelectric control of the spin-charge interconversion in GeTe investigated by spin-pumping FMR. (a) Setup for the study of the ferroelectric switching of the spin-charge interconversion in GeTe. Above, electrical circuit for ferroelectric switching monitored by resistance changes. Below, sketch of the contacts used to measure the lateral voltage proportional to the charge current production in the same experiment. Negative (positive) voltage pulses were applied by a source-measure unit (SMU) to set the ferroelectric polarization direction ($P_{in}$ or $P_{out}$). (b) Hysteresis loop of the conductance versus $V_{write}$ of a Au(3 nm)/Fe(20 nm)/GeTe(15 nm)/Si sample. In the inset, I-V curves of the heterostructure after the application of two saturating voltage pulses at $V_{write}$= -30 V and +60 V. Ferromagnetic resonance (FMR) spectra (c-d) and normalized current production (e-f) at 300 K for the slab oriented along the ZA and ZU direction versus ferroelectric polarization. Dashed curves correspond to $P_{in}$ ($V_{write}$< 0) and dotted curves to $P_{out}$ ($V_{write}$> 0). The spin pumping peak is positive (negative) for $P_{in}$ and negative (positive) for $P_{out}$. The green curve in panels e and f refers to the pristine (unpoled) states. The relatively small amplitude of the spin pumping signal in the unpoled state is associated to a multi-domain ferroelectric configuration. (g) Temperature dependence of the charge current production [530].*

Electrically tunable Rashba states at interfaces between a ferroelectric and a material with large spin-orbit coupling has also been explored. Mirhosseini et al. [544] predicted a Rashba state at the interface between $BaTiO_3$ and an ultrathin film of Bi, with a modest dependence on polarization direction. This system was later explored experimentally and a spin splitting was observed [545]. A fully switchable, giant Rashba coefficient was predicted in oxide heterostructures combining $BaTiO_3$ with $BaRuO_3$, $BaIrO_3$ or $BaOsO_3$ [546] and in $BiInO_3/PbTiO_3$ heterostructures [547].

Experimentally, interfacial systems have been used to achieve a ferroelectric control of spin-charge interconversion. A remarkable result from Fang et al [548] is reported in **Figure 40**c-d. Working with a $La_{0.7}Sr_{0.3}MnO_3/PbZr_{0.2}Ti_{0.8}O_3/Pt$ heterostructures, the authors inject a spin-polarized current from $La_{0.7}Sr_{0.3}MnO_3$ by tunneling through the thin (5 nm) $PbZr_{0.2}Ti_{0.8}O_3$ ferroelectric layer, which gets converted into a charge current through ISHE by the Pt. Depending on the ferroelectric polarization direction, the sign of the ISHE signal is reversed. These experiments were reported at low T only, due to the low spin polarization of $La_{0.7}Sr_{0.3}MnO_3$ at higher temperatures [549], but could probably be extended to room temperature by replacing $La_{0.7}Sr_{0.3}MnO_3$ by another material.

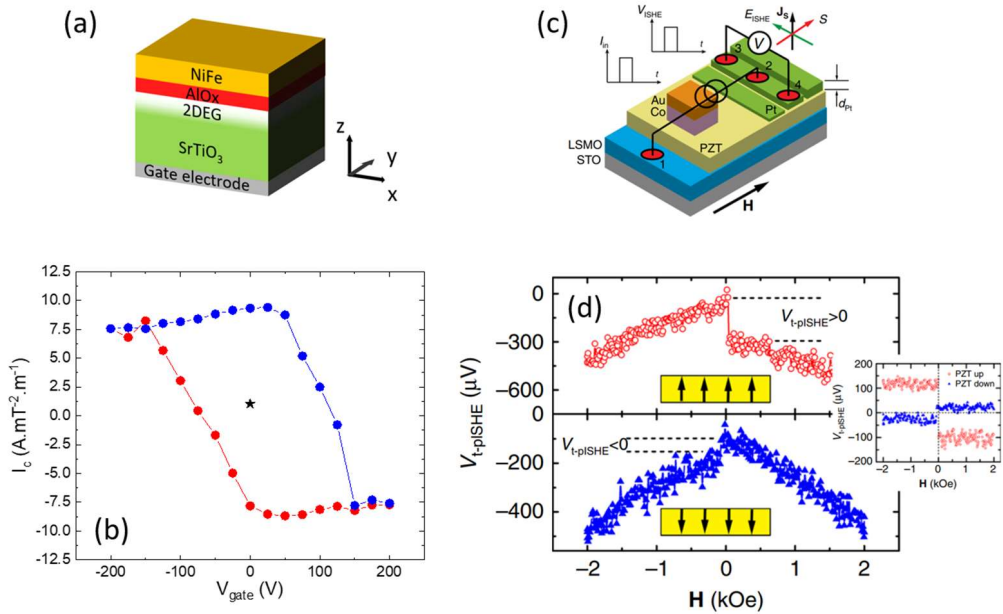

*Figure 40. (a) Sketch of the sample for detecting the ferroelectric control of the IEE (b) Gate voltage dependence of the current produced by spin-charge interconversion through the IEE in $NiFe/AlO_x/SrTiO_3$ heterostructures after applying a large electric field to the $SrTiO_3$ to induce a ferroelectric-like state [520]. T=7 K. (c) Schematic illustrations of tunneling pulsed ISHE measurements in the ISHE-type based on a $La_{0.7}Sr_{0.3}MnO_3/Pb(Zr,Ti)O_3/Pt$ stack. The injected pulsed tunneling current ($I_e$) generates a flow of pulsed spin current ($J_S$) in the Pt metal, which produces a transverse pulsed ISHE voltage ($V_{t-pISHE}$) at T=10 K [548].*

The large IEE reported in $SrTiO_3$ 2DEGs [392,393,550] make them an appealing system for ferroelectric control of spin-charge interconversion. This is all the more true that $SrTiO_3$ is on the verge of ferroelectricity: $^{18}O$ substitution for $^{16}O$ [551], minute Ca substitution for Sr [552], epitaxial strain [553], the application of fs light pulses [554] or the application of a large electric field [555–557] all induce a ferroelectric (or ferroelectric-like) state in STO.

**Figure 40**a-b presents spin-charge interconversion experiments in $SrTiO_3$ 2DEGs formed by the deposition of a thin Al layer, after applying a large electric field (of 5-10 kV/cm). The produced charge

current displays a strong hysteretic dependence on the applied gate voltage, reminiscent of the ferroelectric loops observed in this system [520]. Remarkably, two different remanent states, with opposite produced current signs are obtained, as sketched in **Figure 38**c. This strong gate dependence and sign change is likely connected with the multi-orbital nature of the 2DEG electronic structure, with competing bands having different effective Rashba coefficients. Also important is the very large spin-charge interconversion figure of merit in this system, with $\lambda_{IEE}$~30 nm.

Finally, we mention several recent predictions of ferroelectric Rashba systems in 2D or monolayer materials. This includes $Ag_2Te$ monolayers [558], $MX_2$ monolayers (M=Mo, W ; X=S, Se, Te) [559] and $WO_2Cl_2$ [560].

### 4.3. Electric control of STT and SOT

Starting with STT-based devices, Wang et al. [561,562] first reported the combined effect of VCMA and STT in a MTJ with PMA, consisting of a CoFeB/MgO/CoFeB stack (see **Figure 41**a). In such MTJ, $H_C$ of the free CoFeB layer shows a dramatic change under different bias voltages due to the VCMA (**Figure 41**b). By applying consecutive negative pulses with alternating amplitude, the free CoFeB layer is reversibly switched as monitored with low-voltage TMR measurements (**Figure 41**c). The explanation of the unipolar switching is sketched in **Figure 41**d. All in all, the strong reduction of $H_C$ at negative voltage allows the STT switching to occur at a current density of ~$10^4$ A.cm$^{-2}$, much smaller than the expected ~$10^6$ A.cm$^{-2}$. Using also the combination of VCMA and STT and the same MTJ type, Kanai et al. [563] apply a switching scheme with two voltage pulses: whereas the first pulse induces magnetization precession by the electric-field effect on magnetic anisotropy (see Section 0), the second pulse stabilizes the magnetization direction by STT. This way, a faster and more reliable switching can be obtained. Theoretical simulations show that, in this system, when combining E-field and STT with a single pulse, a deterministic switching is achieved with a current density above ~5x$10^5$ A.cm$^{-2}$, leading to a decrease in the power consumption by 2 orders of magnitude when compared to the switching by STT only [564,565].

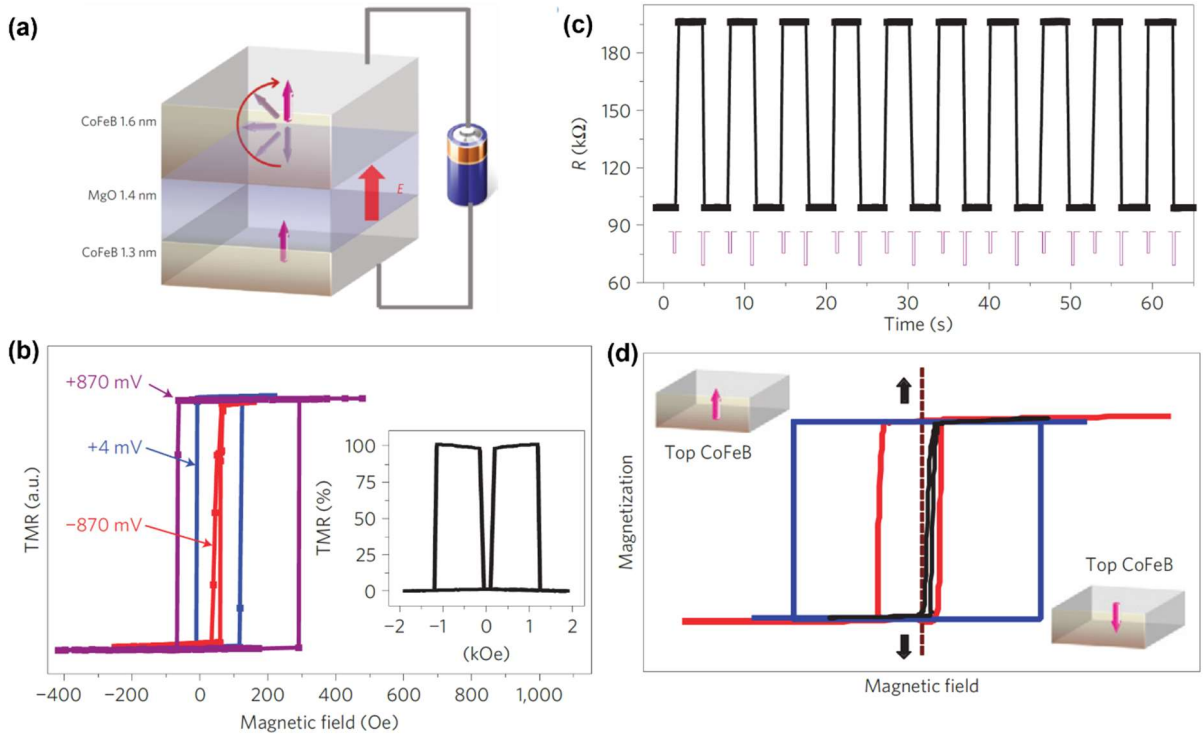

*Figure 41. Electric control of STT. (a) Sketch of a perpendicularly magnetized MTJ and the effect of electric field through a voltage to the free CoFeB layer. (b) Normalized minor loops of the TMR curve at different bias values applied. Inset: The full TMR curve measured at low bias. (c) Unipolar switching of the MTJ by a series of negative pulses (schematically shown in purple at the bottom) with alternating amplitudes of -0.9 V and -1.5 V. A constant biasing magnetic field ($H_{bias}$) of 55 Oe in favor of the antiparallel state at -0.9 V was applied. (d) Sketch of the hysteresis loops of the top CoFeB layer showing the unipolar switching process: magnetization-down → up switching at $V=V_1$ (red) through STT with greatly reduced energy barrier; magnetization-up → down switching at $V=V_2$ (black) by another negative electric field, where $|V_2|>|V_1|$. The loop for V=0 is shown in blue. The vertical dotted line represents the position of the constant $H_{bias}$. The moment of the bottom CoFeB is fixed pointing down. From [561].*

Once the interest of the community shifted from STT to SOT, so did the possibility of combining the effect of E-field with SOT through, e.g., the E-field control of charge-spin conversion. By using the prototypical Pt/Co/Al$_2$O$_3$ stack for SOT, Liu et al. [566] observed the modulation of the field-like torque with an E-field, caused by the enhancement of the interfacial Rashba effect. A modulation of interfacial spin–orbit fields by directly applying an E-field has been confirmed in Fe/GaAs (100) interface by Chen et al. [567].

Although, in these cases, the E-field directly affects the charge-spin conversion, in general it influences the charge-spin conversion in a more indirect way, for instance through oxygen ion migration. By replacing Al$_2$O$_3$ with GdO$_x$, a non-volatile, voltage-control of the oxidation state in the Co/GdOx interface was achieved, leading not only to the expected decrease in the magnetic anisotropy of Co, but also to an enhancement of the damping-like torque, although the later origin could not be addressed [568]. With this same system, Mishra et al. [569] observed not only a change in the magnitude but also in the direction of the SOT, which they attributed to the transport of oxygen ions ($O^{2-}$) modifying the interfacial Rashba SOT at the Pt/Co interface. In a similar stack, Pt/Co/HfO$_x$, and by using ionic liquid gating, Yan et al. [570] reported the modulation of the damping-like torque, in this case attributed to the variation of the spin transparency of the Pt/Co interface with the E-field. Also using HfO$_x$ as a gate insulator, Hirai et al. [571] studied the voltage control of SOT in an in-plane magnetized Pd/Co/Pd/HfO$_x$ stack, in which O$^{2-}$ migration at the top Co/Pd interface is at the origin of the modulation of both the damping-like and field-like torque through different mechanisms. By using oxygen-incorporated Pt in a stack, Pt(O)/FeNi/SiO$_2$, where the damping-like torque is claimed to arise from the Pt(O)/FeNi interfacial SOC, An et al. [572] achieved a voltage control of such SOT through reversible migration of O$^{2-}$ towards or away from that interface. Another indirect way in which an E-field can modulate the charge-spin conversion is through strain, that has also been shown by Filianina et al. [573] to influence the SOT in perpendicularly magnetized W/CoFeB/MgO stacks grown on piezoelectric Pb(Mg$_{1/3}$Nb$_{2/3}$)O$_3$–PbTiO$_3$ (PMN-PT) through a combination of SOC, crystal symmetry, and orbital polarization. Moving from metals to more exotic systems such as TIs, the E-field can change the Fermi level position within the gap of the material. Fan et al. [574] reported E-field control of SOT in a single layer of Cr-doped (Bi,Se)$_2$Te$_3$, a magnetically doped TI. By voltage gating the TI, the SOT strength could be modulated up to a factor of 4, and was attributed to the variation of the carrier density of the topologically protected surface states, which are the source of the charge-spin conversion.

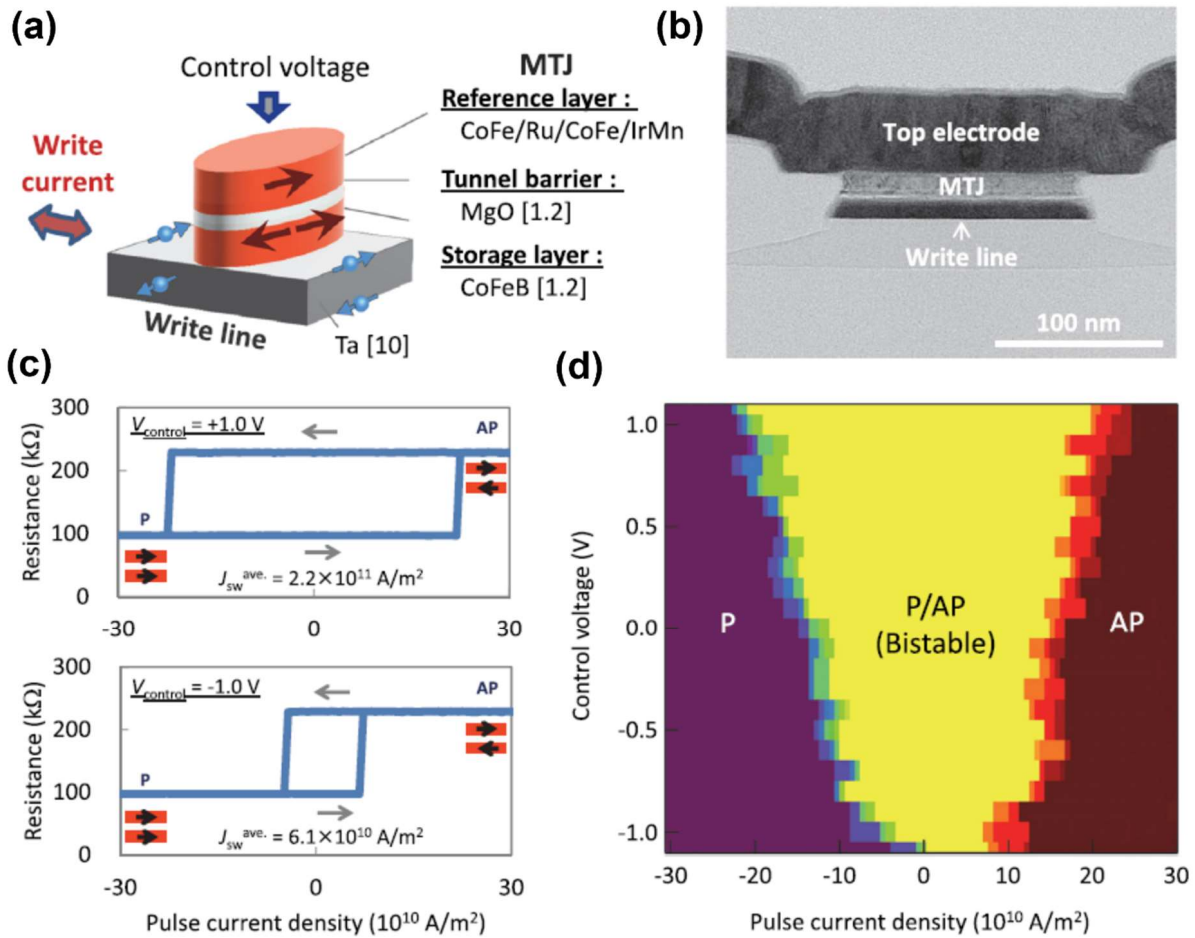

*Figure 42. Electrical control of SOT. (a) Sketch and (b) cross sectional TEM image of the device, a single MTJ consisting of Ta/CoFeB/MgO/CoFeB/Ru/CoFe/IrMn fabricated on a thermally oxidized Si wafer. (c) MTJ resistance as a function of the write pulse current density while applying a control voltage pulse of +1.0 V (top) and -1.0 V (bottom). The width of both write current pulse and control voltage pulse was 50 ns. No H$_{bias}$ was applied during measurement. (d) Switching phase diagram obtained by taking the resistance-write pulse current density curves [575].*

A second possibility is that the E-field directly controls the VCMA, which is the case reported by Inokuchi et al. [575], where the switching current is reduced up to 3.6 times in in-plane magnetized Ta/CoFeB/MgO/CoFeB/Ru/CoFe/IrMn stacks by changing the control voltage from -1.0 V to +1.0 V (see **Figure 42**). In many recent works, though, the E-field effect has been shown to modulate both the VCMA and the charge-spin conversion. For example, Xu and Chien [576] report an efficient voltage control of SOT in a W/CoFeB/MgO stack with PMA that arises from both a decrease in the coercivity field of the ferromagnet and increase in the damping-like torque efficiency. In contrast, by using perpendicularly magnetized IrMn/CoFeB/MgO stacks, Li et al. [577] observe that, while VCMA helps reducing the switching current, the damping-like torque decreases with applied voltage, becoming detrimental for the switching current reduction.

SOTs produced by charge-spin conversion were also showed to be tunable by ferroelectricity. In Pt/CoNiCo/Pt/PMN-PT heterostructures, by switching the in-plane ferroelectric polarization of the PMN-PT substrate, the chirality of the current-induced magnetization switching curves is reversed [578]. The ferroelectric polarization is argued to generate an additional, switchable SOT in the CoNiCo.

# 5. Devices

## 5.1. Spintronic devices for logic and memory based on electrical control of magnetism

### 5.1.1. From Toggle MRAM to SOT-MRAM

Nowadays, with the growing demand for big-data storage and processing, a highly efficient and low power processing of massive data becomes a major challenge which is difficult to reach with conventional electronic components. The separation of memory and processor units in conventional Von-Neumann architectures causes long memory access latency, limited memory bandwidth and large power dissipation known as "Memory Wall" and "Power Wall" [579–582]. Therefore, to break this bottleneck, processing in memory has reignited great interest and is stimulated by the development of nonvolatile memories such as the spintronic MRAM and the Magneto-Electric Spin-Orbit (MESO) devices. The STT-MRAMs, in production since a few years in major electronic companies, already begin to contribute to some reduction of the huge energy consumption and significant contribution to global warming by all the information and communication technologies (about 10% of the worldwide electricity production today, about 20% expected for 2030 [16], cf **Figure 4**).

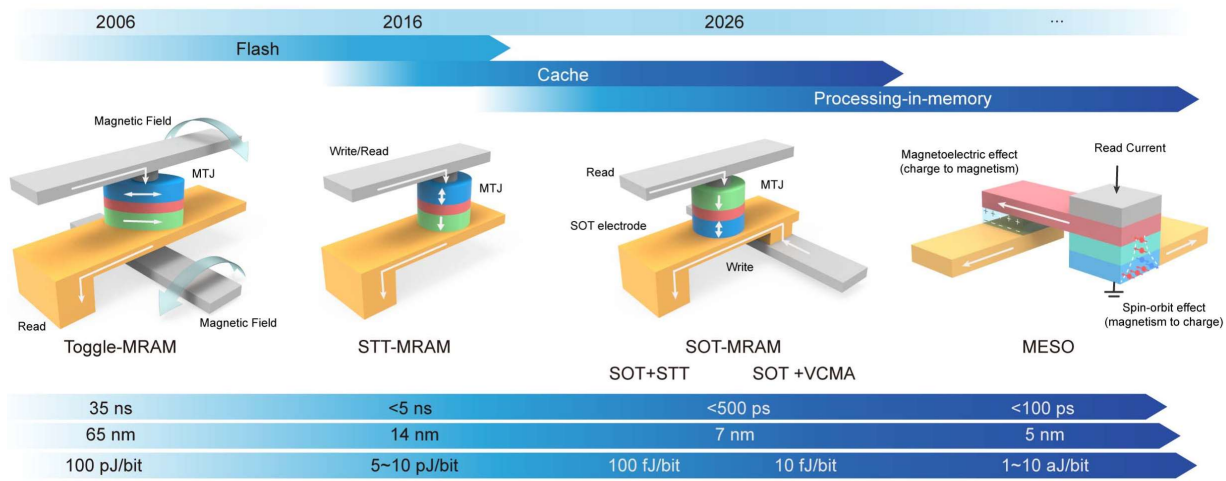

*Figure 43. Roadmap for spintronic logic and memory devices and advances to higher write speed, smaller size, lower power dissipation in the direction of processing-in-memory from the Toggle-MRAM on the market since 2006 to the STT-MRAM in production today and the SOT-MRAM or MESO devices expected for the next generations. Adapted from [579].*

A roadmap for spintronic logic and memory devices is displayed in **Figure 43**. In almost all MRAMs, the memory is associated with the relative orientations of the magnetization in the free layer and reference layer of a MTJ, and the main differences are in the writing process. Toggle-MRAM [583], on the market since 2006, are written by the magnetic field generated by currents in additional lines. The increase of the critical switching field with downsizing of the Toogle-MRAM and the resulting increase of driving currents degrades the power consumption performance at small sizes. However, due mainly to their radiation hardness and wide temperature range, the Toggle-MRAM have been significantly adopted in some technologies for avionics, space and defense.

The memory of a STT-MRAM is written by the action of the STT generated by a vertical current in the structure, as discussed in Section 3.2. The magnetizations can be in-plane (IMA) or out-of-plane (PMA). IMA requires a shape anisotropy (ellipse or rectangle) to generate an easy axis of magnetization and the resulting thermal stability. However, at small sizes, the shape anisotropy is not large enough to provide enough thermal stability. Consequently, STT-MRAMs with PMA are more adapted for downsizing and low dissipation [579]. It is the type of STT-MRAM developed today by the electronic industry. The STT-MRAMs are of high interest to replace embedded Flash and DRAM memories. In addition, with technology nodes of STT-MRAM scaling down to 10 nm and write speed reaching the ns range, they have also a possible interest to replace the relatively large SRAM in logic circuits (Processing-in-memory).

The promising MRAMs for the next generation are the SOT-MRAMs with writing by SOT (damping-like torque). As described in section 3.3.2, the advantage of the SOT with PMA is the time scale for switching and writing which can be in the ns range or shorter. Both heavy metals (Pt, Ta, W, etc.) and 2DEGs at Rashba interfaces or surface/interfaces states of TI/Dirac semi-metals have been tested as spin source. As discussed in section 3.3.2, the generation of spin current by 2DEGs can be more efficient than with heavy metals, at least if the shunting by the magnetic layers or the bulk part of the spin-orbit coupling material can be controlled at a low level. Growth by molecular beam epitaxy (MBE) can give interface of better quality but sputtering ($\alpha$-Sn) has also led to good results.

However, with PMA, a difficulty for switching by SOT is the generation of the needed in-plane field. In section 3.3.4, we have described how field-free switching can be achieved by exchange bias coupling with an antiferromagnetic material or by combining SOT with STT. Very recently, it has been theoretically and experimentally demonstrated that the combination of SOT and STT enables sub-ns ultrafast and low-power magnetization switching through a proper timing scheme [584,585].

Another solution is SOT with VCMA in which a voltage pulse changes the interfacial magnetic anisotropy [575,586,587]. The reorientation of the magnetization and field-like torque induces precessions between the two stable magnetization states and allows the magnetic switching. In addition, with no current trough the MTJ, this solution is of interest for dissipation reduction. Recently, Grimaldi et al. [588] showed that the combination of SOT, STT and VCMA leads to reproducible sub-ns switching with a narrow distribution of the switching times. The study was performed in a perpendicularly magnetized MTJ (top-pinned CoFeB/MgO/CoFeB free layer) deposited on a β-phase W current line by simultaneously applying a bias in the MTJ and a current in the W line [588]. Such a combination reaches an energy efficiency comparable to that of STT, with the main advantage of SOT for switching in the sub-ns range [589]. Finally, spintronic reconfigurable logic gates based on SOT and VCMA have been also proposed and tested for several types of logic operations [590].

Other efforts were devoted recently to the introduction of concepts of two-terminal devices having advantages on the three-terminal device displayed in **Figure 43** in term of downscaling of the structure. An example of two-terminal SOT-MRAM using an in-plane current not only to write by the SOT induced by the SHE of Pt but also to read by in-plane current and GMR was reported by [591].

A comparison between the properties of current volatile devices (DRAM, SRAM) and perpendicular STT-MRAM and SOT-MRAM is presented in **Table 3**.

|  | Volatile | | | | Non-volatile | |
|---|---|---|---|---|---|---|
|  | **DRAM 10x** | **HP-SRAM 5nm** | **HD-SRAM 5 nm** | **HD-SRAM 7 nm** | **pSTT 35 nm WER** | **SOT 35 nm** |
| Techn/node | 10x | 5 nm | 5 nm | 7 nm | 5 nm | 5 nm |
| Write energy/bit (fJ) | 89 | 19 | 76 | 70 | <500/375 | 75 |
| Read energy/bit (fJ) | 58 | 17 | 55 | 50 | 60/52 | 15 |
| Write latency (ns) | 10 | >1 | 2.75 | 2.5 | >10/7.5 | 1.2 |
| Read Latency (ns) | 10 | >1 | 2.5 | 2.2 | 3.5/3.5 | 1 |
| Cell size (μm$^2$) | 0.0026 | 0.034 | 0.0267 | 0.0422 | 0.014/0.009 | 0.0282 |

*Table 3. Comparison of the properties of volatile memory technologies and perpendicular STT-MRAM and SOT-MRAM at advanced CMOS technology modes (7 nm and 5 nm). The numbers for SRAM and DRAM are for current technologies and those for STT-RAM (pSTT 35nm WER column) and SOT-RAM are extrapolated to optimized devices. Adapted from Dieny et al. [592].*

We end this section by pointing out that MRAMs are commercial products that are entering the consumer electronics market. For instance, Sony's CXD5605 GPS receiver uses a 8 MB MRAM chip manufactured by Samsung (28 nm node) and is used in Huawei's GT2 smartwatch. Another example is Ambiq's Apollo, a system-on-a-chip for the internet of things, that uses one 2 MB and one 1 MB MRAM chips [593]. A much larger market may open for MRAM if they can scale beyond 22 nm which is believed to be the limit for embedded Flash memories [594].

### 5.1.2. Multiferroic junctions

Parallel to MTJs, another type of tunnel device, consisting of an ultrathin ferroelectric layer sandwiched between two metallic electrodes [595] was more recently investigated [596,597]. In such ferroelectric tunnel junctions, the reversal of the ferroelectric polarization by an external electric field can produce a large change of the tunnel transmission due to electrostatic effects (if there is any asymmetry between the two interfaces) [598], an effect called tunnel electroresistance [599–601]. Merging ferroelectric and MTJs, in so-called multiferroic tunnel junctions consisting of a ferroelectric tunnel barrier sandwiched by two ferromagnetic electrodes, gives rise to a four-resistance state memory due to the combined tunnel electroresistance and tunnel magnetoresistance effects related to the two ferroic orders.

The existence of a four-state memory was first experimentally reported using a multiferroic (ferroelectric and ferromagnetic) tunnel barrier of $La_{0.1}Bi_{0.9}MnO_3$ sandwiched between $La_{0.7}Sr_{0.3}MnO_3$ and Au electrodes [100]. Resorting to pure ferroelectric and ferromagnetic materials is probably more adequate for this type of multiferroic devices as it should in principle allow room-temperature operation (high ordering temperature in traditional ferroelectric materials), as well as a more efficient magnetic decoupling between the barrier and the magnetic electrode. In addition, interfacial magnetoelectric coupling between the ferroelectric tunnel barrier and the ferromagnetic electrode can be detected by measuring the variations of tunnel magnetoresistance induced by ferroelectric polarization reversal. For instance, large interfacial magnetoelectric coupling was predicted as a result of a modification of the bonding at the $Fe/BaTiO_3$ interface, with sizeable changes of the Fe and Ti-induced magnetic moments when reversing the ferroelectric polarization [233]. Experiments using $Fe/BaTiO_3$ (1.2 nm)/$La_{0.7}Sr_{0.3}MnO_3$ tunnel junctions confirmed these predictions with large changes of the tunnel magnetoresistance (of up to 450%) depending on the ferroelectric polarization state of the tunnel barrier (**Figure 44**a) [236]. The tunnel magnetoresistance is high (low) when the $BaTiO_3$ polarization points towards Fe ($La_{0.7}Sr_{0.3}MnO_3$), in agreement with electric-field induced modifications of the spin polarization at the $Fe/BaTiO_3$ interface [235]. Thus, the electric-field control of the polarization of the ferroelectric tunnel barrier provides a way to control the spin-polarization in a non-volatile way and with low energy.

Radaelli *et al.* demonstrated that ferroelectric polarization reversal at the $Fe/BaTiO_3$ interface controls the magnetic interaction of the interfacial ultrathin FeO [602], suggesting an alternative scenario for the large changes of tunnel magnetoresistance reported in $Fe/BaTiO_3/La_{0.7}Sr_{0.3}MnO_3$: when the ferroelectric polarization points toward Fe, ferromagnetism in FeO promotes a significant spin-polarization while when it points away from Fe, antiferromagnetism in FeO results in a low effective spin-polarization. Later on, it was shown that the sign of the tunnel magnetoresistance can even be reversed by switching the ferroelectric polarization in $Co/PbZr_{0.2}Ti_{0.8}O_3$ (3.2 nm)/$La_{0.7}Sr_{0.3}MnO_3$ tunnel junctions [603]. Although, the tunnel magnetoresistance is not large in these particular devices, its relative variation with the ferroelectric polarization reaches -230%. The ferroelectric tunnel junction can not only be used as a simple binary non-volatile resistive memory encoded by the two saturated states of polarization, but also as a memristor related to the presence of multiple non-uniform configurations of ferroelectric domains [604]. Consequently, a multilevel state of tunnel magnetoresistance (varying from -3% to -30%) was reported for $Co/PbTiO_3$ (4.8 nm)/$La_{0.7}Sr_{0.3}MnO_3$ junctions, by progressively tuning the ferroelectric domain population under voltage pulses (**Figure 44**b) [605].

In some cases, ferroelectric polarization reversal can even trigger interfacial phase transitions as it was suggested for $La_{0.7}Sr_{0.3}MnO_3/La_{0.5}Ca_{0.5}MnO_3$ (0.8 nm)$/BaTiO_3/La_{0.7}Sr_{0.3}MnO_3$ [199]. The polarization-induced metal/insulator phase transition in $La_{0.5}Ca_{0.5}MnO_3$ is accompanied by a ferromagnetic/antiferromagnetic transition, giving rise to a change of the tunnel magnetoresistance from about 100% when the ferroelectric polarization points towards $La_{0.5}Ca_{0.5}MnO_3$ (ferromagnetic state) to nearly zero when it points away from $La_{0.5}Ca_{0.5}MnO_3$ (antiferromagnetic state). Therefore, driving an interfacial magnetic phase transition with the ferroelectric polarization of the tunnel barrier is an efficient way to control the spin-polarization of the tunnel current. More recently, it was shown that spin reconstructions at the interfaces of a $La_{0.7}Sr_{0.3}MnO_3/BaTiO_3/La_{0.7}Sr_{0.3}MnO_3$ multiferroic tunnel junction result in a spin filtering effect that can be turned on and off by reversing the ferroelectric polarization [606]. This tunable spin filter enables a giant electrical modulation of the tunneling magnetoresistance between 10% and 1000%. Alternatively, multiferroic tunnel junctions including an organic ferroelectric barrier of PVDF were investigated. Interestingly, the tunnel magnetoresistance of these $Co/PVDF/La_{0.6}Sr_{0.4}MnO_3$ junctions changes its sign when the ferroelectric polarization is reversed (**Figure 44**c), which is interpreted by a change of sign of the spin-polarization at the Co/PVDF interface [607].

As all the above-mentioned experiments on multiferroic tunnel junctions use an epitaxial oxide perovskite of $La_{0.7}Sr_{0.3}MnO_3$ as a bottom electrode, a sizeable tunnel magnetoresistance is only limited to low temperature [608,609], preventing their potential for applications. Other material combinations including transition metals and their alloys and new ferroelectric materials (such as HfOx, two-dimensional ferroelectrics, etc.) should be investigated thoroughly to develop efficient ferroelectric control of spin-polarization at room temperature. In this vein, first-principle calculations performed on van der Waals multiferroic tunnel junctions combining two-dimensional ferroelectric $In_2Se_3$ and ferromagnetic $Fe_nGeTe_2$ have recently predicted multiple resistance states with sizeable tunnel magnetoresistance and electroresistance, together with low resistance area products (<1 $\Omega \cdot \mu m^2$) [610].

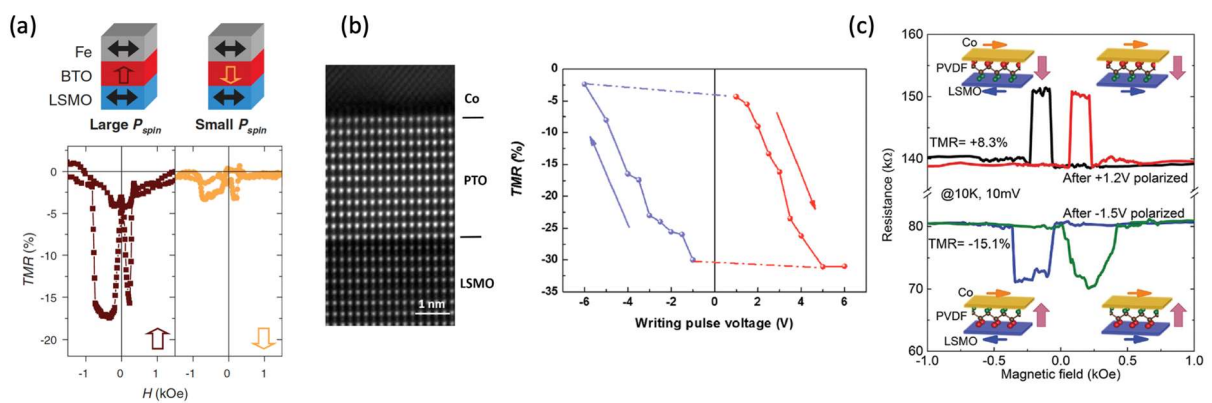

*Figure 44. (a) Ferroelectric control of the tunnel magnetoresistance in Fe/BaTiO₃/La₀.₇Sr₀.₃MnO₃ tunnel junctions. (top) The orientation of the ferroelectric polarization of the tunnel barrier controls the spin polarization at the Fe/BaTiO₃ interface. (bottom) Tunnel magnetoresistance (4.2 K, 50 mV) for both polarization states (after ±1 V, 1 s pulses). From [236]. (b) (left) Annular dark field scanning transmission microscopy cross section image of the Co/PbTiO₃/La₀.₇Sr₀.₃MnO₃ junction. (right) Hysteretic dependence of the tunnel magnetoresistance (10 K, 10 mV) with the polarization state of the PbTiO₃ that is controlled*

*with various pulse voltages (50 μs). From [605]. (c) Tunnel magnetoresistance of a Co/PVDF/La$_{0.6}$Sr$_{0.4}$MnO$_3$ junction (10 K, 10 mV) after polarizing the PVDF downward (+1.2 V) and upward (-1.5 V). From [607].*

### 5.1.3. Magnetoelectric memories (e.g. GMR on top of MR/piezo)

The MRAM outperforms other non-volatile memory technologies in terms of reading/writing speed and endurance. However, writing the magnetic states either by spin-transfer or spin-orbit torques, requires high current densities, which limits the scalability of these devices. Therefore, several schemes of magnetoelectric RAMs (MeRAMs), involving electric-field control of magnetization rather than current-based control, were proposed in the late 2000s.

One of them consisted in applying an electric-field across the antiferromagnetic magnetoelectric $Cr_2O_3$ during a cooling step through its Néel temperature, to tune the exchange bias onto an adjacent Co/Pt multilayer of an MRAM [611]. A simpler concept proposed by Bibes and Barthélémy consisted of using an antiferromagnetic and ferroelectric multiferroic (such as $BiFeO_3$) exchange-coupled to one of the ferromagnetic layers of a spin-valve [32]. In this three-terminal device, the electric field applied across the multiferroic thin film switches the ferroelectric polarization and the antiferromagnetic order via the magnetoelectric coupling [57,68]. Switching of the antiferromagnetic multiferroic modifies the exchange coupling to the ferromagnetic layer, and ideally reverses its direction by 180 degrees at zero magnetic field. This magnetization reversal is then probed electrically by the two-terminal current-perpendicular-to-plane giant magnetoresistance. Allibe et al. explored experimentally this concept and reduced the leakage of the multiferroic $BiFeO_3$ film while preserving the exchange bias to a metallic ferromagnet [612], and demonstrated the first electric-field control of the giant magnetoresistance in Co/Cu/CoFeB/$BiFeO_3$ magnetoelectric devices, although the effect was not reversible [613]. By optimizing the quality of the $BiFeO_3$ multiferroic thin films and using an in-plane geometry for the switching of polarization, Heron et al. demonstrated, in a two-step process for the switching of polarization, a deterministic switching of ferromagnetism and detected a hysteretic variation of the resistance of a Pt/Co$_{0.9}$Fe$_{0.1}$/Cu/Co$_{0.9}$Fe$_{0.1}$ spin valve as a function of the voltage applied to the $BiFeO_3$ (**Figure 45**a) [157], see also **Figure 13**.

Another approach proposed by Pertsev and Kohlstedt consisted in using strain resulting from the voltage applied across the piezoelectric ferroelectric to control the magnetization direction of a magnetostrictive electrode of a MTJ [614]. Using phase simulations, Hu et al. further extended the concept of a strain-mediated MeRAM and simulated low write energy (0.16 fJ/bit) together with potentially high memory density (88 Gb inch$^{-2}$) on MRAMs composed of magnetostrictive Ni coupled to relaxor lead magnesium niobate-lead titanate [615]. Lei et al. demonstrated that voltage-driven strain effects from a Pb(Zr,Ti)$O_3$ gate can be used to pin the domain wall propagation in a magnetostrictive CoFeB magnetic wire [616]. The resulting coercive field change of this free CoFeB magnetic layer is then probed by the modifications of the giant magnetoresistance of IrMn/Co/Cu/CoFeB as a function of voltage (**Figure 45**b). The butterfly hysteretic voltage loop of the propagation magnetic field of the CoFeB layer is correlated to capacitance vs voltage hysteresis loops of the Pb(Zr,Ti)$O_3$, supporting that strain-driven magnetoelectric effects are controlling the spintronic device.

The same kind of geometry was used to control the giant magnetoresistance of Co/Cu/Fe spin valves on $BaTiO_3$ single crystals [617]. Using an IrMn/CoFeB/AlO$_x$/CoFeB MTJ on Pb(Mg$_{1/3}$Nb$_{2/3}$)$_{0.7}$Ti$_{0.3}$O$_3$, Li et al. demonstrated a volatile 90-degree rotation of the free CoFeB layer by applying a vertical electric field to the (011) ferroelectric substrate, which resulted in modifications of the tunnel magnetoresistance under electric field [618]. A similar volatile strain-mediated MeRAM was then proposed with CoFeB/MgO/CoFeB MTJs on $Pb(Mg_{1/3}Nb_{2/3})_{0.7}Ti_{0.3}O_3$ by using a local gating scheme [619]. More recently, Chen et al. demonstrated a large (55%), reversible and non-volatile change of the tunnel magnetoresistance of CoFeB/MgO/CoFeB on $Pb(Mg_{1/3}Nb_{2/3})_{0.7}Ti_{0.3}O_3$, without the need for a magnetic field (**Figure 45**c) [620]. This was achieved by the electric-field induced remanent magnetization rotation by 90 degrees of the CoFeB top free layer via strain-mediated magnetoelectric coupling (sketch in **Figure 45**c). Using a similar stack but combining two pairs of in-plane electrodes on the ferroelectric (**Figure 46**a), Chen et al. later demonstrated a full control of the in-plane magnetic anisotropy of the CoFeB free layer by the electric-field-induced in-plane strain (**Figure 46**b) [621]. By combining voltage sequences to the different gate electrodes, they achieved a complete non-volatile 180-degree rotation of the free magnetic layer, accompanied with 200% resistance contrast without any external magnetic field (**Figure 46**c).

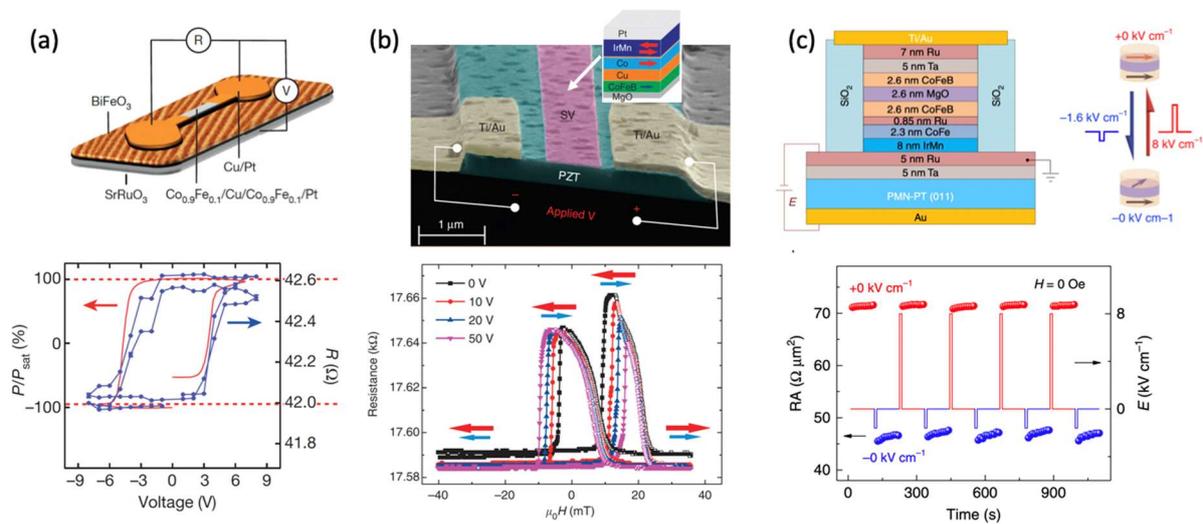

*Figure 45.* (a) (top) Schematic of the magnetoelectric device consisting of a $Co_{0.9}Fe_{0.1}/Cu/Co_{0.9}Fe_{0.1}$ spin valve on $BiFeO_3$. (bottom) Two R(V) loops under zero magnetic field along with a ferroelectric loop (red line) from a neighboring device. From [157]. (b) (top) Sketch of the spin-valve (SV) stack and cross section of the device measured by scanning electron microscopy. (bottom) Giant magnetoresistance loops with different applied voltages, which starts from a depolarized state of the $Pb(Zr,Ti)O_3$ layer. From [616]. (c) (top) Schematic of the MTJ device structure deposited on $Pb(Mg_{1/3}Nb_{2/3})_{0.7}Ti_{0.3}O_3$. (bottom) Repeatable bistable remanent resistance states modulated by 8 kV cm$^{-1}$ and –1.6 kV cm$^{-1}$ electric field pulses in the absence of a bias magnetic field. From [620]. RA is the resistance-area product.

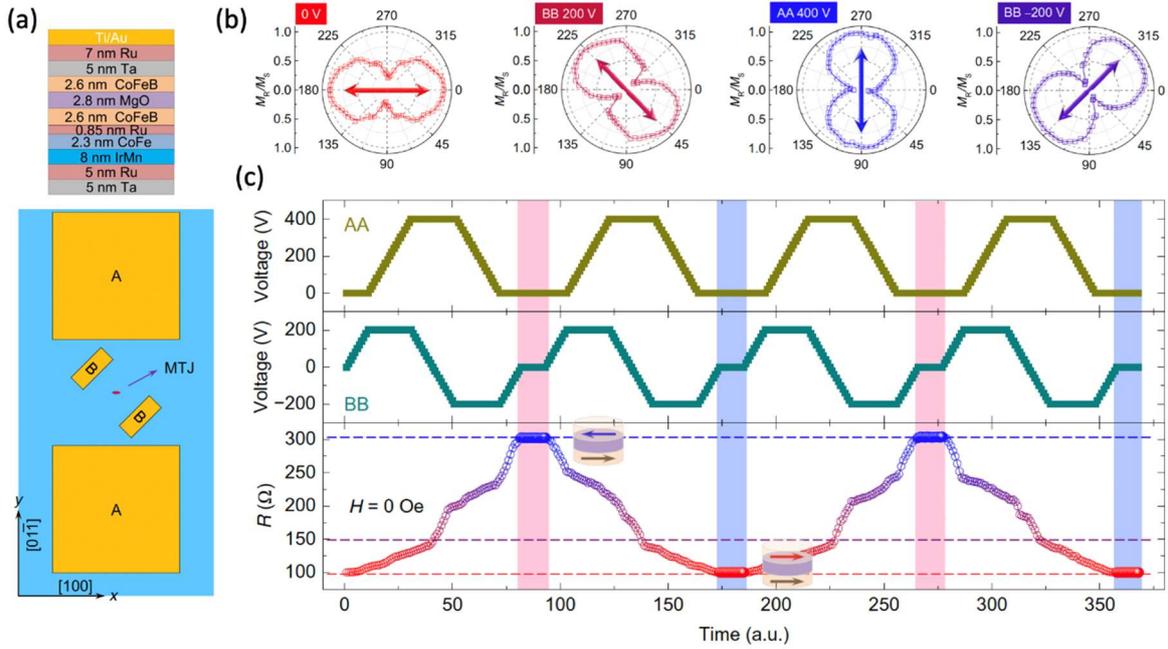

*Figure 46.* (a) Detailed structure of the MTJ and schematic top view of the sample structure with two pairs of AA and BB electrodes. The major axis of the elliptical device was along the x axis. The pinning direction of the MTJ was along the [100] direction of the PMN-PT substrate (+x axis). (b) Polar curves of the angular-dependent $M_R/M_S$ of a CoFeB layer, when the applied voltages were 0 V, BB 200 V, AA 400 V, and BB −200 V. The [100] direction of the $Pb(Mg_{1/3}Nb_{2/3})_{0.7}Ti_{0.3}O_3$ substrate corresponds to 0. The double-headed arrows indicate the direction of the magnetic easy axis. (c) Dependence of the resistance of the tunnel junction on voltage synergistically applied to the AA and BB electrode pairs at H = 0 Oe. The reversible resistance switching between high- and low-resistance states corresponds to the antiparallel and parallel magnetization configurations of the MTJ, as illustrated by the insets, which indicates the 180-degree magnetization switching of the free layer driven by voltage. From [621].

### 5.1.4. MESO devices

In 2019, Intel proposed a new concept of logic device coined MESO (for Magneto-Electric Spin-Orbit) [8] which they argue could result in 10 to 30 times higher efficiency and 5 times higher logic density compared to CMOS. MESO is expected to strongly reduce power consumption for computation by harnessing ferroic materials that have embedded non-volatility and by relying on a voltage rather than a current to switch the ferroic order parameter [2,622]. A sketch of MESO is shown in **Figure 47**. The core of MESO is a ferromagnetic element whose magnetization is switched thanks to a magnetoelectric element at the input. The output comprises a spin-orbit element that converts a spin current injected into it from the ferromagnet into a charge current (through the ISHE or the IEE), allowing to read the information stored by the magnetization state in the ferromagnet. MESO is a logic-in-memory concept and individual MESO elements are concatenable, i.e., the output line of one element can be used as the input line of the next one. This is possible because MESO operates with and generates bipolar currents (with positive or negative signs), unlike CMOS devices. For MESO-based architectures to benefit from concatenation, the SO module must generate an output voltage of at least 100 mV, while the ME module must switch with 100 mV or less. To satisfy both these conditions is extremely challenging. In particular, the scarcity of multiferroic materials practically imposes using $BiFeO_3$ (or slightly modified or doped versions of it) for the ME module. For the SO module to generate >100 mV, the SO element must not only possess a very high spin-charge interconversion efficiency but also a high resistance [623].

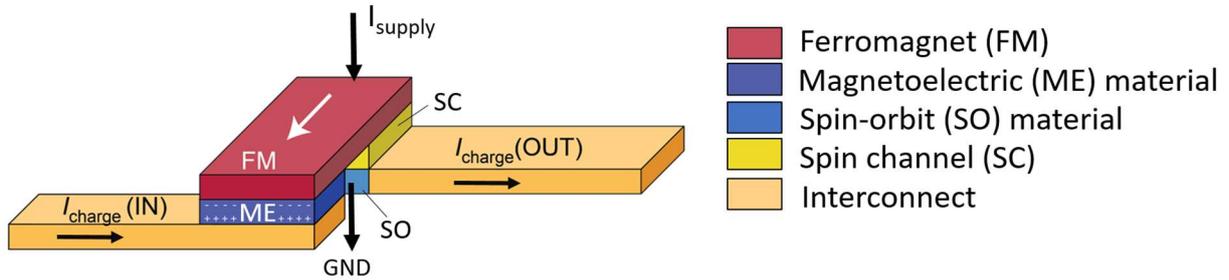

*Figure 47. Sketch of a MESO device, adapted from [8].*

Efforts towards a first proof-of-concept MESO have involved optimizing devices [623–625] with a T-shaped geometry. A prototype combining $BiFeO_3$, CoFe and Pt has been recently presented [248,626], cf. **Figure 48**. As visible in **Figure 48**b, the output resistance of the Pt element displays two different levels depending on the magnetization of the CoFe ferromagnetic element. Applying a voltage to the ME element (**Figure 48**c) switches the magnetization of the CoFe, which results in two different output voltage levels in the Pt **Figure 48**d).

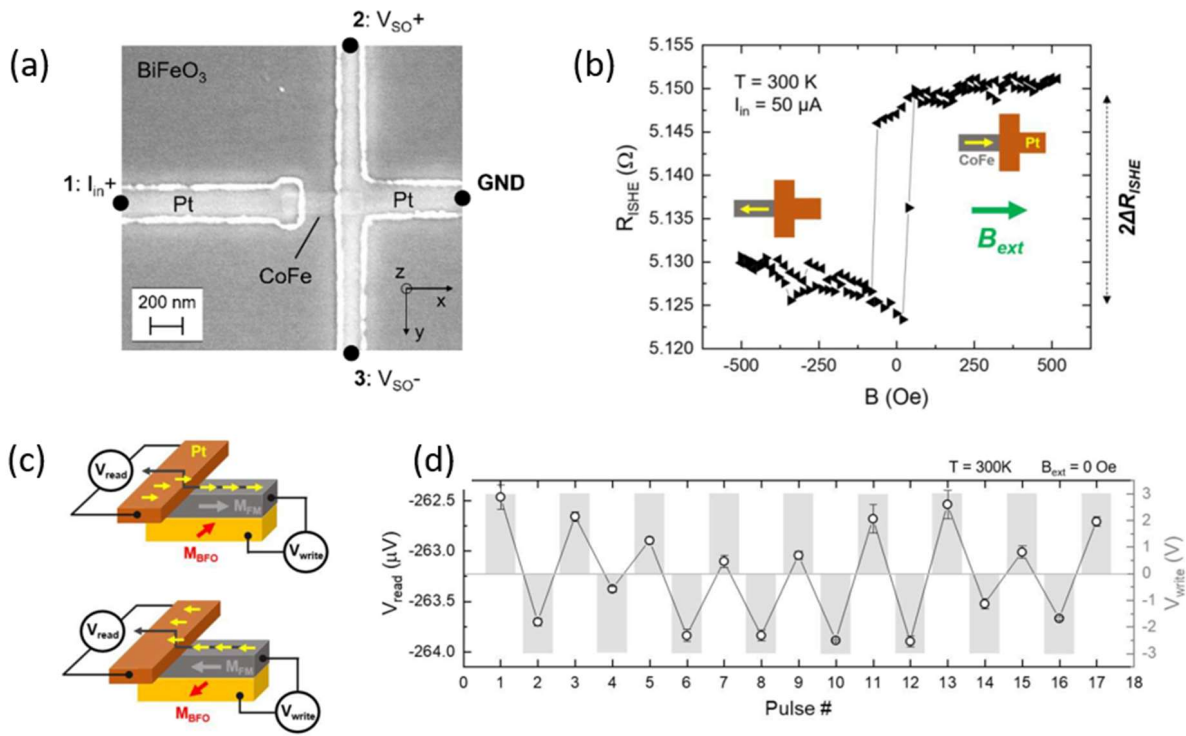

*Figure 48. (a) Scanning electron microscope image of the SO module device region. The CoFe element dimensions are 500 nm × 100 nm × 2.5 nm (length, width, thickness). An applied charge current $I_{in}$, between contact 1 and GND, becomes spin-polarized and is injected in the T-shaped Pt structure through a 100 nm × 100 nm junction. Due to the ISHE, an output voltage $V_{SO}$ is detected between contacts 2 and 3. (b) Output signal of the SO module, obtained from the transverse resistance $R_{ISHE}$ as a function of an external magnetic field $B_{ext}$. The two magnetization states of the CoFe element, with an amplitude of $2\Delta R_{ISHE}$, are depicted in the inset by the yellow arrows. (c) Sketch for the full MESO operation at room temperature, without any external magnetic field applied, shown in panel e. Voltage pulses $V_{write}$ drives $BiFeO_3$ magnetization switching ($M_{BFO}$) and subsequent magnetization $M_{FM}$ reversal of the ferromagnetic*

*element. $M_{FM}$ is electrically read through ISHE in the Pt element. (d) The output signal $V_{read}$ changes by ~1.5 μV for $V_{write}$ = ±3 V, reflecting opposite $M_{FM}$ orientations. After each pulse, the magnetization state is read 3 times (with intervals of 1 second) and averaged (from [248]).*

## 5.2. Spin-torque nano-oscillators and spin diodes

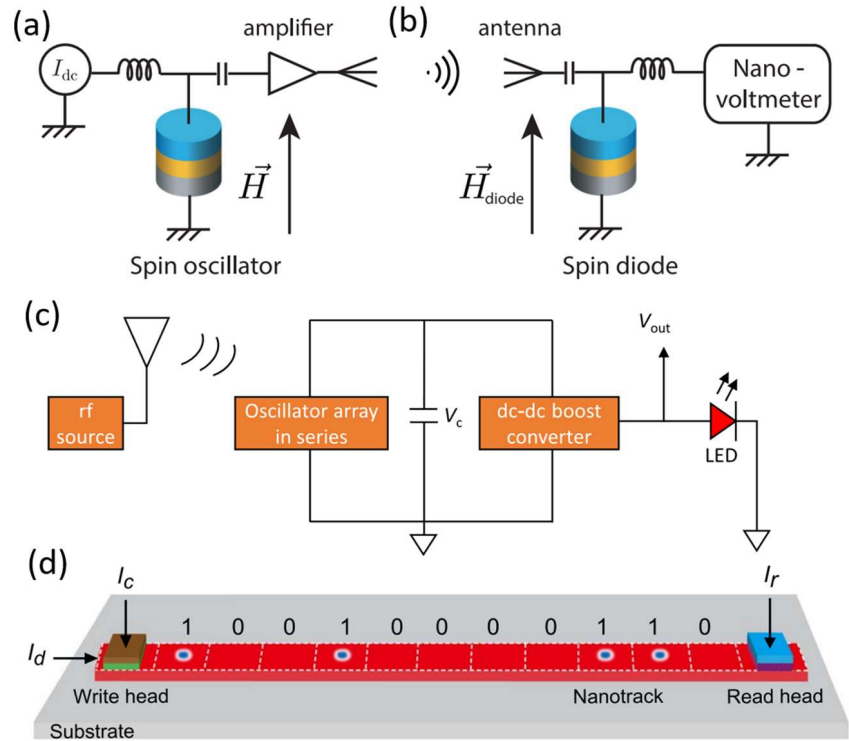

*Figure 49. Applications of spin-torque nano-oscillators (STNO). (a-b) Schematics of STNO in the function of spin oscillator for rf emission in (a) and spin-diode for conversion from rf to dc in (b) [627]. (c) Schematic of circuit with arrays of eight spin-diodes used for energy harvesting and lightning the LED in the right [628]. (d) Schematic of a skyrmion-based racetrack memory [629].*

Spin-torque nano-oscillators (STNO) based on today's-standard MTJs can be used in two ways as illustrated by **Figure 49**a-b. They can be efficient nanoscale rf emitters, as described in Section 3.2, and they can also act as spin-diodes, that is nanoscale transducers from RF to DC in which an input rf signal, rf field or rf spin-torque, induces magnetization oscillations that are in turn converted into a DC voltage via a magnetoresistive effect [630]. The recent advances have led to active developments of communication and signal processing systems exploiting the frequency tunability, the nanoscale size and the multifunctionality of the STNO [631–636]. The RF detection bandwidth of the MTJ based spin diode devices make them comparable or even better in performance in comparison with the semiconductor Schottky diode. A first approach is based on resonant passive approach with sensitivity approaching 1000 V/W [633], a conversion efficiency larger than state-of-the art Schottky diodes. The sensitivity can be further amplified through dc spin transfer effects [634] or spin bolometer effect reaching sensitivity up $4.4 \times 10^6$ V/W in the sub-GHz region [635]. Another strategy has been to harness magnetic configurations showing larger susceptibility [636] and/or non-linear response [637] that results into broadband rectification effect, up to a few GHz, an important feature for their use for energy harvesting [628,637], as illustrated by **Figure 49**c. It is also important to mention the development of arrays of nanoscale STNO in which the emission by a given STNO can be detected by other STNOs [627,638], an interesting result for the design of circuits and chips based on STNO communication through microwave. A promising development is the exploitation of such arrays of STNO to the development of spintronic neural networks [639,640].

### 5.3. Devices based on skyrmions and DWs

Many devices harnessing magnetic skyrmions have been proposed during the last decade. The best known, illustrated by **Figure 49**d, is the skyrmion racetrack memory [465,470,641] based on the same principle as the racetrack memory with magnetic domain walls proposed by Parkin [642]. The information can be encoded by a sequence of individual skyrmions which can be moved in a magnetic track between write head (injector) where the skyrmions are injected and read head (detector) where there are detected (**Figure 49**d). The diameter of skyrmions can be as small of 10 nm or less and, in addition, can be compressed by decreasing the track width [465]. As the spacing between neighboring skyrmions in a track can be of the order of the skyrmion diameter, one can expect a higher density with skyrmions than with DW in a racetrack memory [643].

The most convenient way to put the skyrmions into motion is the SOT generated by SHE in, for example, a heavy metal layer [465,470,641]. Velocities up to the order of 100 m/s can be obtained with realistic current densities. The lateral component of the velocity (skyrmion Hall effect) can be suppressed by working with coupled skyrmions in antiferromagnetic arrangements of layers [475]. It can be also obtained in sufficiently narrow tracks when the repulsion by the edges keeps the skyrmion in the center of the track. Another advantage of skyrmions is that their motion by spin torques will be similar in straight tracks or in curved ones as they are guided by the confinement from the edges, whereas the motion of DWs will be affected in curved parts of the racetrack because the torques will act differently in the wall at the inner and the outer parts of the track.

The skyrmions can be injected in the track by current pulses through nanocontacts or also deleted by opposite pulses [643,644]. They can be detected at the read head by sensing the change of Hall voltage induced by the skyrmion (Anomalous Hall Effect or Topological Hall Effect) [473], through the TMR of a tunnel junction deposited on the track or by transport effects specific associated to the topological nature of skyrmions, e.g. non-collinear magnetoresistance. Interestingly, this concept of skyrmion racetrack can be easily transformed and adapted to become a nanoscale voltage gate skyrmion transistor. This new function has been proposed by X. Zhang et al [645] by adding a gate in a given part of track in order to locally modify though the application of an electric field, the magnetic properties of the magnetic media, being the perpendicular anisotropy or the DMI and thus controlling the passing or not of a skyrmion equivalent of the "on/off" switch of a transistor.

Finally, it can be noted that skyrmions have been proposed not only for conventional storage of information in racetrack memories but also to implement reservoir computing models in recursive neural networks of neuromorphic computer [643].

In addition to devices based only on skyrmions, the transformation of skyrmions into domain walls and vice-versa in track of varying width has been proposed for concepts of logic gate for conventional computing [646]. Finally, another type of application of skyrmions is the magnonic crystal based on a periodic and reconfigurable arrangement of skyrmions [647].

# 6. Perspectives

Electric-field control of the magnetization direction at room temperature is now clear with the voltage required to accomplish this dropping down to 0.5 V. To get to an aJ switch, it is critical to reduce these switching voltages down even further (100 mV and below) in conjunction with a switching charge density of ~10 µC/cm$^2$. How robust can this be, especially with respect to repeated cycling of the electric and magnetic states? In this regard, as in the field of ferroelectric thin films [648] for memory applications, it appears that we need to increase the focus on the nature of the ferromagnet and its interface to the multiferroic. Prior experience with ferroelectric capacitors has shown that a conducting oxide contact yields a very robust capacitor; in a similar vein, we expect an oxide ferromagnet to form a more robust contact to the oxide multiferroic or piezoelectric. Thus, there is an urgent need to discover and interface an oxide ferromagnet that couples magnetically to the multiferroic at room temperature. A template for this is already available from the work on La$_{0.7}$Sr$_{0.3}$MnO$_3$/BiFeO$_3$ interfaces, which display robust electric-field control of the magnetization direction, albeit at 100 K. Can double perovskites, such as Sr$_2$(Fe,Mo)O$_6$ [649,650] or Sr$_2$(Cr,Re)O$_6$ [651] be possible alternatives to the La$_{0.7}$Sr$_{0.3}$MnO$_3$ system? In the same vein, it is highly important to discover more room-temperature multiferroics so that one can explore multiple pathways to use these novel functionalities. Computational discovery platforms such as the Materials Genomics approach driven by machine learning pathways [652] should be particularly valuable in this endeavor. The confluence of crystal chemistry, computational discovery and atomically precise synthesis is a potent combination that has already shown to lead to unexpected phenomena [653].

In this sense, tremendous progress has been made in understanding chemistry-structure-property relationships, and in engineering specific atomic architectures, so that an era of "multiferroic materials by design" is already underway. In particular, targeted functionalities, such as large magnetization and polarization and even exotic polarization topologies, are now within reach. For magnetoelecric devices to be technologically competitive will therefore require precise growth of ultra-thin films guided by theoretical studies to exactly define the chemical compositions needed to optimize the polarization and coercive field. This will require improved fundamental understanding, which can be facilitated by improved first- and second-principles methods. Even with such a low-field-switching breakthrough, scale-up and integration, in particular compatibility with existing silicon processing methods, and integration with the appropriate peripheral electronics are key challenges.

| Science | Technology |
|---|---|
| • Room temperature multiferroics with robust coupling between magnetism and ferroelectricity and high remanent magnetic moment | • Thermal stability of ferroelectric and magnetic order parameters, as well as robust coupling between them, in 10nm length-scales at room temperature |
| • New magnetoelectric coupling mechanisms and understanding and approaching the limits of such phenomena. | •Reducing the voltage required for ferroelectric / magnetoelectric switching to ~100mV |
| • Quantitative measurements of magnetoelectric and multiferroic coupling at 10nm length scales<br><br>• Reaching the theoretical Landauer limit for switching (kT(ln2)) would be desirable and will require significant effort<br><br>• Atomic-scale design and layer-by-layer growth to discover and synthesize new multiferroics<br><br>• Understanding the limits, controlling and exploiting dynamics<br><br>• Are there convergences between multiferroics and other correlated electron materials/phenomena?<br><br>• Search for materials with efficient conversion from charge to spin current by SHE or IEE at room temperature.<br><br>• Better control of Rashba interfaces and surfaces/interfaces of topological insulators or Dirac semimetals.. Mastering a simple and efficient way for field-free switching of perpendicular magnetization by SOT.. Better understanding and control of nucleation and current-induced motion of skyrmions.<br><br>• Mastering the synchronization of large assemblies of STNOs for additive outputs.<br><br>• Developing reliable methods to raise the ordering temperature of 2D magnets well above room temperatures.<br><br>• Exploring the advantages for spin-orbitronics coming for the combination of spin-orbit coupling and broken inversion symmetry in single layers or at interfaces of van der Waal stacks.<br><br>• Extension of experiments of magnetization switching by SOT to magnetic insulators, TmIG and others. | • AttoJoule switch: designing proper ferroelectric multiferroics with small but stable spontaneous polarizations of ~1-5 µC/cm$^2$<br><br>• Integration and scale-up of synthetic approaches to enable manufacturing would be valuable.<br><br>• Speeding up the development of SOT-RAMs, (SOT + STT)-RAMs, (SOT + VCMA)-RAMs and devices integrating logic and memory functions.<br><br>• Development of logic and memory devices combining ferroelectric and ferromagnetic materials.<br><br>• Development of STNO-based devices for harvesting of ambient rf energy<br><br>• Developments of STNO-based devices for neuromorphic computing.<br><br>• Development of racetrack memories based on DW or skyrmions.<br><br>• Development of the application of skyrmions for logic and memory devices as well as for elements for neuromorphic computing.<br><br>• Development of application of arrangements of skyrmions in magnonic devices.<br><br>• Development of high-speed light-induced SOT-RAMs |
| • Better understanding of the generation of light-induced spin currents for their exploitation for current-induced torques.<br><br>• Better understanding of light-induced terahertz emission from magnetic materials and multilayers.<br><br>• Exploring the potential of pure orbital currents for the control of magnetization in the emerging field of orbitronics. | |

*Table 4. Challenges for the science and technology of multiferroic and magnetoelectric architectures.*

The recent discovery of polar vortices and skyrmions in ferroelectric superlattices presents another tantalizing opportunity to create analogous, coupled spin-charge textures out of multiferroics such as BiFeO$_3$ [304,654,655]. This could present a unique pathway to overcome the antiferromagnetic ground state through such curling patterns spin/dipolar patterns, as illustrated for the case of polar vortices and skyrmions in PbTiO$_3$/SrTiO$_3$ superlattices [304]. A first set of studies have been carried out to explore the possibility of forming polar textures in the BiFeO$_3$ system [655]. Imposing electrostatic boundary conditions by interfacing to a lattice matched, non-polar La- BiFeO$_3$ however, leads to the formation of an array of 109° domains as well as stabilizing an anti-polar structure in the BiFeO$_3$ layer [656]. These results seem to suggest that while the idea of imposing electrostatic boundary conditions, does work in a general sense, the consequences are governed more by the structural details, particularly the octahedral tilts, that are such a key component of the crystal structure of BiFeO$_3$. The rather surprising outcome of the formation of the anti-polar structure can be rationalized through the fact that the electrostatic energy is more than sufficient to raise the free energy of the polar phase above that of the antipolar phase. Indeed, this seems to be a hallmark of the BiFeO$_3$ system, where a number of phases are within a close proximity in energy scale to the ground state [657].

An aspect that would benefit from a detailed crystal chemistry based phase equilibrium study is the stabilization of metastable phases; for example, one could be looking for polytypoids (phases that have the same crystal structure but different chemical/stacking sequence, for example Y-Si-Al-O-N's or the polytypes in SiC) [658] of the BiFeO$_3$ composition or chemically distinct derivatives thereof. Two examples of this could be : (i) based on the hexagonal BaM type layered ferrites [122], (ii) the Ruddelsen-Popper type perovskites or the Aurivillius type phases [659]. This magnetoeletric behavior has been demonstrated in the hexagonal ferrites [122]. Further, chemically substituted Aurivillius phases have been known to exhibit magnetoelectric behavior, although the magnetic state is not a robust ground state (more like a spin glass) [325]. On this note, it seems worthwhile to start with ferrimagnets (such as the layered hexaferrites) and attempt to induce a robust ferroelectric state into them, through chemical substitution or epitaxy. Charge ordering transitions, such as the Vervey transition in Fe$_3$O$_4$, were thought to lead to breaking inversion symmetry [660]; demonstrating a robust magnetoelectric effect in such systems should be a focus for research in the coming years. 2D materials represent a huge space of opportunities for magnetoelectricity, either by combining 2D magnets with 2D ferroelectrics [661], or by designing 2D multiferroic materials [662]. A possible route to reach efficient control of magnetization with an electric field at room-temperature is also by using hybrid magnetoelectric multiferroics, with superlattices made of ferroelectric and ferromagnetic materials. Combining strain-driven improper ferroelectrics with ferrites is an interesting material choice to achieve this goal.

What are the limits on the length scales of the spin-charge coupling? For example, can we manipulate the spin state of a single ion using an electric field? Recent work in this direction is poised to impact not only the fundamental physics of spin-orbit coupling and its coherent manipulation with an electric field, but also has the potential to impact the field of quantum computing in which all of the operations are carried out using an electric field [663].

We expect dynamical effects in multiferroics to increase in importance over the next years, driven by new experimental capabilities such as ultrafast X-ray sources [664], and we expect that fundamental limits on the dynamics of spin-charge-lattice coupling phenomena will be established. Theoretical proposals of dynamical multiferroic phenomena, in which a time-dependent polarization induces a magnetization in the reciprocal manner from that in which spin spirals induce polarization [330] should be validated by careful experiments. At the same time, more work on antiferromagnetic resonance in multiferroics is required; while many studies were carried out in the 1960s [318] and 1970s on conventional antiferromagnets, activity with modern multiferroics, which typically have higher resonance frequencies (~700 GHz in $BiFeO_3$) [70,72,665], compared with ~350 GHz in other perovskite orthoferrites [318]), has been scarce.

It is clear that the field of multiferroics and magnetoelectrics is poised to make further significant breakthroughs and we hope that this article motivates additional research on this fascinating class of materials and their applications. While scientific interest in the field is beyond question, our community needs to identify market niches and enable pathways to products, so that multiferroics go beyond being an "area to watch" and address contemporary technological challenges. To achieve this, a shift of focus from fundamental materials discoveries to translational research and development will be needed, similar to that which occurred in the field of GaN-based light-emitting diodes two decades ago. The complexity of oxide-based material systems raises particular additional challenges, as we have seen for example in the colossal magnetoresistive manganites, making the active engagement of applied physicists and device engineers early in the research and development process even more essential. In this vein, the recent engagement of large microelectronic companies in the field of multiferroics [8] is particularly encouraging. While basic research in multiferroics is vibrant, the field would benefit from an injection of focused programs that address the transition to devices, in particular scale-up and integration issues.

For the control of magnetism by current-induced torques, the advances have been very fast during the recent years, especially on the manipulation of magnetization by SOT [417]. The market entry of high-performance components of the SOT-MRAM type can be expected soon, first at the cache level, later in processing-in-memory structures, as described in Section 5.2. Some last questions must be solved, related to the field-free switching of perpendicularly magnetized layers by SOT (see Section 3.3.4), the combined use of electric field and current-induced torques in VCMA devices (see Section 4.3) or the combined use of magnetoelectric effects and spin-charge interconversion in MESO devices (see Section 5.1.4). Although the results of **Figure 48** demonstrate the feasibility of MESO, much work is needed to increase the output voltage difference. In particular, it appears that optimizing the output signal based on heavy metals such as Pt or Ta will not be enough owing to their low resistivity. Rather, working with two-dimensional systems such as 2DEGs, surface states of topological insulators, or graphene/TMD van der Waals heterostructures is a more promising route owing to their large spin-charge interconversion efficiency ($\lambda_{IEE}$ or $\lambda_{SHE}$) as well as high resistivity [623]. In parallel, in the field of neuromorphic computing, several concepts of nanoscale neuron or synapse components based on SOT have been successfully tested recently and their development as devices by the electronic industry can be expected in the next decade.

Although the next generation of devices will probably use heavy metals as the source of spin current, better performances can be expected in a second stage, again by the use of 2DEGs at the surface of topological insulators or Dirac semimetals and at Rashba interfaces, as well from the introduction of 2D materials. Some results on topological insulators and Dirac semimetals are very promising (see **Table 2**) but their integration into devices can be a long way, after a better control of the interplay between bulk and surface/interface contributions to the production of spin current and improvements in the fabrication/integration processes. On the fundamental research side, advances can come from the use of magnetic materials other than transition metal and associated alloys (Co, CoFeB, etc.) or alloys combining rare-earth and transition metals (TbFe, etc). In these classical magnetic materials, the conduction is by s and d electrons, and mainly by s electrons which do not have SOC. Recently, record DW velocities have been obtained in magnetic alloys with p carriers as nitrides of Mn [454]. Other types of magnetic materials with p conduction could be explored, as, for example, the transition metal dichalcogenides. The use of antiferromagnets as the magnetic material is another promising direction, with the advantage of having no net magnetization, which makes them insensitive to spurious magnetic fields and thus very robust as memory elements, while they can be written by current-induced torques (or electric fields).

These recent years have also seen the demonstration of the remarkable properties of the 2D materials, particularly the 2D magnets, as described in Section 3.6. The control of magnetism in layered magnets with an electric field has a strong potential, since the atomically thick materials can be more sensitive to electric field than normal thin films, with the additional advantage to obtain almost ideal interfaces when stacking them with other van der Waals materials (such as the 2D materials with efficient spin-charge interconversion mentioned in the previous paragraph). Regarding voltage control of magnetism present in these atomically thick materials, some attempts have been performed [666,667] to integrate the voltage-induced switching of the magnetic order of $CrI_3$ (see section 2.5) in a device that shows non-volatility and could be an alternative in MRAM applications. Regarding current-induced torques, the performance of 2D-magnet-based devices requires small current density and small applied fields (a comparison between the potential of 3D and 2D magnets for switching by SOT can be seen in **Figure 35**d), although the small electrical signal for reading the magnetic state of the semiconducting 2D magnet (based on spin Hall magnetoresistance) will need to be improved. So far, the obvious drawback of 2D magnets is their ordering temperature, generally below room temperature. However, recent works have shown that this temperature in some systems can be raised by proximity effect with another 2D material [482] or by electric fields [255]. If this possibility becomes more largely accessible, the 2D magnets will become also promising materials for the electrical control of magnetization.

Another emerging direction for the current-induced control of magnetization is the possibility of exploiting orbital currents which carry orbital angular momentum rather than the usual spin currents carrying the intrinsic angular momentum. They can be generated by the orbital Hall effect, which is expected to be larger than the spin Hall effect, even in transition metals with weak spin-orbit coupling [668,669]. Likewise, the orbital equivalent of the Edelstein effect (orbital Edelstein effect) is predicted to generate a current-induced orbital magnetization [670–674]. The orbital currents generated in a nonmagnetic material could efficiently exert a torque when injected on a ferromagnet. For this to occur, spin-orbit coupling is needed to convert the orbit current into a spin current. For this purpose, one could use a middle layer with strong spin-orbit coupling between the nonmagnetic metal and the ferromagnet [675] or could directly use a ferromagnet with strong spin-orbit coupling [676]. This new field of research, called *orbitronics*, might open the door to a plethora of materials and interfaces, not considered before because of their lack of SOC, to be used to achieve large current-induced torques. Recently, light induced orbit currents have also been used for efficient terahertz emission [677,678].

Finally, although this review has been devoted to the control of magnetization by electric field and electrical currents, it is quite probable that we will see soon an interplay of these performant electrical controls and additional controls by light to go in the direction of faster speeds and better energy efficiency. The most recent experiments show that the magnetization of a magnetic layer can be controlled by a ultra-short laser pulse. The magnetization can be switched with a single non-polarized laser pulse in specific ferrimagnetic materials such as GdCo, GdFeCo [679] or Tb/Co multilayer [680]. Moreover, a large variety of materials (ferrimagnetic, ferromagnetic, synthetic antiferromagnets, granular media...) can also be switched by circularly polarized laser pulses [681]. Those types of all optical switching effects could be applied, for example, to switch the magnetization of one layer in a MTJ stack to change the magnetic state of a MRAM in which one of the electrodes is made with one of these ferrimagnetic materials. More recently however, it was demonstrated that the out-of-plane magnetization of a standard ferromagnetic layer (such as Co, Co/Ni, Co/Pt) can be electronically switched by the transmission of the spin-polarized current generated by a light pulse on a GdFeCo layer (without switching the magnetization of GdFeCo) [682]. This hybrid way, combining the generation of spin-polarized ultra-short current pulse by light in a first magnetic layer and the switching of a second magnetic layer by spin current injection, could be used for the writing of MRAM based on the optimized materials of today. Anyway, by using direct or indirect control of magnetization by light, it turns out that future generations of ultrafast devices will probably combine the performant electrical controls we have described in this review with direct or indirect controls by light.

**Acknowledgements**

We have written this article on behalf of many collaborators, co-workers, students and postdocs worldwide and acknowledge their intellectual participation and contribution. The rapid pace of development in this field means that it is impossible to acknowledge and cite each of them independently. We encourage the interested reader to look at the review articles cited in this paper as well as reach out to us if we can be of further assistance. We also warmly thank Frédéric Nguyen Van Dau for proofreading the article and making many valuable suggestions to improve it. Our work would not have been possible without the sustained support of federal and industrial funding agencies. In particular, we acknowledge the support by Intel Corporation through the "FEINMAN" Intel Science

Technology Center. R.R. would like to acknowledge the sustained support of the U.S. Department of Energy, Basic Energy Sciences Office, the Semiconductor Research Corporation's JUMP Initiative, the National Science Foundation, specifically the MRSEC program, the Army Research Office. M.B. acknowledges support from the European Research Council through AdG n°833973 "FRESCO", the French Agence Nationale de Recherche project "CONTRABASS", the M-ERANet project "SWIPE" and Intel Corp. F.C. thanks the support from the Spanish MICINN under Project RTI2018-094861-B-100 and under the Maria de Maeztu Units of Excellence Programme (MDM-2016-0618 and CEX2020-001038-M), from the European Union H2020 under the Marie Slodowska Curie Actions (955671-SPEAR), and from the 'Valleytronics' Intel Science Technology Center.